\documentclass[twocolumn,amsmath,amssymb,10pt,superscriptaddress,a5paper,letterpaper,fleqn]{revtex4-1}
\usepackage{amssymb}
\usepackage{epsfig}
\usepackage{graphicx}
\usepackage{dcolumn}
\usepackage{array}
\usepackage{bm}
\usepackage{fancyheadings}
\usepackage{longtable}
\usepackage{multirow}
\usepackage{float}
\usepackage{afterpage}  
\usepackage{color}

\usepackage[warnundef]{jabbrv}
\DefineJournalAbbreviation{Test}{Tes} % <-- Note that there is no period!
\setlongtables

\begin{document}
\setcounter{page}{1}
%%
%% ----------------------------------------  Start to modify the template here ----------------------------------
%%   Change title, authors, affiliations and abstract as they should appear on the title page of your paper.
%%
\title{
     \qquad \\ \qquad \\ \qquad \\  \qquad \\  \qquad \\ \qquad \\ 
 IAEA Photonuclear Data Library 2019
}

\author{T.~Kawano}
   \email[Corresponding author: ]{kawano@lanl.gov}
   \affiliation{Theoretical Division, Los Alamos National Laboratory, Los Alamos, NM 87545, USA}

\author{Y.~S. Cho}
   \affiliation{Nuclear Data Center, Korea Atomic Energy Research Institute, Daedeok-Daero 989-111, Yuseong-gu, Daejeon, Korea}

\author{P.~Dimitriou}
   \affiliation{NPAC-Nuclear Data Secion, International Atomic Energy Agency, PO Box 100, 1400 Vienna, Austria}

\author{D.~Filipescu}
   \affiliation{Horia Hulubei National Institute for Physics and Nuclear Engineering (IFIN-HH), 30 Reactorului, Bucharest-Magurele 077125, Romania}

\author{N.~Iwamoto}
   \affiliation{Nuclear Data Center, Japan Atomic Energy Agency, Tokai-mura, Ibaraki 319-1195, Japan}

\author{V.~Plujko}
   \affiliation{Nuclear Physics Department, Taras Shevchenko National University, Kyiv, Ukraine}

\author{X.~Tao}
   \affiliation{China Nuclear Data Center, China Institute of Atomic Energy, P.O. Box 275(41), Beijing 102413, China}

\author{H.~Utsunomiya}
   \affiliation{Department of Physics, Konan University, Okamoto 8-9-1, Higashinada, Kobe 658-8501, Japan}

\author{V.~Varlamov}
   \affiliation{Lomonosov Moscow State University, Skobeltsyn Institute of Nuclear Physics, Moscow 119991, Russia}

\author{R.~Xu}
   \affiliation{China Nuclear Data Center, China Institute of Atomic Energy, P.O. Box 275(41), Beijing 102413, China}

\author{R.~Capote}
   \affiliation{NPAC-Nuclear Data Secion, International Atomic Energy Agency, PO Box 100, 1400 Vienna, Austria}

\author{I.~Gheorghe}
   \affiliation{Horia Hulubei National Institute for Physics and Nuclear Engineering (IFIN-HH), 30 Reactorului, Bucharest-Magurele 077125, Romania}

\author{O.~Gorbachenko}
   \affiliation{Nuclear Physics Department, Taras Shevchenko National University, Kyiv, Ukraine}

\author{Y.L. Jin}
   \affiliation{China Nuclear Data Center, China Institute of Atomic Energy, P.O. Box 275(41), Beijing 102413, China}

\author{T. Renstr{\o}m}
   \affiliation{University of Oslo, Sem Saelands vel 24, P.O. Box 1048, Oslo 0316, Norway}

\author{K.~Stopani}
   \affiliation{Lomonosov Moscow State University, Skobeltsyn Institute of Nuclear Physics, Moscow 119991, Russia}

\author{Y. Tian}
   \affiliation{China Nuclear Data Center, China Institute of Atomic Energy, P.O. Box 275(41), Beijing 102413, China}

\author{G. M. Tveten}
   \affiliation{University of Oslo, Sem Saelands vel 24, P.O. Box 1048, Oslo 0316, Norway}

\author{J.M.~Wang}
   \affiliation{China Nuclear Data Center, China Institute of Atomic Energy, P.O. Box 275(41), Beijing 102413, China}

\author{T.~Belgya}
   \affiliation{Centre for Energy Research, Hungarian Academy of Sciences, Konkoly Thege Miklos 29-33, 1525 Budapest, Hungary}

\author{R.~Firestone}
   \affiliation{University of California, Berkeley CA 94720, USA}

\author{S.~Goriely}
   \affiliation{Institut d'Astronomie et d'Astrophysique, Universit\'e Libre de Bruxelles, Campus de la Plaine, CP 226, 1050 Brussels, Belgium}

\author{J.~Kopecky}
   \affiliation{JUKO Research, Kalmanstraat 4, Alkmaar 1817, The Netherlands}

\author{M.~Krti\v{c}ka}
   \affiliation{Charles University, V Hole\v{s}ovi\v{c}k\'{a}ch 2, 18000 Prague, Czech Republic}

\author{R.~Schwengner}
   \affiliation{Helmholtz Zentrum Dresden-Rossendorf, Bautzner Landstrasse 400, 01328 Dresden, Germany}

\author{S.~Siem}
   \affiliation{University of Oslo, Sem Saelands vel 24, P.O. Box 1048, Oslo 0316, Norway}

\author{M.~Wiedeking}
   \affiliation{iThemba LABS, P.O. Box 722, Somerset West, 7129, South Africa}
\date{\today}
%% Do not touch the line below, it serves editorial purposes. 
   \received{xx July 2018; revised received xx September 2018; accepted xx October 2018}

\begin{abstract}
Photo-induced reaction cross section data are of importance for a
variety of current or emerging applications, such as radiation
shielding design and radiation transport analyses, calculations of
absorbed dose in the human body during radiotherapy, physics and
technology of fission reactors (influence of photo-reactions on neutron
balance) and fusion reactors (plasma diagnostics and shielding),
activation analyses, safeguards and inspection technologies, nuclear
waste transmutation, medical isotope production and astrophysical
applications.

To address these data needs the IAEA Photonuclear Data library was
produced in 2000, containing evaluated photo-induced cross sections
and neutron spectra for 164 nuclides which were deemed relevant for
the applications.

Since the release of the IAEA Photonuclear Data Library however, new
experimental data as well as new methods to assess the reliability of
experimental cross sections have become available. Theoretical models
and input parameters used to evaluate photo-induced reactions have
improved significantly over the years. In addition, new measurements
of partial photoneutron cross sections using mono-energetic photon
beams and advanced neutron detection systems have been performed
allowing for the validation of the evaluations and assessments of the
experimental data. Furthermore, technological advances have led to the
construction of new and more powerful gamma-beam facilities, therefore
new data needs are emerging.

We report our coordinated efforts to address these data needs and
present the results of the new evaluations of more than 200 nuclides
included in the new updated IAEA Photonuclear Data Library, where
the photon energy goes up to 200~MeV. We discuss
the new assessment method and make recommendations to the user
community in cases where the experimental data are discrepant and the
assessments disagree. In addition, in the absence of experimental
data, we present model predictions for photo-induced reaction cross
section on nuclides of potential interest to medical radioisotope
production.

\end{abstract}
\maketitle

%% Running header appears after the \maketitle command, make sure to modify as suggested below. 
\lhead{IAEA Photonuclear Data $\dots$}          % Use 2-3 initial words of the title of your paper
\chead{NUCLEAR DATA SHEETS}                             % Do not touch this line
\rhead{T. Kawano \textit{et al.}} % Put the lead author of your paper here
\lfoot{}                                                           
\rfoot{}                                                          
\renewcommand{\footrulewidth}{0.4pt}
\tableofcontents{}

%%
%%-------------------------------------------------------- Body of the paper follows -------------------------------------------------------------------
%%

\section{INTRODUCTION}
\label{sec:introduction}
Photonuclear data describing interactions of photons with atomic
nuclei are important for a range of applications such as (i) radiation
shielding and radiation transport analyses, (ii) calculation of
absorbed doses in the human body during radiotherapy, (iii) activation
analyses, (iv) safeguards and inspection technologies, (v) nuclear
waste transmutation, (vi) fission and fusion reactor technologies, and
(vii) astrophysical nucleosynthesis.

Photons can be produced as bremsstrahlung radiation by electron
accelerators which are commonly used in hospitals, industries and
laboratories.  Significant technological advances led to the
production of quasi-monochromatic beams using the positron
annihilation in flight technique at the national facilities of the
Lawrence Livermore National Laboratory (LLNL, USA) and Centre
d'\'{E}tudes Nucl\'{e}aires de Saclay (France) in the 1960's. As a
result, a large number of measurements of photonuclear cross sections
was performed at these two facilities for almost three decades (1962
-- 1987). This pioneering work was captured by B.L. Berman in his
comprehensive compilation of photonuclear cross sections ``Atlas of
Photoneutron Cross-sections obtained by Monoenergetic Photons''
published in 1975~\cite{Berman1975b}, and later in the follow-up
produced in 1988~\cite{Dietrich1988}. Note that because these
institutes changed their names several times, we herewith abbreviate them
simply to Livermore and Saclay.

Despite those experimental efforts, there was still a need for evaluated
photonuclear data since (a) it is not possible to produce a complete
photonuclear data files based on measured cross sections alone, (b)
often the experimental data suffer from systematic discrepancies which
are not easy to resolve, and (c) there is a lack of data in a number
of cases. These deficiencies in the experimental data can be addressed
and often resolved by performing an evaluation which consists of three
systematic operations: compilation of experimental data, critical
assessment of the measurement techniques used and theoretical
calculations based on reliable nuclear models. As a result of the
evaluation, one can obtain production cross sections as well as energy
and angular distributions of the emitted particle for a wide range of
incident and outgoing energies, which is useful for the applications.

To address the growing needs for photonuclear data, the IAEA held a
Coordinated Research Project (CRP) under the title Compilation and
Evaluation of Photonuclear Data for Applications between 1996 and
1999. This CRP produced the IAEA Photonuclear Data Library which is
described in the Handbook on Photonuclear Data for
Applications~\cite{IAEAPhoto1999} and is available at the IAEA
web-site (http://www-nds.iaea.org/photonuclear). The library includes
photon absorption data, total and partial photo-neutron reaction cross
sections and neutron spectra for 164 isotopes, primarily for
structural, shielding, biological and fissionable materials.
In this paper we refer this previous library to IAEA 1999.

The list of 164 isotopes included in the 1999 Photonculear Data
Library~\cite{IAEAPhoto1999} can be broken down in four categories:
\begin{itemize}
\item Structural, shielding and bremsstrahlung target materials: Be,
  Al, Si, Ti, V, Cr, Fe, Co, Ni, Cu, Zn, Zr, Mo, Sn, Ta, W, and Pb.
\item Biological materials: C, N, O, Na, S, P, Cl, and Ca;
\item Fissionable materials: Th, U, Np, and Pu; and
\item Other materials: H, K, Ge, Sr, Nb, Pd, Ag, Cd, Sb, Te, I, Cs,
  Sm, and Tb.
\end{itemize}
Of the above four categories, the most important for the applications
are the 40 major isotopes of the 29 elements in the first three
groups.

Although this database has been extremely useful to a broad user
community, it has become evident that it needs to be revised since
\begin{itemize}
\item some of the experimental data measured with quasi-monochromatic
  photon beams are unreliable and discrepant,
\item data have been measured for 37 isotopes that have not been
  evaluated,
\item improved methods to resolve experimental discrepancies are
  available, and
\item new data measured with modern techniques have been published in
  recent years.
\end{itemize}

New experimental facilities, such as HI$\gamma$S at the Triangle University
National Laboratory (USA) and the Laser Compton Scattering Facility
(NewSUBARU) at University of Hyogo (Japan), offering highly monoenergetic
photon beams in combination with advancements in neutron detector
technologies have opened the field to new possibilities: new
measurements of photo-neutron cross sections with better accuracy that
are expected to help  resolve the long-standing discrepancies
observed between the data measured using quasi-monoenergetic beams.

Furthermore, the needs for evaluated photonuclear data are growing. In
the field of medical isotope production, photonuclear reactions are
being explored for the production of medical radionuclides. With the
advent of new facilities producing brilliant photon beams with
extremely high activity, photo-production of some important
radionuclides could become a competitive alternative to the
traditional methods using neutrons produced at highly-enriched Uranium
reactors or charged-particle beams~\cite{Habs2011, Nichols2014,
INDC0596, INDC0717}. A list of radionuclides proposed for potential
diagnostic and therapeutic applications in nuclear medicine which have
also been identified as suitable candidates for production via
photonuclear reactions is given below:
\begin{itemize}
\item $^{166,170}$Er for the production of $^{165,169}$Er,
\item $^{187}$Re for the production of $^{186}$Re,
\item $^{226}$Ra for the production of $^{225}$Ra which further decays
  to $^{225}$Ac,
\item $^{98}$Ru fort the production of $^{97}$Ru,
\item  $^{194}$Pt for the production of $^{\rm 193m}$Pt,
\item $^{132}$Xe for the production of $^{131}$I,
\item $^{162}$Dy for the production of $^{161}$Tb, and
\item  $^{178}$Hf for the production of $^{177}$Lu.
\end{itemize}
Other important candidates such as $^{100}$Mo (for the production of
$^{99}$Mo via ($\gamma$,n)), $^{48}$Ca, $^{52}$Cr, $^{65}$Cu,
$^{48}$Ti and $^{46}$Ti are already included in the existing
photonuclear data library.

%  total 220
%  older 164
%  new    56
%         37 exp available
%          9 new demand (incl. model calc?)
%         10 model calc

In view of the above developments, it was timely to update the
existing IAEA Photonuclear Data Library to reflect the progress in the
field and the emerging demands for photonuclear data.  A coordinated
research project (CRP) was endorsed by the International Nuclear Data
Committee at the 2014 meeting in Vienna and was initiated by the IAEA
with the title ``Updating the Photonuclear Data Library.'' All the 164
isotopes in the existing library were revisited and evaluated by
considering new data, the results of experimental-based evaluations as
well as the new Giant Dipole Resonance (GDR) parameters from the
recently updated Atlas of GDR parameters~\cite{Plujko2018} and
improved reaction models. In addition to these 164 isotopes,
evaluations were performed for 37 isotopes for which experimental data
are available, as well as for the 9 isotopes identified in the
above-mentioned list as relevant for medical applications. In some
cases where no experimental information is available, a model
prediction is given as the evaluation.  In total, 220 isotopes were
evaluated. All available experimental data for photo-absorption cross
sections, photo-neutron production cross sections and yields, as well
as for partial photo-neutron cross sections and photo-charged-particle
cross sections that were available in the EXFOR
database~\cite{Otuka2014} up to the cut-off date of April 2019 were
considered. In addition, all the new measurements obtained with the
new direct neutron-multiplicity sorting
technique~\cite{Utsunomiya2017} that was fully developed and
implemented for the CRP, were also taken into account in the
evaluations. The evaluations were extended to energies of 200~MeV for
use in accelerator-driven transmutation technologies, to complement
the neutron and proton high-energy libraries that are being developed
for radiation transport simulation codes.

The CRP has a second branch under the title ``Generating a Reference
Database for Photon Strength Functions'' which has led to a new
database of experimental and calculated photon strength
functions~\cite{Goriely2019}. The evaluations performed for the
updated photonuclear data library have formed the basis for the
recommended photonuclear photon strength functions in Ref.~\cite{Goriely2019}.
In this paper we adopt the same definition of the photon strength functions
as in Ref.~~\cite{Goriely2019}.

In this report, we present the results of both experimental and
evaluation efforts carried out within the CRP to update the IAEA
Photonuclear Data Library.  The measurements of reliable photo-neutron
production cross sections using the new neutron multiplicity sorting
technique as well as an overview of all available experimental data
can be found in Sec.~\ref{sec:expdata}. In Sec.~\ref{sec:models}, we
present the nuclear reaction models that are used to describe
photonuclear reactions as well as the different codes employed in this
coordinated effort.  In Sec.~\ref{sec:evaluation}, we present the
different evaluations and highlight some comparisons with experimental
data. The contents of the new IAEA photonuclear data library are
described in Sec.~\ref{sec:libcontent}.  Our conclusions are given in
Sec.~\ref{sec:conclusions}. The new Atlas of GDR parameters is
provided in the Appendix.

\section{DEFINITIONS}
\label{sec:crosssec}
Experimental photonuclear reaction data are usually obtained by
directly counting the number of emitted particles or by measuring the
residual nucleus activity. For energies above the multi-particle
emission threshold, more than one combination of emitted
light-particles can accompany the same number of produced neutrons or
can lead to the same residual nucleus, respectively. Often, authors do
not mention clearly what it is they measure which can lead to
confusion when comparing experimental data with evaluations. For
example, an experimental data set that is reported as the $(\gamma,n)$
cross section could have charged particles emitted concurrently with
one-neutron emission, {\it i.e.}
$(\gamma,1n)+(\gamma,1np)+(\gamma,n2p)+\ldots$ depending on the
energy. This can lead to significant differences, especially for light
nuclei, where photo-charged-particle emission is quite strong. Here
we define several cross sections relevant to the photonuclear data
library.

When the charged-particle emission is negligible, the measured
one-neutron emission cross section is identical to the cross section
for the production of $(Z,A-1)$ nucleus, and the two-neutron emission
is equal to the production of $(Z,A-2)$, and so on. However, when
charged-particle emission is non-negligible, then the measured
one-neutron emission cross section $ \sigma_{1nX}$ should read
\begin{equation}
  \sigma_{1nX}
   = \sigma(\gamma,1n) +  \sigma(\gamma,np ) +  \sigma(\gamma, n\alpha) + \ldots \ ,
  \label{eq:sig1nx}
\end{equation}
and {\it ditto} for $\sigma(\gamma,2nX)$, $\sigma(\gamma,3nX)$, {\it
etc.}  Thus, $\sigma(\gamma,inX)$ is understood to be the inclusive
i-neutron emission cross section, where $X$ stands for anything except
for $i$-neutrons. These cross sections are not given explicitly in the
evaluated photonuclear data library, instead one has to reconstruct
$\sigma_{1nX}$ by summing each term in Eq.~(\ref{eq:sig1nx}).

The photo-neutron production cross section is defined in two ways,
either involving the neutron multiplicity or not. We denote the
photo-neutron yield (or production) cross section $\sigma_{xn}$ as
\begin{eqnarray}
  \sigma_{xn}
  &=&  \sigma(\gamma,1n) +  \sigma(\gamma,np ) +  \sigma(\gamma, n\alpha) + \ldots \nonumber \\
  &+& 2\sigma(\gamma,2n) + 2\sigma(\gamma,2np) + 2\sigma(\gamma,2n\alpha) + \ldots \nonumber \\
  &+& 3\sigma(\gamma,3n) + 3\sigma(\gamma,3np) + 3\sigma(\gamma,3n\alpha) + \ldots, \nonumber \\
  &=& \sum_i i \sigma_{inX} \ ,
  \label{eq:sigN}
\end{eqnarray}
and the total photo-neutron cross section $\sigma_{Sn}$ as
\begin{eqnarray}
  \sigma_{Sn}
  &=& \sigma(\gamma,1n) + \sigma(\gamma,np ) + \sigma(\gamma, n\alpha) + \ldots \nonumber \\
  &+& \sigma(\gamma,2n) + \sigma(\gamma,2np) + \sigma(\gamma,2n\alpha) + \ldots \nonumber \\
  &+& \sigma(\gamma,3n) + \sigma(\gamma,3np) + \sigma(\gamma,3n\alpha) + \ldots \nonumber \\
  &=& \sum_i \sigma_{inX} \ .
  \label{eq:sigS}
\end{eqnarray}
When the photo-charged particle reaction cross sections ---
$\sigma(\gamma,p)$, $\sigma(\gamma,\alpha)$, {\it etc} --- are negligible,
and no photo-fission occurs, $\sigma_{Sn}$ is the same as the
photo-absorption cross section $\sigma_{abs}$. Otherwise, the sum of
this cross section $\sigma_{Sn}$ with the photo-charged-particle
cross section will give the total photo-absorption cross section as
\begin{eqnarray}
 \sigma_{\rm abs}
 &=& \sigma_{Sn} + \sigma(\gamma,p) + \sigma(\gamma,2p) + \ldots + \nonumber \\
 &+& \sigma(\gamma,d) + \sigma(\gamma,dp) + \dots + \sigma(\gamma,\alpha) + \dots
 \label{eq:siga}
\end{eqnarray}

Usually $\sigma_{xn}$ is explicitly given in the photonuclear data
library as the neutron multiplicity, while $\sigma_{Sn}$ is
implicit. When the target photo-fissions, Eqs.~(\ref{eq:sigN}) and
(\ref{eq:sigS}) have an extra term;
\begin{eqnarray}
  \sigma_{xn}
  &=& \sigma(\gamma,n) + 2\sigma(\gamma,2n) + \ldots + \overline{\nu}\sigma(\gamma,f) \ , \\
  \label{eq:sigNF}
  \sigma_{Sn}
  &=& \sigma(\gamma,n) + \sigma(\gamma,2n) + \ldots + \sigma(\gamma,f) \ ,
  \label{eq:sigSF}
\end{eqnarray}
where $\overline{\nu}$ is the average number of neutrons per fission.

Varlamov {\it et al.} proposed an $F_i$ value~\cite{Varlamov2010}
\begin{equation}
  F_i = \frac{\sigma(\gamma,inX)}{\sigma(\gamma,xn)} \ ,
  \label{eq:Fi}
\end{equation}
to facilitate the assessment and evaluation of the experimental
$\sigma(\gamma,inX)$ cross sections. $F_i$ has a maximum value for
each i-neutron reaction channel: $F_1 < 1$ for $i=1$, $F_2 < 1/2$ for
$i=2$, and $F_3 < 1/3$ for $i=3$. When $F_i$ exceeds these limits then
there are issues in the measurement of $\sigma(\gamma,inx)$ data by
the same group. Since the above-mentioned limits are automatically
satisfied in the theoretical calculations, it is essential to apply a
model-based approach in the evaluation of the available photonuclear
experimental data to avoid such inconsistencies.

\section{AVAILABLE EXPERIMENTAL DATA}
\label{sec:expdata}
The experimental photonuclear reaction data have been obtained in
various types of measurements, with bremsstrahlung and
quasi-monoenergetic photons from positron annihilation in flight and
more recently from laser Compton scattering (LCS). While the
photo-neutron yield cross sections are obtained by counting the total
number of neutrons emitted, the determination of partial photo-neutron
cross sections requires neutron-multiplicity sorting. 
Below we provide a brief description of the experimental procedures most
frequently used in the past~\cite{IAEAPhoto1999} and most recently
developed.

  \subsection{Experiments}
  \label{subsec:experiments}
To determine the photonuclear reaction cross section, one needs to
measure on an absolute scale both the flux of the incident photons and
the number of reaction products. The incident photon flux can be
determined directly with an ionization chamber, scintillator or
solid-state detector or indirectly by normalizing to known reaction
cross sections (for bremsstrahlung beams). The number of reaction
products is determined either by counting the emitted particles or by
measuring the activity of radioactive residual nuclei. We summarize
the characteristics of various $\gamma$-ray sources and photonuclear
cross sections measurements below.

\subsubsection{Bremsstrahlung}
\label{subsubsec:bremsstrahlung}

Bremsstrahlung beams were used in the first measurements of
photo-nuclear cross sections. The continuous bremsstrahlung spectra
were produced by striking a radiator target with an electron beam from
an accelerator (initially betatrons and synchrotrons, and later,
linear accelerators). Several laboratories around the world (mostly in
Russia, Canada and Australia) developed this type of photon beams.

As the spectrum of photon energies is continuous, only the yield of
the reaction can be measured:
\begin{equation}
  Y(E_0) =
     N_R \int_{E_{\rm th}}^{E_0} 
    \frac{\sigma(E_\gamma)}{E_\gamma} W(E_0,E_\gamma) d E_\gamma \ ,
  \label{eq:bremss_yield_reaction}
\end{equation}
where $E_\gamma$ is the photon energy,
$\sigma(E_\gamma)$ is the reaction cross section, $W(E_0,E_\gamma)$ is
the bremsstrahlung energy spectrum, $E_0$ is the end-point energy of
the bremsstrahlung spectrum which is equal to the electron beam
energy, $E_{\rm th}$ is the threshold energy, and $N_R$ is the
normalization coefficient. Changing $E_0$ by small steps allows one to
measure a yield curve and then by applying an ``unfolding'' procedure to
obtain the photo-nuclear reaction cross section.

Several unfolding methods for spectrum-averaged cross sections have
been developed, among which the most widely used ones are:
\begin{itemize}
\item the Photon Difference Method~\cite{Bogdankevich1966} in which
  the difference of two bremsstrahlung spectra near the slightly
  differing end-point energies is interpreted as an almost
  quasi-monoenergetic spectrum; a modified version uses the linear
  combination of three bremsstrahlung spectra~\cite{VanCamp1981} which
  improves the shape of the resulting spectrum;

\item the Penfold-Leiss method~\cite{Penfold1959} in which the
  integral Eq.~(\ref{eq:bremss_yield_reaction}) is replaced by a set
  of linear equations for finite analysis bins; some modifications of
  this method~\cite{Bramanis1972} vary the analysis bin
  depending on the yield accuracy;

\item the regularization method~\cite{Tikhonov1970} in which the mean
  square difference between the reaction cross section and the model
  value is minimized assuming different approaches for smoothing the
  cross section; modifications of this method are based on various
  regularizators~\cite{Cook1963, Geller1963, Turchin1970}.

\end{itemize}

The typical apparatus functions (effective photon spectra) for these
methods have different line shapes localized in photon energy though not close to a 
Gaussian line shape, allowing experimentalists to obtain the
information on the reaction cross section at energy $E_\gamma$ with
energy resolution dependent on the data processing.

The advantage of bremsstrahlung measurements is the large photon beam
intensity, which allows one to obtain reasonable counting statistics
even for relatively small reaction cross sections. However, there are
several disadvantages of using such a technique. First, one needs to
know the bremsstrahlung spectrum sufficiently well for all electron
energies.  Second, measuring a reaction yield curve in small energy
steps requires a stable accelerator and large counting
statistics. Third, the process of subtracting the yield curves in the
unfolding procedure may introduce correlations between the
experimental data points that can lead to unphysical fluctuations in
the unfolded cross sections.

\subsubsection{Positron Annihilation in Flight}
\label{subsubsec:PAiF}

While the unfolding of the experimental reaction yield resulting from
bremsstrahlung measurements may involve a ``mathematical'' method of
obtaining the ``quasi-monoenergetic'' photons, positron annihilation in
flight offered an ``apparatus method'' of producing them with variable
energies. The method~\cite{Tzara1957} was realized in several
laboratories around the world, mostly in LLNL (USA) and Saclay
(France)~\cite{Berman1975}. An intense beam of high-energy electrons
from a linear accelerator hits a thick high-$Z$ converter and produces
bremsstrahlung which undergoes electron-positron pair production in
the converter. The fast positron beam then impinges on a thin, low-$Z$
target and produces a bremsstrahlung spectrum and a peak of
annihilation photons. The Livermore and Saclay facilities were nearly
the same except that while in the Livermore system positrons from the
converter were re-accelerated before undergoing annihilation, in the
Saclay system the positrons were simply separated without
re-acceleration.

To separate the bremsstrahlung component from the annihilation
photons, this kind of measurement proceeded in the following three
steps:
\begin{enumerate}
\item similar to the bremsstrahlung measurement, the reaction
  yield, $Y^+(E_0)$, is obtained with the $\gamma$-ray spectrum that
  consists of the positron annihilation and bremsstrahlung components.
  See Ref.~\cite{Fultz1962b} for the 15.6~MeV positrons on the 0.0060
  in. thick LiH target. The peak energy of the annihilation $\gamma$-rays
  is given by $E_0 + 3mc^2/2$ with the kinetic energy of
  positrons $E_0$ and the electron rest mass energy $mc^2$;

\item the reaction yield $Y^-(E_0)$ is given by bremsstrahlung of electrons with 
  the energy $E_0$; and

\item the reaction cross section is deduced as the difference 
  between the positron and electron reaction yields
  assuming that the bremsstrahlung spectra from positrons
  and electrons are identical to first order approximation; 
  the subtraction is supposedly
  subject to the stability of the accelerator parameters.
\end{enumerate}

Hence the photonuclear cross section reads
\begin{equation}
  \sigma(E_\gamma) = Y^+(E_0)-Y^-(E_0) \ .
  \label{eq:PAiF_sigma}
\end{equation}

The effective photon spectrum for this method is essentially the
positron-annihilation line shape, which may be asymmetric with respect
to the low-energy tail because of the subtraction procedure described
above. The disadvantage of this method, as compared to bremsstrahlung,
is the low intensity of the photon beam attributed to the small cross
sections of the electron-positron pair production and positron
annihilation processes.

\subsubsection{Bremsstrahlung Tagging}
\label{subsubsec:bremsstrahlung_tagging}

As an alternative to the complex task of unfolding the reaction cross
section from the experimental reaction yield by solving the inverse
integral equation of Eq.~(\ref{eq:bremss_yield_reaction}), the method
of producing (quasi)monoenergetic photons by tagging was
proposed~\cite{Cardman1983}. It was implemented at the  University of
Illinois (USA) using the beam from a high duty-cycle electron linear
accelerator. The idea was to produce photons by letting electrons
impinge on a very thin radiator with an energy $E_0$ and then to tag the photons
with scattered electrons with an energy $E_\mathrm{scatt}$. Thus,
photons are tagged with an energy $E_\gamma$,
\begin{equation}
  E_\gamma = E_0 - E_\mathrm{scatt} \ .
  \label{eq:bremss_tagg_egamma}
\end{equation}

The scattered electron is deflected by the magnetic spectrometer to an
array of detectors placed in its focal plane. The energy spread
(resolution) of the photon beam depends on the number of detectors
mounted in focal plane. A thin-sliced part of the bremsstrahlung
spectrum represents the typical effective photon spectrum from this
method, which is close to a Gaussian line shape. partial reaction
cross sections were obtained using the tagged bremsstrahlung photons.

The disadvantage of this method is the low-intensity photon beam extracted from
the very thin bremsstrahlung-producing radiator that makes coincidence
measurements difficult.

\subsubsection{Laser Compton-scattering}
\label{subsubsec:LCS}

\begin{figure}
  \begin{center}
    \resizebox{\columnwidth}{!}{\includegraphics{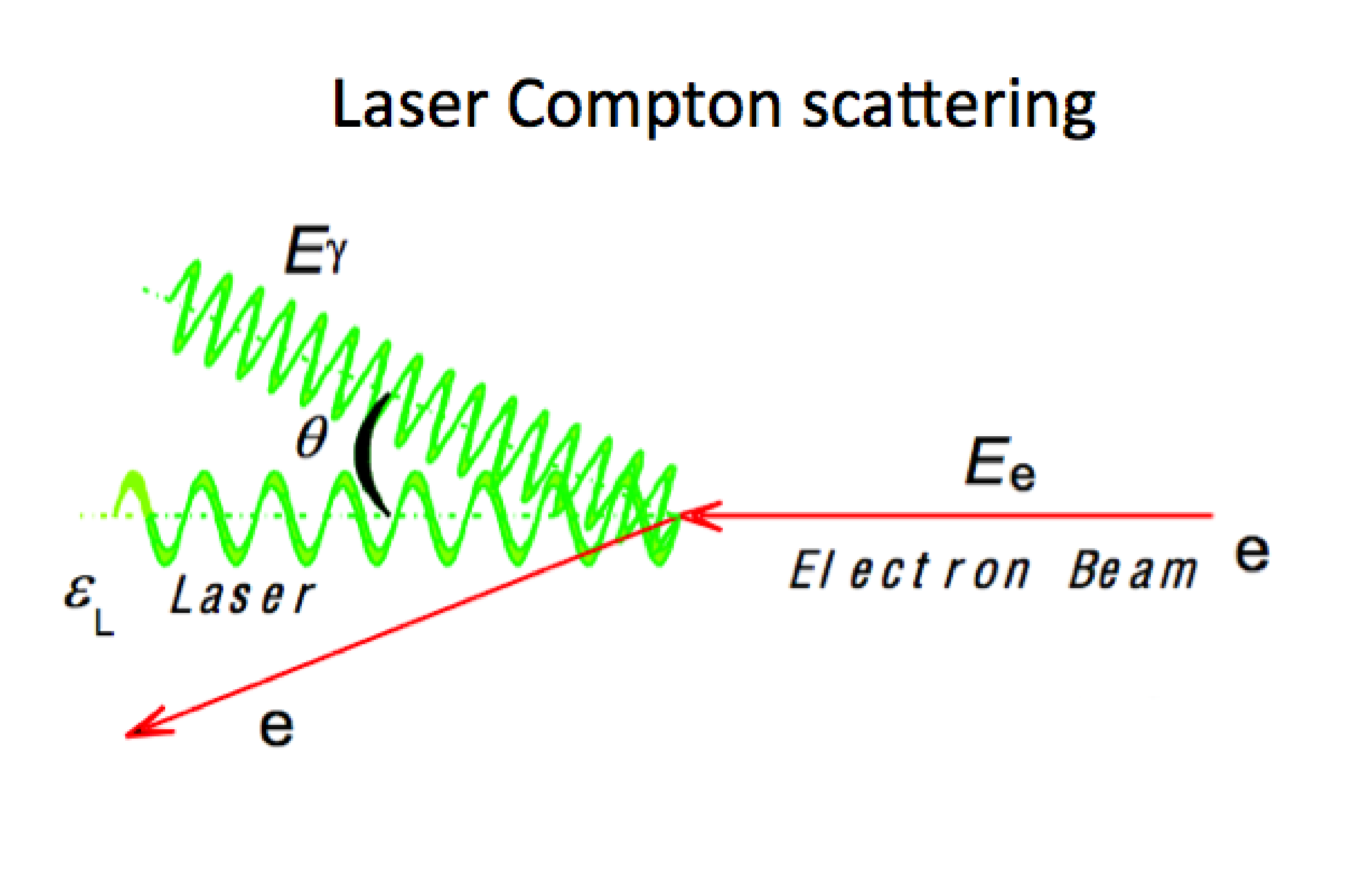}}
  \end{center}
  \caption{(Color online) Laser Compton scattering of laser photons from relativistic electrons.}
  \label{fig:LCS_scatt}
\end{figure}

\begin{figure*}
 \begin{center}
  \resizebox{0.9\textwidth}{!}{\includegraphics{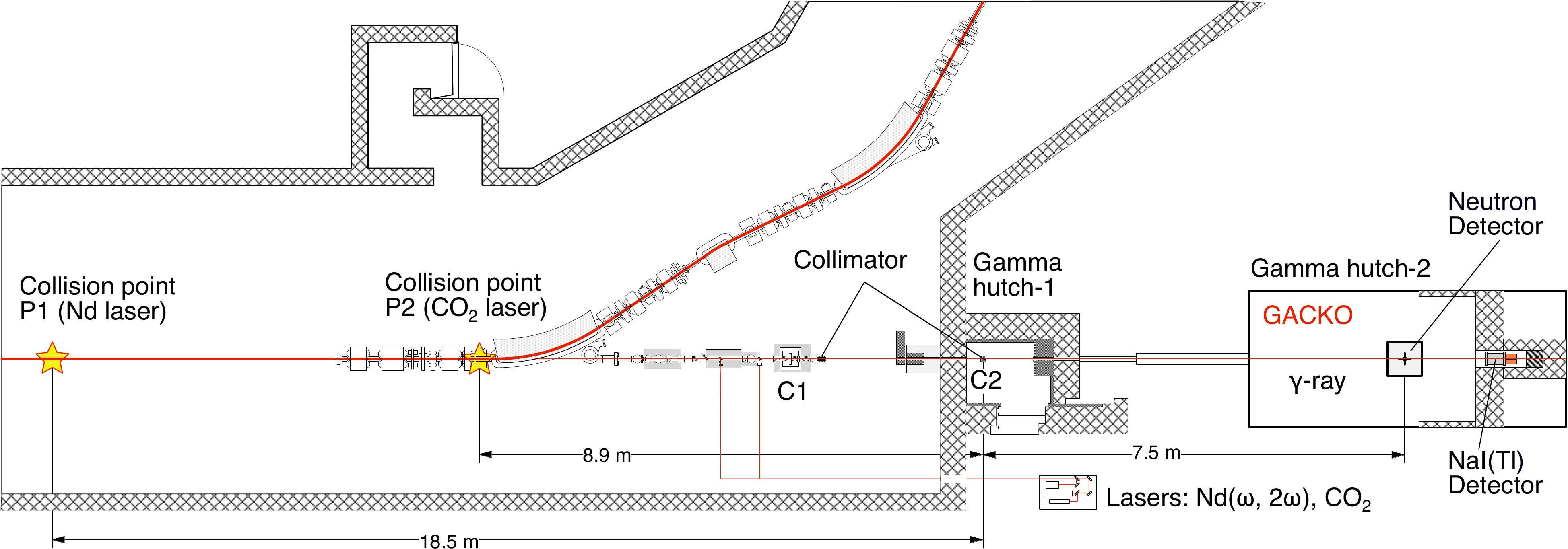}}
 \end{center}
 \caption{(Color online) The $\gamma$-ray beam-line BL01 of the NewSUBARU synchrotron radiation facility.}
 \label{fig:NS_BL1}
\end{figure*}

Following the development of high duty-cycle electron accelerators and
high-power lasers, new quasi-monoenergetic $\gamma$-ray beams have been
produced in collisions of laser photons with relativistic
electrons which is referred to as laser Compton scattering
(LCS)~\cite{Ohgaki2000, Amano2009, Horikawa2010, Utsunomiya2015} as
illustrated in Fig.~\ref{fig:LCS_scatt}. The LCS $\gamma$-ray
beam-line of the NewSUBARU synchrotron radiation
facility~\cite{Utsunomiya2015} is shown in Fig.~\ref{fig:NS_BL1}. An
electron beam at 1.0~GeV in nominal energy is injected from a linear
accelerator to the storage ring and either decelerated to 0.5~GeV or
accelerated to 1.5~GeV.

Pulse $\gamma$-ray beams are produced with Q-switch lasers; low-energy
beams are produced typically at 20~kHz with the INAZUMA laser (1064~nm
wavelength and 60~ns pulse width) for $(\gamma,1n)$ cross section
measurements below $(\gamma,2n)$ threshold, while high-energy beams at
1~kHz with the Talon laser (532~nm, 40~ns) for partial cross section
measurements above $(\gamma,2n)$ threshold. The frequency and pulse
width of the electron beam in the NewSUBARU storage ring are 500~MHz
and 60~ps, respectively, so that the frequency and width of the
$\gamma$-ray pulse are the same as those of the laser.

The electron beam energy has been calibrated with the accuracy of the
order of 10$^{-5}$ \cite{Utsunomiya2014} by using low-energy LCS
$\gamma$-ray beams produced with a grating-fixed CO$_2$ laser with the
central wavelength of the P(20) master transition ($\lambda$ = 10.5915 $\mu$m
$\pm$ 3 $\mathrm{{\AA}}$) with the bandwidth 1.3 $\mathrm{{\AA}}$ in
the full width at half maximum. The $\gamma$-ray beam energy is
determined from the calibrated electron beam energy.

\begin{figure}
  \begin{center}
    \resizebox{0.9\columnwidth}{!}{\includegraphics{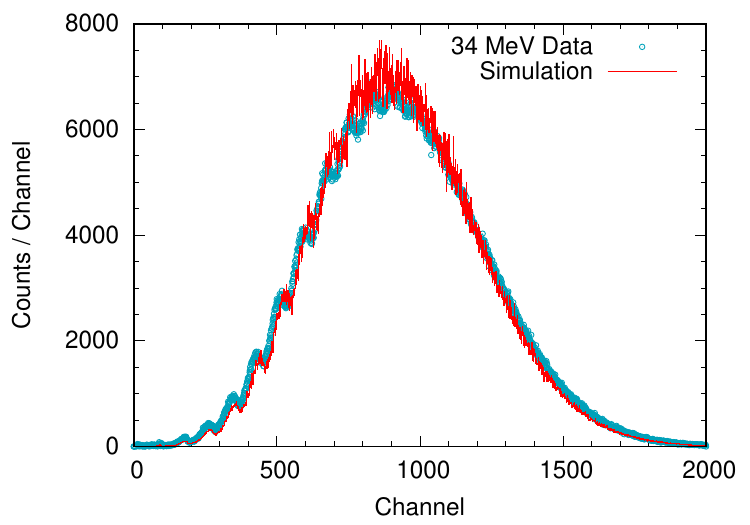}} 
  \end{center}
  \caption{(Color online) Experimental multi-photon spectrum at 34~MeV in comparison
     with the best-fit Poisson distribution.}
  \label{fig:LCS_MP}
\end{figure}

The flux of pulsed $\gamma$-rays is determined with the pile-up or
Poisson-fitting method~\cite{Kii1999, Kondo2011, Utsunomiya2018}. 
Large numbers of laser photons and electrons are involved in the
collision with a small probability of the laser Compton
scattering. As a result, the number of photons involved in a $\gamma$-ray pulse
follows the Poisson distribution~\cite{Kii1999}. Figure~\ref{fig:LCS_MP} shows an
experimental multi-photon (pile-up) spectrum for 34-MeV $\gamma$-rays
measured with a 8\rq\rq $\times$ 12\rq\rq NaI(Tl) detector along with
the best-fit Poisson distribution. The average number of photons per
$\gamma$-pulse is determined with the intrinsic accuracy less than
0.1$\%$~\cite{Utsunomiya2018}. The $\gamma$-ray flux is determined
with the experimental formula~\cite{Kondo2011, Utsunomiya2018} for a
product ($m^{\rm exp} \times N_\gamma^{\rm pulse}$) of the average number,
\begin{equation}
  m^{\rm exp} = \frac{\langle N_m \rangle}{\langle N_s \rangle} \ ,
\end{equation}
and the number of $\gamma$-ray pulses
\begin{equation}
  N_\gamma^{\rm pulse} = \sum_i N_m(i) \ ,
\end{equation}
where $N_m(i)$ and $N_s(i)$ are the number of photons at channel $i$ in the multi- and
single-photon spectra, respectively, and $\langle N \rangle$ stands
for their average channel number~\cite{Utsunomiya2018}.

\begin{figure}
  \begin{center}
    \resizebox{0.9\columnwidth}{!}{\includegraphics{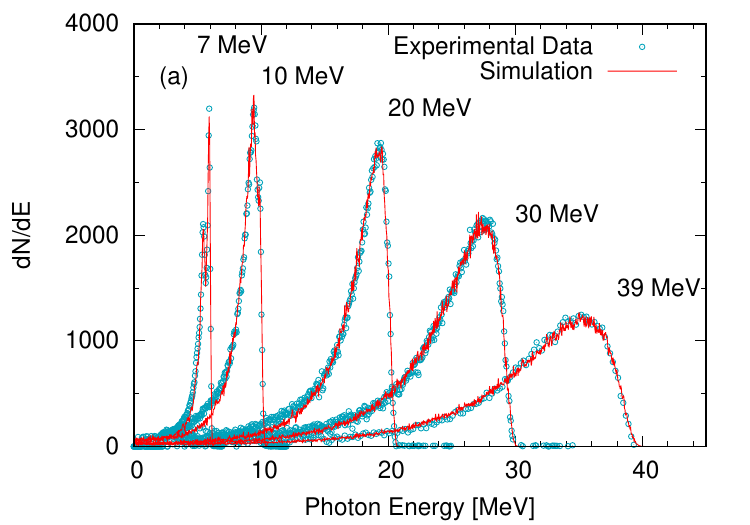}}\\
    \resizebox{0.9\columnwidth}{!}{\includegraphics{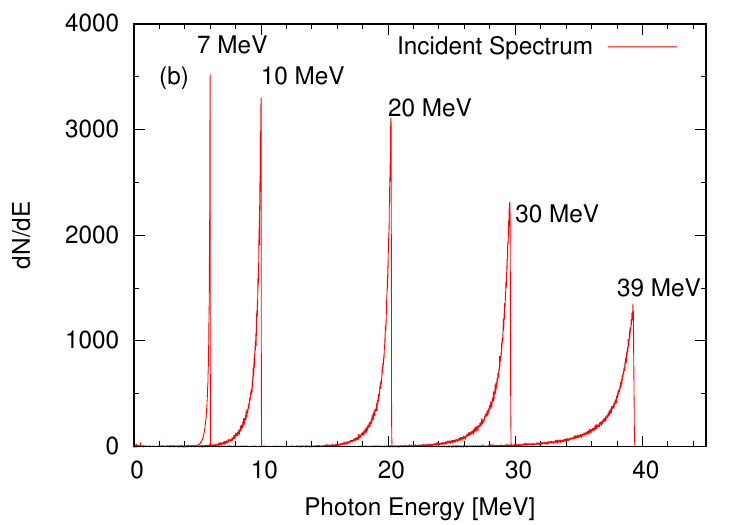}} 
  \end{center}
  \caption{(Color online) (a) Experimental and simulated response functions of a
    3.5\rq\rq $\times$ 4.0\rq\rq LaBr3(Ce) detector to LCS
    $\gamma$-ray beams, and (b) energy spectra of incident LCS $\gamma$-ray
    beams obtained by Monte Carlo simulations with the GEANT4
    code. The energy spread is a few $\%$ in the full width at half
    maximum: 68, 247, 325, 578, and 965 keV for 6.0, 10, 20, 30, and
    39 MeV LCS $\gamma$-ray beams, respectively.}
\label{fig:LCS_spectra}       
\end{figure}

The energy-profile of the LCS $\gamma$-rays is determined by best
reproducing response functions of an energy-profile monitor with a
Monte Carlo code GEANT4 that incorporates the kinematics of the laser
Compton scattering, transportation of LCS $\gamma$-rays through the
collimators (C1 and C2) to the energy-profile monitor mounted in the
experimental hutch GACKO (GAmma Collaboration hutch of KOnan
university) in the $\gamma$-ray beam line of the NewSUBARU facility,
and electromagnetic interactions of $\gamma$-rays inside the monitor
detector. Figure~\ref{fig:LCS_spectra} shows experimental $\gamma$-ray
spectra for the 3~mm C1 and 2~mm C2 collimators measured with a
3.5"~$\times$~4.0" LaBr3(Ce) detector, best-fit Monte Carlo
simulations, and incident $\gamma$-ray spectra at 6, 10, 20, 30 and
39~MeV. As shown in Fig.~\ref{fig:LCS_spectra}, the energy-profile of
the LCS $\gamma$-rays produced at the NewSUBARU facility is
characterized by a sharp cut-off at the maximum energy corresponding
to the head-on collision and a low-energy tail with an energy spread
determined by the electron beam divergence and the collimator
size. The energy spread is a few $\%$ in the full width at half
maximum.

The synchrotron radiation accompanies the LCS $\gamma$-rays as
background which is considerably lower in both intensity and energy
than the positron bremsstrahlung which accompanied the
positron-annihilation-in-flight $\gamma$-rays. The laser was turned on
and off at 10~Hz for 80~ms and 20~ms, respectively, to measure the
background neutrons resulting from the synchrotron radiation.

  \subsection{Partial Reaction Measurement}
  \label{subsec:PRM}
One can determine the neutron yield cross section from the total
number of neutrons detected using Eq.~(\ref{eq:sigN}) virtually in all
measurements with the different $\gamma$-ray sources summarized in
Sec.~\ref{subsec:experiments}. However, it is not straightforward at
all to determine partial photoneutron cross sections, $(\gamma,1n)$,
$(\gamma,2n)$, $(\gamma,3n)$, {\it etc.}  In this section, we describe
how partial photoneutron cross sections are derived using the various
measurement techniques, and then focus our discussions on the
long-standing discrepancies between the Livermore and Saclay data and
the new direct neutron-multiplicity sorting technique that was a major
topic of the present CRP.

\subsubsection{Bremsstrahlung}
\label{subsubsec:PRM_bremss}

Since the effective photon spectrum is continuous, a special
correction that is based on the statistical theory of nuclear
reactions is applied to $\sigma(\gamma, xn)$ to obtain the total
photo-neutron reaction cross section of Eq.~(\ref{eq:sigS}). After the
correction is applied, $(\gamma,2n)$ and $(\gamma,1n)$ cross sections
below the $(\gamma,3n)$ reaction threshold $B_{3n}$, can be obtained
by subtraction as follows
\begin{eqnarray}
  \sigma(\gamma, 2n) &=& \sigma_{xn} - \sigma_{Sn}   \ ,\\
  \sigma(\gamma, 1n) &=& \sigma_{Sn} - \sigma( \gamma, 2n) \ ,\\ 
  \sigma(\gamma, 1n) &=& \sigma_{xn} - 2\sigma(\gamma, 2n) \ .
\end{eqnarray}

\subsubsection{Quasi-monoenergetic Annihilation Photons}
\label{subsubsec:QAP}

Highly-efficient 4$\pi$ neutron detectors for measuring multi-neutron
coincidences and neutron multiplicity sorting techniques for
separating partial reactions were developed in
Livermore~\cite{Berman1975}, Saclay, and other laboratories, to deal
with the competition between various partial reactions $(\gamma,1n)$,
$(\gamma,2n)$, $(\gamma,3n)$, {\it etc.} over a wide range of energies.

Large arrays of $^{10}$BF$_3$ tubes embedded in a paraffin or
polyethylene matrix in several concentric rings were used at
Livermore. For example, the array of 48 $^{10}$BF$_3$ tubes in a
paraffin moderator had a neutron detection efficiency of 45 -- 30\% in
the energy range up to 5~MeV~\cite{Berman1967}. The ring-ratio
technique was developed to determine the average neutron kinetic
energy~\cite{Berman1967}. The ring-ratio, i.e. the ratio of the
neutron count of the outer-ring detector to that of the inner-ring
detector, is a monotonically-increasing function with neutron
energy. It allows to distinguish, for example, between $(\gamma,2n)$
and two $(\gamma,1n)$ events. Thus, it became possible to directly
determine/probe the total photo-neutron (Eq.~(\ref{eq:sigS})) and
neutron yield (Eq.~(\ref{eq:sigN})) cross sections.

A large tank of Gd-loaded liquid scintillator was used at Saclay. The
high neutron detection efficiency of over 90\% that was achieved in
the energy range up to 5~MeV allowed direct measurements of partial
photoneutron reaction cross sections which were used to determine the
total and neutron-yield cross sections, Eqs.~(\ref{eq:sigS}) and
(\ref{eq:sigN}), respectively.  The discrimination of various partial
reaction cross sections was more complex and less reliable with the
Livermore detector than with the Saclay detector, even with the help
of the ring-ratio technique. The reason was that the detection
efficiency of the Livermore detector was lower and varied more with
energy than that of the Saclay detector. The Saclay detector, on the
other hand, suffered from pile-up and background events as well as
dead-time caused by a $\gamma$-flash from the
target~\cite{Berman1975}.

The results obtained at the two laboratories for the same nuclei show
intricate discrepancies in magnitude not only for the total but for
partial cross sections as well. For example, it was
found~\cite{Wolynec1984, Wolynec1987, INDC0433} that, in general, the
Saclay $(\gamma,1n)$ cross sections are larger than the Livermore
data, whereas the $(\gamma,2n)$ cross sections are smaller than the
corresponding Livermore data. This is discussed in detail in
Sec.~\ref{subsec:evaldata}.

\subsubsection{Bremsstrahlung Activation}
\label{subsubsec:bremssACT}

A series of activation measurements was performed using high-intensity
bremsstrahlung. The partial photoneutron reaction cross sections
$(\gamma,1n)$, $(\gamma,2n)$, $(\gamma,3n)$, {\it etc.} were
determined by measuring the activity of the residual unstable nuclei.

After irradiating a target by bremsstrahlung with various end-point
energies, $\gamma$-ray counting was carried out for residual
nuclei. The $\gamma$-lines unique to residual nuclei were clearly
identified in the $\gamma$-ray spectra measured with an energy- and
efficiency-calibrated high-purity germanium (HPGe) detector
\cite{Belyshev2014, Belyshev2015, Naik2016}. Photonuclear reaction
cross sections were deduced from the $\gamma$-lines.

\subsubsection{Direct neutron-multiplicity sorting with a flat-efficiency detector}
\label{subsubsec:DNMSFED}

The partial photoneutron cross section $\sigma(\gamma,in)$ with
neutron multiplicity $i = 1, 2, 3, \dots$ can typically be determined
experimentally from the number of $(\gamma,in)$ reactions $N_i$ by
\begin{equation}
  N_i = N_\gamma N_T \sigma(\gamma,in) \ ,
 \label{eq:EXPFORMULA}
\end{equation}
where $N_\gamma$ is the number of $\gamma$-rays incident on a target,
and $N_T$ is the number of target nuclei per unit area.

The problem with using Eq.~(\ref{eq:EXPFORMULA}) lies in the fact that
$N_i$ is not a direct experimental observable. The neutron detection
efficiency of moderator-based neutron detectors depends on the neutron
kinetic energy. The ring ratio technique
\cite{Berman1967} was developed at Livermore to determine the average
neutron energy. However, this technique cannot be applied to $N_i$,
but to the experimental observable, i.e. multi-neutron coincidence
events. This may be a source of uncertainties associated with the
Livermore partial photoneutron cross sections.

A novel technique~\cite{Utsunomiya2017} was developed to overcome the
shortcomings of the neutron-multiplicity sorting of Livermore. This
technique, referred to as direct neutron-multiplicity sorting with a
flat-efficiency detector (FED), combined with the LCS $\gamma$-ray beam
provides an experimental opportunity to obtain partial photoneutron
cross sections with improved reliability.

In the case of neutron detection, for example, between $(\gamma,3n)$
and $(\gamma,4n)$ thresholds, using a pulsed $\gamma$-ray beam,
$j$-fold events for single ($j=1$), double ($j=2$), and triple ($j=3$)
neutron coincidences are observed.

The single neutron event corresponds to observing only one neutron
during the time interval of two successive $\gamma$-ray pulses. There
are three contributions from $(\gamma,1n)$, $(\gamma,2n)$, and
$(\gamma,3n)$ reactions to the single neutron event as
\begin{eqnarray}
  N_s &=& N_1\varepsilon(E_1) \nonumber \\
      &+& N_2\ {}_2C_1\varepsilon(E_2) \left\{ 1-\varepsilon(E_2) \right\} \nonumber \\
      &+& N_3\ {}_3C_1\varepsilon(E_3) \left\{ 1-\varepsilon(E_2) \right\}^2 \ .
  \label{eq:DNMSFED_NS}
\end{eqnarray}

The first term means simply that one neutron emitted in the
$(\gamma,1n)$ reaction is observed with detection efficiency
$\varepsilon(E_1)$ for neutron kinetic energy $E_1$. The second term
means that one of two neutrons emitted in the $(\gamma,2n)$ reaction
is observed with detection efficiency $\varepsilon(E_2)$ for neutron
kinetic energy $E_2$ and that the other neutron is not observed with
unobserved efficiency $(1-\varepsilon(E_2))$.  The third term
corresponds to the observation of one of three neutrons emitted in the
$(\gamma,3n)$ reaction. It is noted that there is no way to know
$E_1$, $E_2$, and $E_3$ because the ring-ratio technique is applied to
the experimental observable $N_s$, not to the number of reactions
$N_1$, $N_2$, and $N_3$, individually.  Furthermore, the neutron
kinetic energy depends on the emission order from an excited
nucleus. Therefore, the second term of Eq.~(\ref{eq:DNMSFED_NS})
should be written as
\begin{eqnarray}
 &~& N_2\ {}_2C_1\varepsilon(E_2) \left\{ 1-\varepsilon(E_2) \right\} \nonumber \\
 &=& N_2 \varepsilon(E_{21}) \left\{ 1-\varepsilon(E_{22}) \right\} \nonumber \\
 &+& N_2 \varepsilon(E_{22}) \left\{ 1-\varepsilon(E_{21}) \right\} \ ,
\end{eqnarray}
using kinetic energies $E_{21}$ and $E_{22}$ of the first neutron and
second neutron emitted, respectively.

The concept of the novel technique is to make the detection efficiency
independent of neutron kinetic energies. Thus, using a constant
efficiency $\varepsilon$, we can rewrite Eq.~(\ref{eq:DNMSFED_NS}) as
\begin{eqnarray}
  N_s &=& N_1\varepsilon \nonumber \\
      &+& N_2\ {}_2C_1\varepsilon(1-\varepsilon) \nonumber \\
      &+& N_3\ {}_3C_1\varepsilon(1-\varepsilon)^2 \ .
  \label{eq:DNMSFED_NSFED}
\end{eqnarray}
Similarly, the double and triple neutron coincident events are written
as
\begin{equation}\label{eq:DNMSFED_NDFED}
  N_d = N_2\varepsilon^2 + N_3\ {}_3C_2\varepsilon^2(1-\varepsilon) \ ,
\end{equation}
and
\begin{equation}\label{eq:DNMSFED_NTFED}
  N_t = N_3\varepsilon^3 \ ,
\end{equation}
respectively.

One can solve the set of Eqs.~(\ref{eq:DNMSFED_NSFED}) --
(\ref{eq:DNMSFED_NTFED}) with known $\varepsilon$ to obtain $N_1$,
$N_2$, and $N_3$ from which then the partial cross sections
$(\gamma,1n)$, $(\gamma,2n)$, and $(\gamma,3n)$ are determined as in
Eq.~(\ref{eq:EXPFORMULA}).

\begin{figure}
  \resizebox{0.9\columnwidth}{!}{\includegraphics{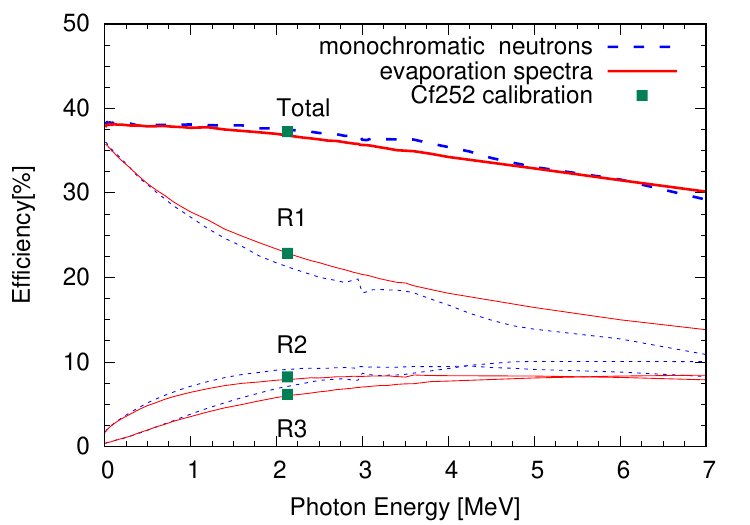}} 
  \caption{The total detection efficiency and efficiencies of three rings.}
  \label{fig:DNM_EFF}
\end{figure}

The FED consists of three concentric rings of 4, 9, and 18~He counters
embedded in a 46~cm (horizontally) $\times$ 46~cm (vertically)
$\times$ 50~cm (along the beam axis) polyethylene moderator at the
distances of 5.5, 13.0 and 16.0~cm from the $\gamma$-ray beam axis,
respectively. The moderator is shielded by additional 5~cm-thick
borated polyethylene plates for background neutron
suppression. Figure~\ref{fig:DNM_EFF} shows the total detection
efficiency and efficiencies of the individual rings of the
FED. Results of the calibration with a $^{252}$Cf source are shown by
the filled symbols. Results of the MCNP Monte Carlo simulations for
monochromatic neutrons~\cite{Utsunomiya2017} are shown by the broken
lines, while those for the neutron-evaporation spectra by the solid
lines.

There are two issues that make the data reduction to obtain partial
photoneutron cross sections rather complicated. First, double firings
of photo-reactions can be induced by multi-photons involved in a single
$\gamma$-ray pulse. For example, double firings of $(\gamma,1n)$
reactions can be identified as a $(\gamma,2n)$ reaction. Indeed, a
small amount of non-zero $(\gamma,2n)$ events attributable to such
double firings was often observed below $(\gamma,2n)$
threshold. Second, the electromagnetic interactions (pair production,
Compton scattering, photo-electric absorption) of high-energy
$\gamma$-rays in thick high-$Z$ target material can generate secondary
$\gamma$-rays which can induce the giant dipole resonance most
effectively in the peak region which is mostly governed by the
$(\gamma,1n)$ channel. Thus, the secondary gamma rays produce extra
neutrons which may mistakenly be assigned to reaction neutrons of the
$(\gamma,1n)$ channel associated with the primary gamma rays. Such
effect was observed when a 10mm-thick $^{209}$Bi target was irradiated
with a 40~MeV LCS $\gamma$-ray beam.

\begin{figure}
  \begin{center}
    \resizebox{0.9\columnwidth}{!}{\includegraphics{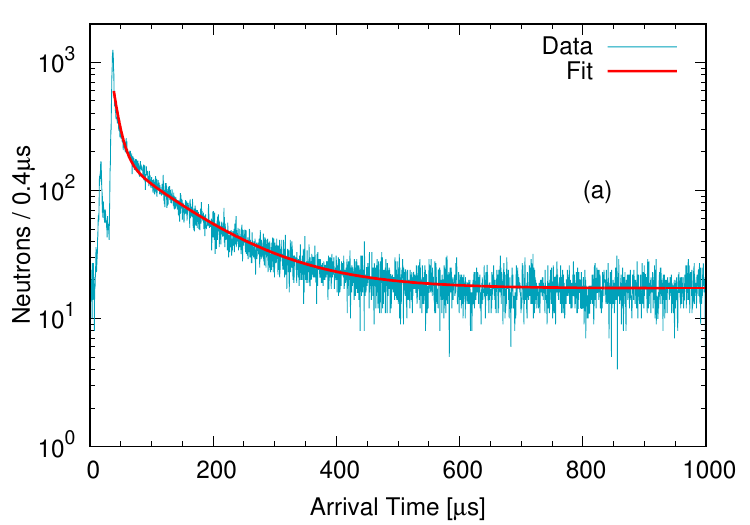}}\\
    \resizebox{0.9\columnwidth}{!}{\includegraphics{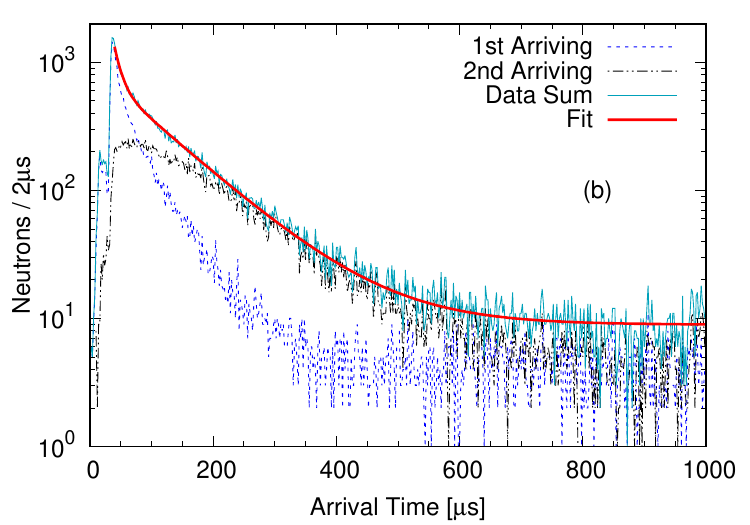}}\\
    \resizebox{0.9\columnwidth}{!}{\includegraphics{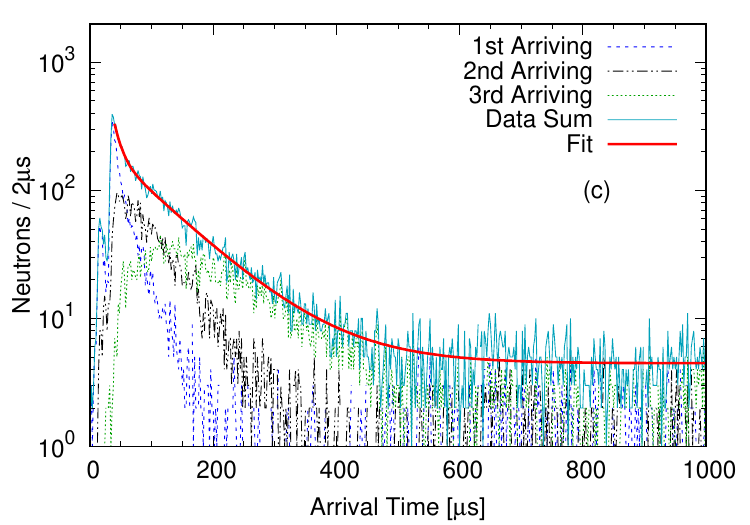}} 
  \end{center}
  \caption{(Color online) Neutron arrival-time distributions for (a) the single,
    (b) double, and (c) triple neutron events in the
    $^{209}$Bi$(\gamma,in)$ reactions at 34~MeV.}
  \label{fig:DNM_TimeSpectra}       
\end{figure}

The procedure of unfolding the photoneutron cross section which is
convoluted with the energy distribution of the LCS $\gamma$-ray beam
(Fig.~\ref{fig:LCS_spectra}) is rather straightforward because of the
monochromaticity of the LCS $\gamma$-ray beam with low
background. Previously, the Taylor expansion method
\cite{Utsunomiya2006} and the least-square fitting method
\cite{Itoh2011} were developed. These methods were used to mostly
deduce $(\gamma,1n)$ cross sections at the average $\gamma$-ray energy
below the $(\gamma,2n)$ reaction threshold~\cite{Utsunomiya2003,
Utsunomiya2010, Kondo2012, Utsunomiya2013}. More recently, a new
unfolding method has been developed~\cite{Renstrom2018} that deduces
partial photoneutron cross sections at the maximum $\gamma$-ray energy
corresponding to the maximum $\gamma$-ray yield
(Fig.~\ref{fig:LCS_spectra}) over a wide energy range. The unfolded
cross sections resulting from the Taylor expansion method and the new
method were cross-checked in the case of
$^{209}$Bi~\cite{Gheorghe2017} partial cross sections and were found
to be in agreement.

  \subsection{Experimental Databases and Resources}
  \label{subsec:CompiledData}
\subsubsection{Bibliographic Data}
\label{subsubsec:Bibliographic}

The list of photonuclear reaction bibliographies has remained
unchanged since the publication of the previous IAEA Photonuclear Data
Library in 2000 and can be found in Ref.~\cite{IAEAPhoto1999}. The
list includes mainly compilations of total and partial reaction cross
sections for photo-induced disintegration and fission, photon
absorption and scattering processes up to 1999. The list includes, for
example, ``photonuclear reactions,'' 10$^{\rm th}$ IAEA
Bibliographical Series at IAEA~\cite{Antonescu1964}, ``Photonuclear
data - abstract sheets 1955 -- 1982,'' at National Institute of
Standards and Technology (NIST)~\cite{Fuller1983}, ``Photonuclear data
index 1976 -- 1995,'' at Centr Dannykh Fotoyadernykh Eksperimentov
(CDFE)~\cite{Varlamov1996}, and ``Bibliographic Index to photonuclear
reaction data (1955 -- 1992)'' at Japan Atomic Energy Research
Institute (JAERI)~\cite{JAERI-M93-195}. More detailed information is
given in Ref.~\cite{IAEAPhoto1999}. These indices are available at
CDFE~\cite{cdfe.sinp.msu.ru}.

\subsubsection{Compilation of Photonuclear Cross Sections and GDR Parameters}
\label{subsubsec:GDRparameters}

The photonuclear cross sections and GDR parameters compiled prior to
this CRP, have been reviewed in Ref~\cite{IAEAPhoto1999}. Below we
briefly summarize the important publications and the most recent
compilation of GDR parameters performed within this CRP.

\begin{itemize}
\item Dietrich and Berman~\cite{Dietrich1988} published 
  ``Atlas of photoneutron cross sections obtained with mono-energetic
  photons,'' which contains the photonuclear data measured with
  quasi-monoenergetic photons by annihilation in flight of fast
  positrons and by bremsstrahlung tagging;

\item ``Handbook on Nuclear Activation Data'' compiled by Forkman and Petersson
  and published from IAEA~\cite{TechReport273};

\item ``Plots of the Experimental and Evaluated Photonuclear Cross-Sections'' 
  by Blokhon and Nasyrova~\cite{INDC0337} gives graphs of experimental
   and evaluated photo-neutron reaction data for some selected nuclides;

\item Varlamov {\it et al.}~\cite{INDC0394} published ``Atlas of GDR,''
  which is a compilation of many photonuclear reaction data.  This
  report covers all the reaction data compiled by
  Antonescu~\cite{Antonescu1964}, and is given in the Annex of
  previous IAEA photonuclear data library report~\cite{IAEAPhoto1999};

\item the GDR parameters for heated atomic nuclei are determined from
  the $\gamma$-decay data. Compilation and parametrization of the GDR
  resonances built on excited states are given by Schiller and
  Thoennessen~\cite{Schiller2007}; and

\item the most recent comprehensive databases of the GDR parameters
   with their uncertainties from ground-state photo-absorption are
   compiled by Plujko {\it et al.}~\cite{Plujko2011, Plujko2018}.
\end{itemize}

In the recent databases~\cite{Plujko2011, Plujko2018}, Plujko {\it et
al.} applied the least-squares technique to the updated experimental
databases to obtain the GDR parameters and their uncertainties.  The
theoretical photo-absorption cross section, which is fitted to the
experimental data, consists of both the GDR component with the SLO and
SMLO models~\cite{RIPL3} and the QD contribution, which are given
later.  These databases~\cite{Plujko2011, Plujko2018} were constructed
based on the experimental data reported before Jan. 2010 and June
2017, respectively. Further updated GDR parameter tables, where the
data published before Dec. 2018 are involved, are given in
Appendix~\ref{appendix:GDR}.

\subsubsection{EXFOR Database}
\label{subsubsec:EXFOR}

The EXFOR database~\cite{Otuka2014, BNL-NCS-63330-00}, maintained by
the international network of Nuclear Reaction Data Centers
(NRDC)~\cite{INDC0401}, is the main foundation for producing 
evaluated nuclear data libraries nowadays. The database as well
as the Web data retrieval system~\cite{Zerkin2018} facilitates 
access to published experimental data in a computer-readable format.

\section{NUCLEAR MODELS AND CODES}
\label{sec:models}

  \subsection{Theory for Photonuclear Reactions}
  \label{subsec:model}
\subsubsection{Photo-Absorption Cross Section}

To calculate and evaluate the photonuclear reaction cross sections in the 1 --
200~MeV photon energy range, we adopt the independent hypothesis of a
compound nucleus (CN) reaction, namely that the decay of the CN is independent
of how it is formed, whereas the energy, spin, and parity are
conserved. A photon is absorbed by a nucleus through two distinct
nuclear reaction mechanisms; the Giant Dipole Resonance (GDR) and the
quasi-deuteron (QD) photo-absorption process~\cite{Levinger1951,
Levinger1979}. While GDR is the dominant process in the energy range
of typical photonuclear data applications ($E_\gamma = 10 \sim
20$~MeV), the QD model describes the photo-absorption mechanism for
$E_\gamma > 30$~MeV or so. The experimentally observed
photo-absorption cross section $\sigma_{\rm abs}$ includes both of these
mechanisms
\begin{equation}
  \sigma_{\rm abs}(E_\gamma) = \sigma_{\rm GDR}(E_\gamma) + \sigma_{\rm QD}(E_\gamma) \ ,
\end{equation}
where $\sigma_{\rm GDR}$ and $\sigma_{\rm QD}$ are the absorption
cross sections for the GDR and QD mechanism, respectively.

We often apply a phenomenological parameterization to describe the GDR
part of the photonuclear reaction cross sections. Several models can
be used to represent the measured absorption cross section
$\sigma_{\rm GDR}$, such as Lorentzian, Breit-Wigner or Gaussian
functions. However the most simple and suitable function that has been
confirmed by experimental photo-absorption data \cite{Berman1975} is a
Lorentzian shape of
\begin{equation}
  \sigma_{\rm GDR}(E_\gamma) = \sigma_R 
                     \frac{E_\gamma^2 \Gamma_R^2}
                          {(E_R^2 - E_\gamma^2)^2 + E_\gamma^2 \Gamma_R^2} \ ,
  \label{eq:Lorentzian}
\end{equation}
where $\sigma_R$ is the GDR peak cross section, $E_R$ is the resonance
energy, and $\Gamma_R$ is the width.  In the case of deformed nuclei,
the GDR peak splits into two, where each corresponds to the major or
minor axis of ellipsoid, and the photo-absorption cross section is
given by the sum of two GDRs. This is often called as the Standard
Lorentzian (SLO) model for the photon strength function~\cite{RIPL3},
which is also known as the Brink-Axel Lorentzian.

While the standard Lorentzian (SLO) parameterization of
Eq.~(\ref{eq:Lorentzian}) is adequate to describe GDR around the
peak, it fails to reproduce the data at lower energies around the
neutron separation energy. Modified versions of the SLO, such as the
generalized Lorentzian~\cite{McCullagh1981, Kopecky1990, Kopecky1993}
and the Simple Modified Lorentzian (SMLO)~\cite{Plujko2011,
Plujko2018} were developed to account for data both at the GDR peak
and at lower energies around the neutron separation energies. In
addition to the phenomenological GDR models, significant effort has
been devoted to developing more microscopic approaches such as the
quasiparticle random phase approximation for the description of giant
multipole resonances~\cite{Nakatsukasa2007, Peru2014} and even the
shell model has been applied to electromagnetic excitations of light
nuclei (see Ref.~\cite{Goriely2019} for an overview).

For the purpose of updating the photonuclear evaluations in the IAEA
Photonuclear Data Library, the SLO, SMLO and GLO models available in
RIPL~\cite{RIPL3} were used to describe the GDR cross sections.  The
GDR parameters of these models have been fitted to available
experimental data, and the systematic behavior of these parameters, as
a function of $Z$ and $A$ numbers, $E_\gamma$, and/or nuclear
deformation, has been extracted~\cite{RIPL3, Plujko2011b}.  When we
assume the multipolarity of E1 is the largest contribution to the
photo-absorption, the formed CN state will have the spin $J$ of
$|I-1| \le J \le I+1$, where $I$ is the target nucleus spin, and the
parity flips. This will be the initial configuration in the
statistical model calculation.

The QD model of Chadwick {\it et al.}\cite{Chadwick1991} is commonly
implemented in the model codes we employ, where the so-called Levinger
constant $L$ is taken to be 6.5. This is a scaling constant that has
been adjusted to the experimental photo-absorption cross section
$\sigma_{\rm QD}$ by \cite{Chadwick1991}, and is seldom fine-tuned in
practical calculations. The model reads
\begin{equation}
  \sigma_{\rm QD} = L \frac{NZ}{A} \sigma_d p_b \ ,
  \label{eq:sigQD}
\end{equation}
where the photo-disintegration cross section $\sigma_d$ is parameterized as
\begin{equation}
  \sigma_d =
    \left\{
      \begin{array}{ll}
         61.2 (E_\gamma - B_d)^{3/2} E_\gamma^{-3} \ \mbox{mb} &  E_\gamma > B_d \\
         0 & \mbox{otherwise}
      \end{array}
    \right. \ ,
  \label{eq:sigd}
\end{equation}
where the energies are in MeV, and $B_d$ is the deuteron binding
energy (2.22452~MeV). In the energy range $20 \le E_\gamma \le 140$~MeV,
he Pauli blocking factor $P_b$ is given by a polynomial;
\begin{eqnarray}
  P_b &=& 0.083714 - 0.0098343 E_\gamma + 4.1222 \times 10^{-4} E_\gamma^2 \nonumber \\
      &-& 3.4762 \times 10^{-6} E_\gamma^3 + 9.3537 \times 10^{-9} E_\gamma^4 \ ,
  \label{eq:pb}
\end{eqnarray}
and it is extrapolated to the outside as
\begin{equation}
  P_b =
   \left\{
     \begin{array}{ll}
       \exp(-73.3/E_\gamma) & E_\gamma < 20~\mbox{MeV} \\
       \exp(-24.2348/E_\gamma) & E_\gamma > 140~\mbox{MeV} \\
     \end{array}
   \right. \ .
  \label{eq:pbextra}
\end{equation}
Assuming the QD component is negligible
in the GDR energy range, the determination of the GDR parameters
$\sigma_R$, $E_R$, and $\Gamma_R$ can be performed by taking
$\sigma_{\rm abs} \simeq \sigma_{\rm GDR}$.

\subsubsection{Pre-Equilibrium Particle Emission}

Nucleon or composite particle emission during the pre-equilibrium
process has not been studied extensively for photonuclear
reactions. In the GDR photo-absorption mechanism, the initial nuclear
excitation can be interpreted as a superposition of particle-hole
excitations in the shell model space, whereby the 1$p$-1$h$
configuration initiates the pre-equilibrium chain in the ${\cal
Q}$-space as classified in the theory of Feshbach, Kerman, and
Koonin~\cite{Feshbach1980}, and eventually one nucleon is scattered
into the ${\cal P}$-space. The QD model directly leads to the
pre-equilibrium chain by creating the 2$p$-2$h$ configuration.

However, the nuclear reaction codes that were employed to create the
photonuclear data library --- EMPIRE~\cite{Herman2007, INDC0603},
TALYS~\cite{Koning2008}, CCONE~\cite{Iwamoto2016}, and
CoH$_3$~\cite{Kawano2019} --- have implemented relatively simple
pre-equilibrium models for photonuclear reactions. In some cases, the
exciton model for a neutron-induced reaction is considered as a
surrogate for the photo-induced precompound process, and the initial
configuration of 1$p$-0$h$ or 2$p$-1$h$ is assumed.

The pre-equilibrium process implemented in CoH$_3$ is different from
the other codes. A fraction of the pre-equilibrium process in the GDR
absorption is restricted to the bound 1$p$-1$h$ configuration
\begin{eqnarray}
  \sigma^{\rm PE}_a &=& R \sigma_{\rm GDR} \ , \\
  R &=& \frac{\omega_B(1p,1h)}{\omega(1p,1h)}
     \simeq \frac{S_n}{E_\gamma -\Delta} \ ,
  \label{eq:Rmsc}
\end{eqnarray}
where $S_n$ is the neutron separation energy, $\Delta$ is the pairing
energy, $\omega$ is the 1$p$-1$h$ state density, while $\omega_B$ is
1$p$-1$h$ but with all particles in the bound state. When we insert the
state density of Williams'~\cite{Williams1971} for $\omega$, and
Betak and Dobes~\cite{Betak1976} for $\omega_B$, the
right-hand-side of Eq.~(\ref{eq:Rmsc}) becomes simply $S_n /
(E_\gamma-\Delta)$. This is analogous to a correction factor in the
multi-step compound strength~\cite{Chadwick1993, Kawano1999}. The QD
process creates a 2$p$-2$h$ initial configuration, and CoH$_3$ makes
1$p$-1$h$ configuration in both the neutron and proton shells. On the
other hand, the GDR 1$p$-1$h$ is created in the neutron shell. Because
we have two different initial configurations, CoH$_3$ calculates the
pre-equilibrium process twice and sums the results.

\begin{figure}[!htb]
 \begin{center}
   \resizebox{0.9\columnwidth}{!}{\includegraphics{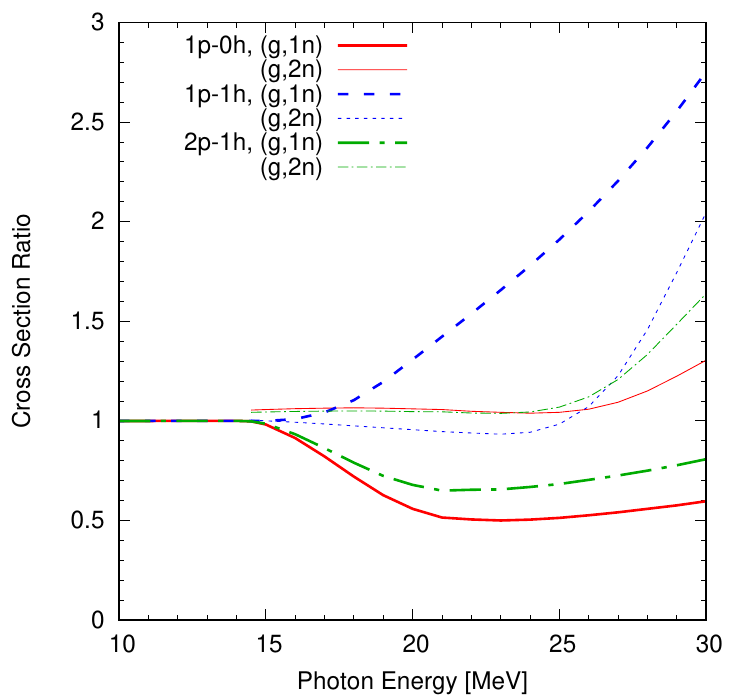}}
 \end{center}
 \caption{(Color online) Calculated $^{181}$Ta$(\gamma,n)$ and
   $(\gamma,2n)$ cross section when the initial exciton configuration is
   1$p$-0$h$ (solid), 1$p$-1$h$ (dashed), and 2$p$-1$h$ (dot-dashed). The calculated
   cross sections are shown by the ratios to the CoH$_3$ default calculation.}
 \label{fig:ta181ratio}
\end{figure}

To illustrate the impact on the pre-equilibrium calculations of the
different treatment of the initial configuration as well as the
pre-equilibrium damping factor in Eq.~(\ref{eq:Rmsc}), we calculate
the $(\gamma,1n)$ and $(\gamma,2n)$ cross sections of $^{181}$Ta with
the initial configuration of 1$p$-0$h$, 1$p$-1$h$, and 2$p$-1$h$. The
results are shown by the ratio to the CoH$_3$ default calculation,
which is the 1$p$-1$h$ for GDR and 2$p$-2$h$ for QD, 
in Fig.~\ref{fig:ta181ratio}. Because
$\sigma_{\rm abs} \simeq \sigma_{\rm QD}$ above 20~MeV, the difference
increases rapidly as the photon energy increases. However, at these
higher energies we are comparing very small cross sections obtained
from Eq.~(\ref{eq:sigQD}). The uncertainty in $\sigma_{\rm QD}$ could
be comparable to the pre-equilibrium model deficiency.

\subsubsection{Decay of Compound Nucleus}

The decay of the CN state formed by the photo-absorption reaction is
treated in the usual Hauser-Feshbach model~\cite{Hauser1952}, the only
difference being the relatively narrow distribution of compound
nucleus $J$ compared to the particle-induced CN reactions. We define a
population $P(E_x J\Pi)$ at the excitation energy of $E_x$, which is
formed by the incoming photon $E_\gamma=E_x$, and is normalized to
$\sigma_{\rm abs}(E_\gamma)$ as
\begin{equation}
  \sigma_{\rm abs}(E_\gamma) = \sum_{J\Pi} P(E_x,J,\Pi) \ .
\end{equation}
The decay of the CN state $(E_x,J,\Pi)$ into a final outgoing channel
is governed by the branching ratio for that channel which is
calculated using the particle and $\gamma$-ray transmission
coefficients. For a nucleon or a composite particle, the transmission
coefficient is denoted by $T_{lj}(\epsilon)$, where $l$ is the orbital
angular momentum, spin $j$, and the center-of-mass energy
$\epsilon$. The transmission coefficient for the $\gamma$-ray emission
is specified by the multipolarity $XL$, where $X=$~E or M.  Hereafter
we drop $\Pi$ since the parity conservation is trivial.

To simplify, we consider a case where no charged particles are
emitted, and all the final states are in the continuum. A population of
$(E_x',J')$ formed after the $\gamma$-ray transition is written as
\begin{eqnarray}
  P(E_x',J') &=& \frac{P(E_x,J)}{N} \sum_{XL} T_{XL}(E_\gamma) \nonumber \\
             &\times&\rho_0(E_x',J') dE \ , \\
  E_\gamma &=& E_x - E_x' \ ,
  \label{eq:popgamma}
\end{eqnarray}
where $N$ is the normalization shown later, $\rho_0(E,J)$ is the level
density of the same nucleus as the target, and $|J-J'| \le L \le J+J'$
is fulfilled in the summation.  When a neutron is emitted, the
population of $(E_x',J')$ in the $(Z,A-1)$ nucleus is
\begin{eqnarray}
  P(E_x',J') &=& \frac{P(E_x,J)}{N} \sum_{lj} T_{lj}(E_n)  \nonumber \\
             &\times&\rho_1(E_x',J') dE \ , \\
  E_n &=& E_x - E_x' - S_n\ ,
  \label{eq:popparticle}
\end{eqnarray}
where $S_n$ is the neutron separation energy, the level density
$\rho_1(E_x',J')$ is for the residual $(Z,A-1)$, and the triangular
relation reads $|J-J'| \le j \le J+J'$. The normalization factor $N$
is given by integrating all the possible final states
\begin{eqnarray}
  N &=& \int_0^{E_x} \sum_{XL} T_{XL}(E_\gamma) \rho_0(E_x',J') dE \nonumber \\
    &+& \int_0^{E_x-S_n} \sum_{lj} T_{lj}(E_n) \rho_1(E_x',J') dE \ .
  \label{eq:popnorm}
\end{eqnarray}

The neutron transmission coefficient $T_{lj}$ is obtained by solving
the Shr\"{o}dinger equation for a complex one-body (optical) potential.
The $\gamma$-ray transmission
coefficients are calculated from the photon-strength function
$f_{XL}(\epsilon)$ for the $\gamma$-decay as
\begin{equation}
  T_{XL}(\epsilon) = 2\pi \epsilon^{2L+1} f_{XL}(\epsilon) \ .
\end{equation}

When the final state is in the discrete part of the level scheme, the
integration in Eq.~(\ref{eq:popnorm}) is replaced by a proper
summation over branching ratios to these discrete levels which are
often taken from the nuclear structure database~\cite{RIPL3}.  An
extension of these formulas to the case where charged-particle
channels are involved should be straightforward. These calculations
are repeated until the initial excitation energy $E_x$ is exhausted by
the emitted particle and $\gamma$-ray energies, as well as the
particle binding energies.

\subsubsection{Energy and Angular Distributions of Secondary Particles}

The secondary particle energy and angular distributions are less known
experimentally, hence this type of information relies strongly on the
nuclear reaction models that are used to calculate the cross
sections. The angle-integrated particle emission spectra consists of
two components --- the pre-equilibrium contribution and the
evaporation component. Many of the reaction codes employ the exciton
model and the Hauser-Feshbach statistical model to calculate the
angle-integrated energy spectra. As mentioned above, different
pre-equilibrium exciton models are implemented in the codes, therefore
the hardness of the calculated energy spectra can be different. These
differences are not expected to be significant, however, since the
most important energy range is near GDR, where the CN decay
dominates giving an isotropic evaporation spectrum and all the codes
describe this part similarly.

A forward-peaked angular distribution is expected at higher incident
photon energies due to the pre-equilibrium process.  This is not so
extensively studied for the photo-induced reaction case, and there is
no general description for the evaluation of the angular distribution.
A practical workaround~\cite{Chadwick1995}, which was also applied to
the previous IAEA photonuclear data library, is to emulate the
neutron-induced reaction, and adopt the double-differential data
systematics by Kalbach~\cite{Kalbach1988}.

\subsubsection{Model Parameters in Photonuclear Reaction Calculation}

As mentioned in the previous section, there are some differences in
the pre-equilibrium models implemented in the nuclear reaction codes
that have been employed to evaluate the photonuclear reaction cross
sections. These differences are, however, notable in the high energy
region only. The dominant photonuclear reaction part is in the GDR
energy range, and the differences in the model calculations mainly
come from the Hauser-Feshbach model parameters.  These parameters
include the neutron and charged-particle optical potentials, the level
density, and the $\gamma$-ray strength function.  When the
photo-absorption cross section is estimated within the reaction code
itself, the built-in GDR parameter library can also be a source of
difference between the various calculations. The global optical
potential of Koning and Delaroche~\cite{Koning2003} is widely used in
these type of calculations, therefore the differences in the
calculations are mostly due to the level density models. Furthermore,
the probability of $\gamma$-ray emission is much smaller than particle
emission, so differences among the $\gamma$-ray strength functions
have a very modest impact on the photonuclear reaction calculations.

  \subsection{Nuclear Reaction Codes for Photonuclear Data Evaluation}
  \label{subsec:code}
Since it is almost impossible to obtain complete information for
photonuclear reactions from experiments only, the nuclear reaction
codes play a significant role in creating the evaluated data library.
For example, the only possible way to generate the evaluated energy
and angular distributions of the emitted particle is to use the model
calculations, which, of course, should consistently describe the
available cross section data. In the previous release of the IAEA
photonuclear data library~\cite{IAEAPhoto1999}, the evaluation was
done using several reaction model codes available at the participating
institutes: GNASH~\cite{LA6947, LA12343} at LANL and KAERI,
ALICE-F~\cite{Fukahori1992} and MCPHOTO~\cite{Kishida1989} at JAERI,
GUNF~\cite{Zhang1998} and GLUNF~\cite{Zhang1999} at CIAE, and
XGFISS~\cite{Blokhin1999} at IPPE.

The modern Hauser-Feshbach nuclear reaction codes,
EMPIRE~\cite{Herman2007, INDC0603}, TALYS~\cite{Koning2008},
CCONE~\cite{Iwamoto2016}, and CoH$_3$~\cite{Kawano2019}, which are
often utilized for nucleon-induced reaction calculations, are also
capable of calculating photo-induced reaction cross sections. As a
result, they are widely used in the evaluation of general purpose
nuclear data libraries, and are combined with the
advanced-capabilities utility codes to create ENDF-6 format data
files.  Albeit these codes are built on common well-established
nuclear reaction models to describe the various reaction mechanisms,
the implementation of the nuclear reaction models in each code may
give rise to differences in the calculated results which are inherent
to the code and do not depend on the model or model parameters. It is
therefore important to understand the characteristics of each code
before comparing evaluations performed with the codes.  Recently, an
inter-comparison of the codes EMPIRE, TALYS, CCONE, and CoH$_3$ was
performed for neutron-induced reactions on actinides~\cite{Capote2017}
by fixing all the input parameters as consistently as possible, so
that any observed difference would be due to the model
implementation. It was concluded that in terms of model
implementation, the implementation of the width fluctuation correction
might give rise to differences in the calculated cross sections by up
to $\approx$15\%.  The ambiguity related to the width fluctuation
correction is not expected to affect the photonuclear reaction case,
since the $(\gamma,n)$, channel is much larger than the photon elastic
scattering (the elastic enhancement does not impact
$(\gamma,n)$~\cite{Kawano2015}.) However, as already discussed in
Sec.~\ref{subsec:model}, the pre-equilibrium models employed in the
different codes are different therefore the pre-equilibrium component
is one of the sources of ambiguity among the codes.

The newly evaluated photonuclear data were produced with the codes
EMPIRE, TALYS, CCONE, and MEND-G~\cite{Cai2011}. Some data files were
produced with ALICE-F as was the case in the previous IAEA
photonuclear data library.  CoH$_3$ has not been used to produce any
evaluations, since it focuses mainly on low-energy nuclear
reactions. However, CoH$_3$ has been considered in the code
comparison. The CPNRM (Combined Photonuclear Reaction Model)
code~\cite{Ishkhanov2007, Ishkhanov2008} developed at SINP/MSU is also
included in the code comparison as it has been used extensively to
correct the experimental data as described in
Sec.~\ref{subsec:evaldata}.

\subsubsection{Special notes on each model code}

As described in Sec.~\ref{subsec:model}, photonuclear reaction
calculations consist of three stages, the photo-absorption process,
the pre-equilibrium particle emission, and the statistical
Hauser-Feshbach decay. The absorption cross section is unambiguously
determined by providing a set of GDR parameters (energy, width and
peak cross section) using a standard Lorentzian or variants such as
the Generalized Lorentzian or Simple Modified
Lorentzian. Alternatively the absorption cross section can be
pre-calculated and stored in an external file.

For the pre-equilibrium emission component, the codes adopt the
one-component or two-component exciton model, however the initial
particle-hole configuration used in each code may be different.
CoH$_3$ assumes the initial configuration is 1$p$-1$h$ created in the
neutron shell for GDR, and 2$p$-2$h$ (1$p$-1$h$ in both neutron
and proton shells) for the QD process. It also includes the damping
factor of Eq.~(\ref{eq:Rmsc}).  EMPIRE begins the calculation with
1$p$-0$h$, unless specified otherwise. TALYS also assumes
1$p$-0$h$. When the two-component exciton model is selected, the
particle-hole pair is given in the neutron shell. In MEND-G, 2$p$-2$h$
is the initial configuration for the QD contribution. CCONE assumes
$1$p-$0$h as the initial configuration.  The particle-hole pair in the
two-component model is given in the neutron shell for photo-reaction.

In the statistical decay stage, the particle transmission coefficients
are calculated using an optical potential. Table~\ref{table:omp}
summarizes the optical potentials employed in each code for the
evaluation. Since often the potential parameters are adjusted to
better reproduce experimental data, we provide a default set of
potentials when not specified. CoH$_3$
does not offer a default option for the potential. However, those
given in the table are typical choices for the photonuclear reaction
calculations.

\begin{table}[!htb]
\begin{center}
 \caption{Optical potential used in the photonuclear cross section calculations.
  The abbreviations are, KD03 (Koning and Delaroche~\cite{Koning2003}), 
  KD03-f (folding KD03 potential by technique of Watanabe~\cite{Watanabe1958})
  AV94 (Avrigeanu {\it et al.}~\cite{Avrigeanu1994}),
  AV09 (Avrigeanu {\it et al.}~\cite{Avrigeanu2009}),
  AV10 (Avrigeanu {\it et al.}~\cite{Avrigeanu2010}),
  AV14 (Avrigeanu {\it et al.}~\cite{Avrigeanu2014}),
  AC06 (An and Cai~\cite{An2006}),
  BO88 (Bojowald {\it et al.}~\cite{Bojowald1988}),
  BG69 (Becchetti and Greenlees)~\cite{Becchetti1969a,Becchetti1969b}
  HA06 (Han {\it et al.})~\cite{Han2006}
  XU11 (Xu {\it et al.})~\cite{Xu2011}}
 \label{table:omp} 
 \begin{tabular}{lcccccc}
  \hline\hline
         & n    & p    &$\alpha$& d      &     t  & $^{3}$He\\
  \hline
EMPIRE   & KD03 & KD03 & AV94 & AC06   & BG69   & BG69 \\
TALYS    & KD03 & KD03 & AV14 & KD03-f & KD03-f & KD03-f \\
CCONE    & KD03 & KD03 & AV10 & HA06   & KD03-f & XU11 \\
MEND-G   & KD03 & KD03 & BG69 & BG69   & BG69   & BG69 \\
CoH$_3$  & KD03 & KD03 & AV09 & BO88   & BG69   & BG69 \\
  \hline\hline
  \end{tabular}
\end{center}
\end{table}

\subsubsection{Photonuclear reaction code inter-comparison}

\begin{figure}[!htb]
 \begin{center}
   \resizebox{0.9\columnwidth}{!}{\includegraphics{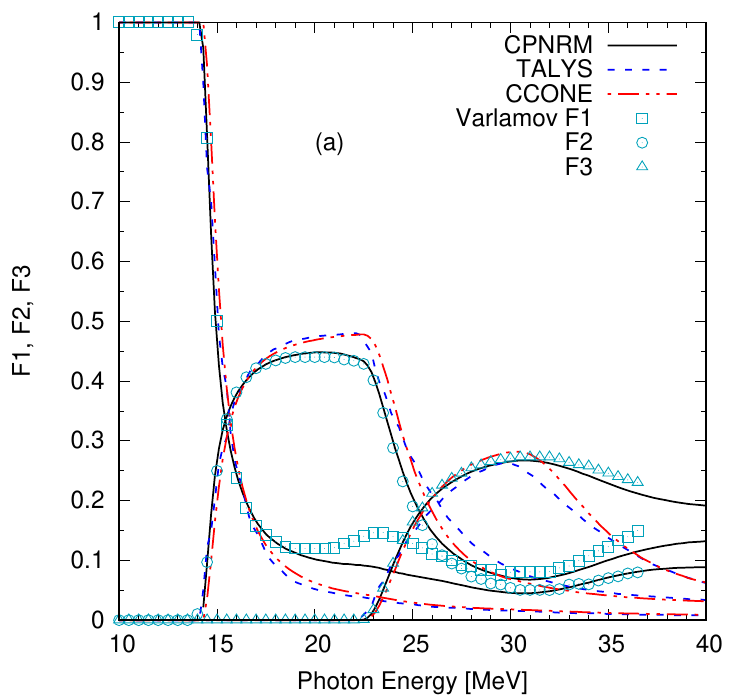}}\\
   \resizebox{0.9\columnwidth}{!}{\includegraphics{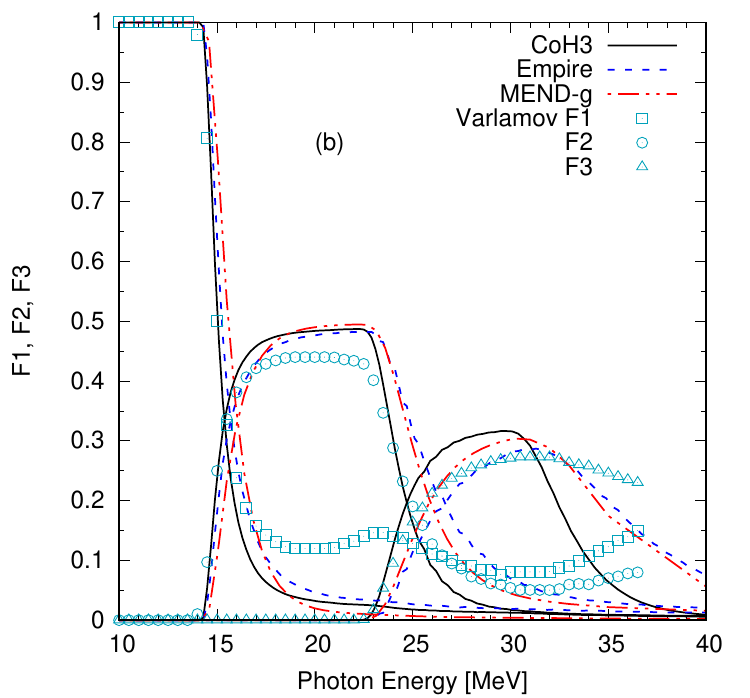}}
 \end{center}
 \caption{(Color online) Calculated $F_i$ values for $^{181}$Ta$(\gamma,inx)$ cross sections
   with (a) CPNRM, TALYS, and CCONE, and (b) CoH$_3$, EMPIRE, and Mend-g. The symbols are
   evaluated experimental data by Varlamov {\it et al.}}
 \label{fig:fivalue}
\end{figure}

The $F_i$ values in Eq.~(\ref{eq:Fi}) were calculated with CPNRM,
TALYS, CCONE, CoH$_3$, EMPIRE, and MEND-G, for the photo-induced
reactions on $^{181}$Ta. We chose all default model parameters in each
code. The results are shown in Fig.~\ref{fig:fivalue}, together with
the evaluated experimental data by Varlamov {\it et
al.}~\cite{Varlamov2014b}.  Large discrepancies amongst the model
codes are seen in the high energy region, where pre-equilibrium
emission is the dominant process, yet it should be noted that the
cross sections in this region are tiny compared to GDR. In general,
CPNRM gives the largest pre-equilibrium contribution, while the
pre-equilibrium component of CoH$_3$ is the weakest. This is
understood as a result of applying the correction factor of
Eq.~(\ref{eq:Rmsc}) in CoH$_3$. The $F_1$ and $F_2$ values by CPNRM
increase above 30~MeV, which is not seen in the other codes.

From Fig.~\ref{fig:fivalue} it is clear that not only do the codes
give different values of $F_i$, but that these differences vary with
energy. The theoretical $F_i$ values are determined by the competition
between the neutron and the $\gamma$-ray emissions in the statistical
decay of CN. Obviously they depend on the neutron and $\gamma$-ray
transmission coefficients, as well as the level density of the final
states.  The conclusion is that if the theoretical $F_i$s are used to
correct problematic or inconsistent experimental data, as proposed in
Sec.~\ref{subsec:evaldata}, then the uncertainties in $F_i$ due to
different codes and model parameters need to be considered. Moreover,
since these uncertainties are shown to vary with energy, the
introduction of an overall constant uncertainty may not be adequate.

\section{EVALUATION}
\label{sec:evaluation}
The new evaluations for the updated Photonuclear Data Library were
performed at five institutes: SINP/MSU (Skobeltsyn Institute of
Nuclear Physics, Lomonosov Moscow State University), NDC/KAERI
(Nuclear Data Center, Korean Atomic Energy Research Institute),
IFIN-HH (Horia Hulubei National Institute for Physics and Nuclear
Engineering), NDC/JAEA (Nuclear Data Center, Japan Atomic Energy
Agency), and CNDC/CIAE (China Nuclear Data Center, China Institute of
Atomic Energy). In the following sections, a summary of the evaluation
procedure, adoption of experimental data, and models used by each
institute is given along with highlights of the resulting
evaluations. The complete list of evaluations with graphical
comparisons will be made available from an IAEA web interface.

  \subsection{Evaluation of Experimental Data}
  \label{subsec:evaldata}
The majority of available experimental total and partial photo-neutron
reaction cross sections were obtained by the quasi-monoenergetic
annihilation photon beam and the photo-neutron multiplicity sorting
technique at Livermore and Saclay~\cite{Berman1975, Berman1975b,
Dietrich1988}.  Significant systematic disagreements in the
experimental $(\gamma,1n)$ and $(\gamma,2n)$ data from these two
institutes have been reported for 19 nuclei~\cite{Wolynec1987,
Varlamov2012, Varlamov2014a}: $^{51}$V, $^{75}$As, $^{89}$Y,
$^{90}$Zr, $^{115}$In, $^{116,117,118,120,124}$Sn, $^{127}$I,
$^{159}$Tb, $^{181}$Ta, $^{197}$Au, $^{208}$Pb, $^{232}$Th, and
$^{238}$U.  It was found that the $(\gamma,1n)$ cross sections from
Saclay tend to be larger than the Livermore data by up to 100\%, while
Livermore gives larger $(\gamma,2n)$ cross sections by up to 100\%. An
average overall estimate of the differences is given by the ratio of
the experimental integrated cross section from Saclay $\sigma_{\rm
S}^{\rm int}$ to that from Livermore $\sigma_{\rm L}^{\rm int}$. The
experimental integrated cross sections are obtained from
\begin{equation}
  \sigma_{\rm S,L}^{\rm int} = 
  \int_{E_{\rm th}}^ {E_{\rm max}} \sigma_{\rm S,L}(E) dE \ ,
  \label{eq:sigint}
\end{equation}
where $E_{\rm th}$ is the threshold energy of $1n$ or $2n$ reaction
and $E_{\rm max}$ is the highest photon energy in the experiments. The
average value $\langle\sigma_{\rm S}^{\rm int} / \sigma_{\rm L}^{\rm
int}\rangle$ is found to be 1.07 for the $(\gamma,1n)$ reaction, and
0.84 for the $(\gamma,2n)$ reaction. The ratio $\sigma_{\rm S}^{\rm
int} / \sigma_{\rm L}^{\rm int}$ fluctuates considerably between 0.76
and 1.34 for the $(\gamma,1n)$ cross section and between 0.71 and 1.22
for all the 19 nuclei mentioned above. Given the large fluctuations in
$\sigma_{\rm S}^{\rm int} / \sigma_{\rm L}^{\rm int}$, the
discrepancies between the Livermore and Saclay data cannot be
rectified by simply applying a constant normalization factor common to
all the 19 nuclei.

On the other hand, the neutron yield cross-section data from these two
institutes (Eq.~(\ref{eq:sigN})) also disagree by up to 10\% on
average~\cite{Varlamov2014a}. This implies that the systematic
inconsistency observed in the measured partial reaction cross sections
may be due to a shortcoming of the neutron multiplicity sorting
method. In order to understand the underlying systematic uncertainties
in the experimental data, Varlamov {\it et al.}~\cite{Varlamov2010}
introduced the $F_i$ function of Eq.~(\ref{eq:Fi}), which provides
objective criteria for assessing the consistency and reliability of
the experimental partial reaction cross-sections obtained in a given
measurement, and allows one to investigate the systematic
uncertainties. As has been mentioned in Sec.~\ref{sec:crosssec},
since $F_1$ is defined as the ratio of $\sigma(\gamma,1n)$ to the
photo-neutron yield cross section $\sigma(\gamma,xn)$, it can never be
larger than 1; analogously $F_2$ cannot exceed 1/2, $F_3$ cannot
exceed 1/3, and so on. In addition, $F_i$ should always be
positive. These constraints on $F_i(E)$ values allow one to identify a
potential problem in the way the neutron yield cross section has been
partitioned into the individual one, two, three, or even more neutron
emission components.

\begin{figure}[!htb]
 \resizebox{\columnwidth}{!}{\includegraphics{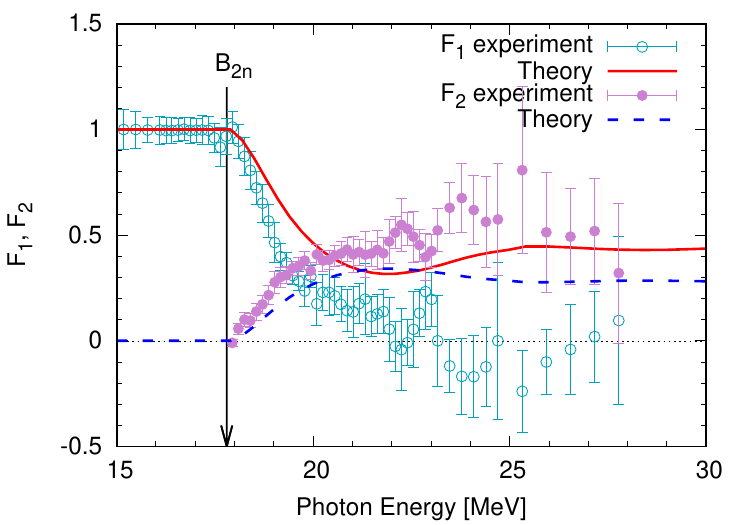}}
 \caption{(Color online) Comparison of the ratios
   $F_i^{\rm exp}$~\cite{Varlamov2016a}
   using the experimental data (triangles; Ref.~\cite{Fultz1964}) and
   the theoretical $F_i^{\rm th}$ functions for
   $^{65}$Cu}
 \label{fig:MSUficomp}
\end{figure}

Varlamov {\it et al.}~\cite{Varlamov2010} has shown that the
constraints imposed by $F_i$ are not satisfied in many
cases~\cite{Varlamov2012, Varlamov2014a, Varlamov2010, Ishkhanov2012a,
Varlamov2014b, Varlamov2015a, Varlamov2015b, Varlamov2016a,
Varlamov2016b}. The problematic experimental data, where negative
cross-sections are reported, include the measurements of photo-neutron
reactions on $^{80}$Se, $^{91,94}$Zr, $^{115}$In, $^{112-124}$Sn,
$^{133}$Cs, $^{138}$Ba, $^{159}$Tb, $^{181}$Ta, $^{186-192}$Os,
$^{197}$Au, $^{208}$Pb, $^{209}$Bi, and some others. In these cases,
the corresponding $F_1^{\rm exp}$ values and/or $F_2^{\rm exp}$ values
are found to be larger than the aforementioned upper limits, or
negative as well. A typical example of the energy dependent $F_i^{\rm
exp}(E)$ values for $^{65}$Cu~\cite{Fultz1964} is given in
Fig.~\ref{fig:MSUficomp}, where it is compared with the theoretical
$F_i^{\rm th}(E)$ curves~\cite{Varlamov2016a} obtained for partial
reactions $(\gamma,in)$ with $i = 1$ and 2 within the framework of the
CPNRM code~\cite{Ishkhanov2007, Ishkhanov2008}.

The partial photo-neutron reactions cross sections should be evaluated
under the condition that the inherent issues in the neutron
multiplicity sorting method as well as the limitations of the
statistical model do not form a bias. In other words, the systematic
experimental uncertainties as well as model-dependent uncertainties
need to be considered in the evaluation.  The evaluation method
proposed by Varlamov {\it et al.}~\cite{Varlamov2010}, combines the
experimental photo-neutron yield data $\sigma^{\rm exp}(\gamma,xn)$
(Eq.~(\ref{eq:sigN})) and the theoretical estimate of $F_i^{\rm th}$
obtained with the CPNRM code~\cite{Ishkhanov2007, Ishkhanov2008} and
is applicable to nuclei in the medium to heavy mass range. In this
method, the physical reliability criteria mentioned above are
automatically satisfied because the competition between the partial
reaction cross sections is imposed by the model calculations performed
with the CPNRM code.  The evaluated partial photo-neutron reaction
cross section $\sigma^{\rm eval}(\gamma,in)$ is obtained by
multiplying the experimental photo-neutron yield cross section
$\sigma^{\rm exp}(\gamma,xn)$ given in Eq.~(\ref{eq:sigN}) by the
theoretical $F_i^{\rm th}$ functions computed with the CPNRM code for
neutron multiplicity $i = 1$, 2, 3, $\ldots$
\begin{eqnarray}
  \sigma^{\rm eval}(\gamma,in)
    &=& F_i^{\rm th}\sigma^{\rm exp}(\gamma,xn) \nonumber\\
    &=& \frac{\sigma^{\rm th}(\gamma,in)}{\sigma^{\rm th}(\gamma,xn)}
        \sigma^{\rm exp}(\gamma,xn) \ .
  \label{eq:sigeval}
\end{eqnarray}

The former term in Eq.~(\ref{eq:sigeval}) satisfies the reliability
criteria as already mentioned, while the second term is not affected
by the issues associated with the experimental neutron multiplicity
sorting. The resulting derived cross section data $\sigma^{\rm
eval}(\gamma,in)$ should therefore be free from the systematic
experimental uncertainties of the Livermore and Saclay data.

\begin{figure}[!htb]
 \begin{center}
   \resizebox{\columnwidth}{!}{\includegraphics{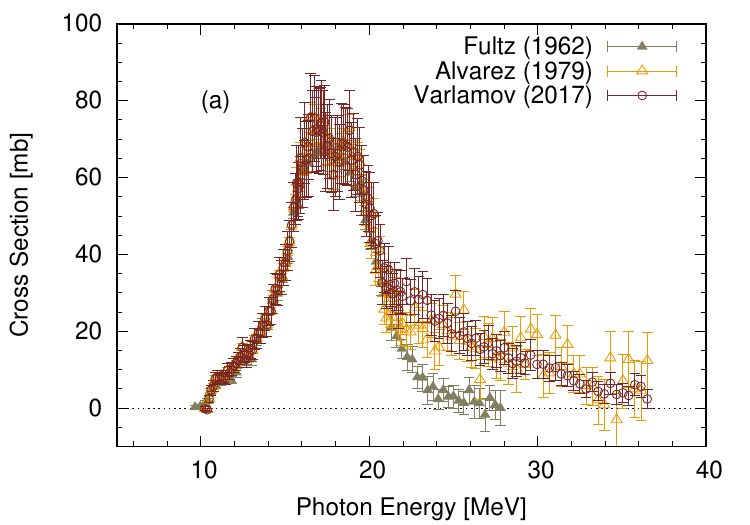}}\\
   \resizebox{\columnwidth}{!}{\includegraphics{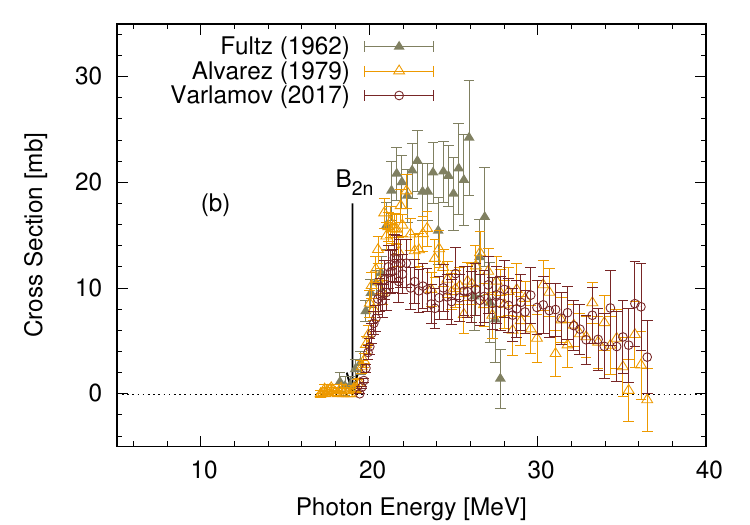}}\\
   \resizebox{\columnwidth}{!}{\includegraphics{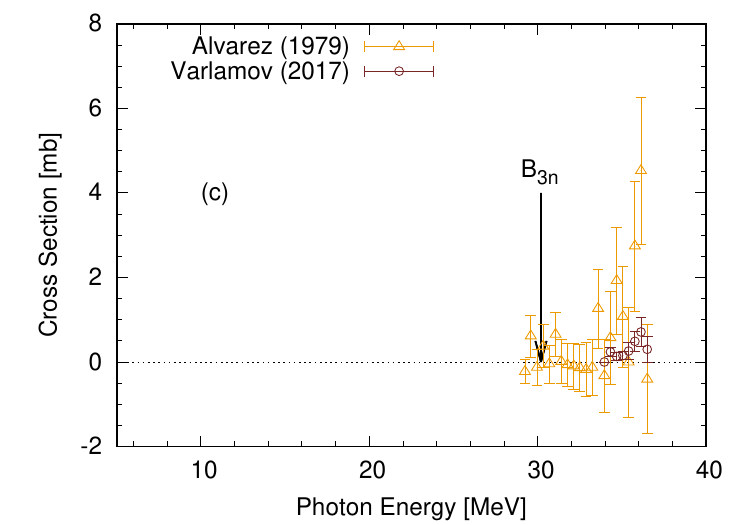}}
 \end{center}

 \caption{(Color online) Comparison of the evaluated (open circles with
   error bars)~\cite{Varlamov2017a} and experimental 
   (filled triangles~\cite{Fultz1962} and open triangles~\cite{Alvarez1979})
   partial photo-neutron cross sections for $^{59}$Co.
   (a) $\sigma(\gamma,1n)$, (b) $\sigma(\gamma,2n)$,
   and (c) $\sigma(\gamma,3n)$.}
  \label{fig:MSUsigcomp}
\end{figure}

In many cases, the evaluated cross-section data $\sigma^{\rm
eval}(\gamma,in)$ obtained from Eq.~(\ref{eq:sigeval}) form a
substantial correction to the experimental data obtained by the
neutron multiplicity sorting method~\cite{Varlamov2012, Varlamov2014a,
Varlamov2010, Ishkhanov2012a, Varlamov2014b, Varlamov2015a,
Varlamov2015b, Varlamov2016a, Varlamov2016b}. A typical example of
this evaluation method is given in Fig.~\ref{fig:MSUsigcomp}, where
the results obtained for $^{59}$Co are compared with the experimental
data.

The neutron multiplicity sorting technique sometimes cannot distinguish
$1n$ and $2n$ events clearly, or $2n$ and $3n$ events. As a result,
 the partial cross sections $\sigma^{\rm eval}(\gamma,in)$ fluctuate
although the total photo-neutron yield cross section $\sigma^{\rm
exp}(\gamma,xn)$ seems to be reasonable. This can be seen clearly
by calculating the differences between the experimental and evaluated
integrated cross sections $\sigma^{\rm int,exp}$ and $\sigma^{\rm
int,eval}$ for the $(\gamma,1n)$ reaction,
\begin{equation}
  \Delta\sigma_1
   = \sigma^{\rm int,eval}(\gamma,1n) - \sigma^{\rm int,exp}(\gamma,1n) \ ,
  \label{eq:delta1}
\end{equation}
and for the $(\gamma,2n)$ reaction,
\begin{equation}
  \Delta\sigma_2
   = \sigma^{\rm int,eval}(\gamma,2n) - \sigma^{\rm int,exp}(\gamma,2n) \ .
  \label{eq:delta2}
\end{equation}

\begin{figure}[!htb]
 \resizebox{\columnwidth}{!}{\includegraphics{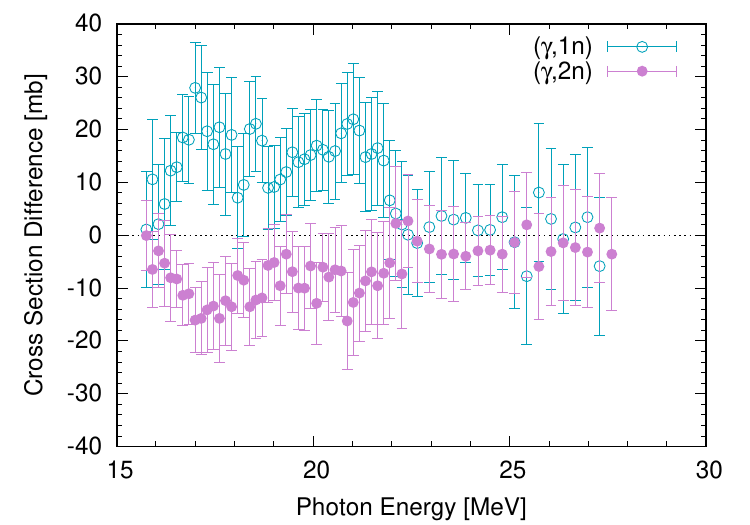}}
 \caption{(Color online) Differences between the
   evaluated~\cite{Varlamov2018} and the experimental [21] cross
   sections for $^{92}$Zr. The open circles are for the $(\gamma,1n)$
   reaction ($\Delta_1$ in Eq.~(\ref{eq:delta1})), and filled circles
   are $(\gamma,2n)$ reaction ($\Delta_2$ in Eq.~(\ref{eq:delta2})).}
 \label{fig:MSUsigdiff}
\end{figure}

An example of the differences is shown for
$^{92}$Zr~\cite{Varlamov2018, Berman1967} in
Fig.~\ref{fig:MSUsigdiff}.  From the figure, it is evident that some
of the $(\gamma,1n)$ events are recorded as $(\gamma,2n)$ events. It
has been shown~\cite{Varlamov2014a, Ishkhanov2012b, Belyshev2015} that
this problem is due to the fact that neutrons were sorted into each
reaction channel according to the measured neutron kinetic
energy. However, the neutron energy spectra for different partial
reactions may overlap with each other leading to cross-talks between
different partial cross sections, whereby some of the neutrons
originating from the $(\gamma,1n)$ reaction are mistakenly assigned to
those from the reaction $(\gamma,2n)$ and vice versa.

To test the evaluation method described above, partial photo-neutron
reaction cross sections evaluated for $^{181}$Ta~\cite{Varlamov2014b}
and $^{209}$Bi~\cite{Varlamov2016b} were compared with those obtained
in activation measurements of reaction yields with
bremsstrahlung~\cite{Ishkhanov2012b, Belyshev2015}. In the activation
measurement, cross sections of various partial reactions were directly
determined by identifying the radioactivity from the residual
nuclei. Ishkhanov {\it et al.}~\cite{Ishkhanov2012b} and Belyshev {\it
et al.}~\cite{Belyshev2015} obtained $(\gamma,in)$ cross sections with
$i = 1$ -- 6 for $^{181}$Ta and $^{209}$Bi as a function of the
bombarding photon energy based on the experimental photo-neutron
yields and the CPNRM calculations. The evaluated data for the
photo-neutron multiplicity sorting obtained agree well with the
activation data. A further confirmation was the agreement found by
making a detailed comparison~\cite{Varlamov2017b} of the evaluated
$^{197}$Au $(\gamma,1n)$ and $(\gamma,2n)$ data with those obtained in
activation measurements with bremsstrahlung~\cite{Naik2016}.

It is therefore, advisable to compare the $F_i^{\rm exp}$ with the
from $F_i^{\rm th}$ and check whether the following listed conditions
are satisfied. In case they are not satisfied, one should investigate
possible issues in the measurement that could be responsible for
unreliable partial photo-neutron cross sections.
\begin{itemize}
\item $F_i^{\rm exp}$ do not exceed the definitive upper limits;
\item $\sigma^{\rm exp}(\gamma,in)$ and corresponding $F_i^{\rm  exp}$
      are positive; and
\item the differences between $F_i^{\rm exp}$ and $F_i^{\rm th}$
      are not significant.
\end{itemize}
Considering the above criteria, the partial and total photo-neutron
reaction cross sections were evaluated using the method described
above for the following nuclei; $^{59}$Co, $^{63,65}$Cu, $^{75}$As,
$^{76,78,80,82}$Se, $^{89}$Y, $^{90,91,92,94}$Zr, $^{98}$Mo,
$^{103}$Rh, $^{115}$In, $^{116,117,118,119,120,122,124}$Sn,
$^{133}$Cs, $^{138}$Ba, $^{139}$La, $^{140,142}$Ce, $^{141}$Pr,
$^{145,148}$Nd, $^{153}$Eu, $^{159}$Tb, $^{160}$Gd, $^{165}$Ho,
$^{181}$Ta, $^{186}$W, $^{188,189,190,192}$Os, $^{197}$Au, $^{208}$Pb,
and $^{209}$Bi. The derived cross-section data were compiled into the
international EXFOR database as ``evaluated'' data and are available to
the nuclear data community.

  \subsection{Data Evaluation at KAERI}
  \label{subsec:evalKAERI}
\begin{figure*}[!htb]
 \begin{center}
   \resizebox{0.8\textwidth}{!}{\includegraphics{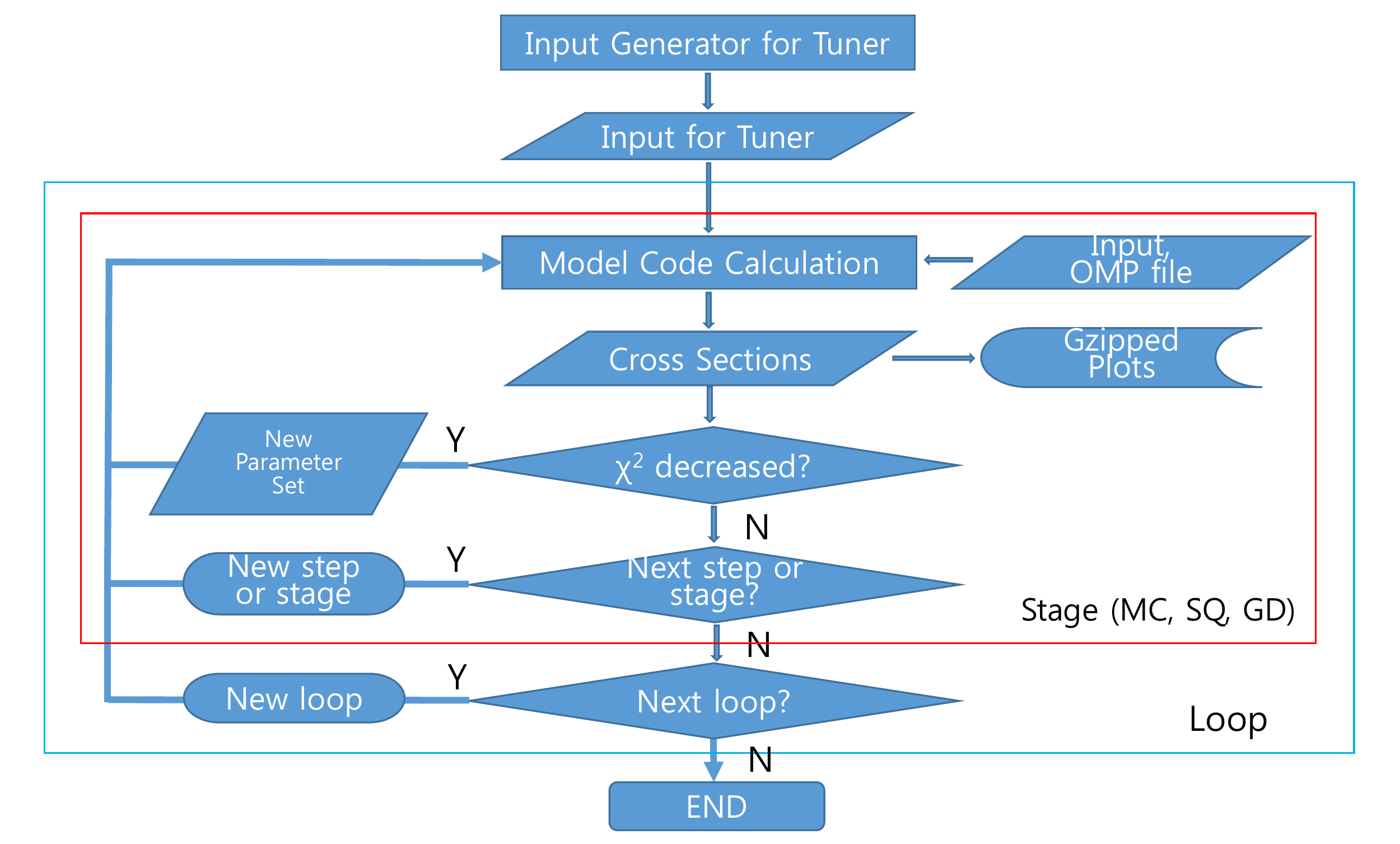}}
 \end{center}
 \caption{(Color online) Schematic flow of the automated model parameter tuning at KAERI.}
 \label{fig:KAERIprocedure}
\end{figure*}

The photonuclear cross sections have been evaluated for 40 nuclides at
KAERI. The experimental data were mainly obtained from the EXFOR
database~\cite{Otuka2014} and used for the evaluation work. The
$F_i$-corrected data and the new measurements produced under this CRP,
if any, were used preferentially over other data.

As for the nuclear reaction model code, the TALYS
code~\cite{Koning2008} was adopted as it was considered more
appropriate for the automatic model parameter tuning system employed
to facilitate the present evaluation work. The optical model potential
parameters, the GDR parameters, the level density-related parameters,
and pre-equilibrium model parameters were adjusted --- most of them
were within 20\% from their default values, and sometimes it goes up
to 50\% --- to reproduce the experimental data or the $F_i$-corrected
data.  For the level densities and the $\gamma$-ray strength function,
the constant temperature and Fermi gas models and the Brink-Axel
Lorentzian, respectively, were used in the present evaluations. For
the pre-equilibrium reaction, the exciton model was used. The list of
adjusted model parameters is summarized below.

\begin{itemize}
\item {\bf Optical model}: local parameters for neutron-induced
  reactions based on the global optical model parameters in the
  Koning-Delaroche form~\cite{Koning2003}. 19 model parameters are
  included.

\item {\bf Level density related parameters}: the level density
  parameter, the asymptotic level density, the damping parameter for
  the shell effects, the pairing correction, the temperature in the
  Gilbert-Cameron formula for the target and some residual
  nuclei. About 35 parameters in total for each evaluation are
  included.

\item {\bf GDR parameters}: the peak cross section, energy and
  width of GDR are adjusted. The $\gamma$-ray transmission coefficient
  calculated from GDR can be re-normalized by a scaling factor.

\item {\bf Pre-equilibrium model parameters}: in TALYS the strength of
the exciton model is determined by the average squared matrix element $M^2$. 
  This is parameterized as 
  \begin{equation}
     M^2 = \frac{C_1}{A^3}
           \left\{ 
              7.48 C_2
              + \frac{4.62\times 10^5}
                     { \left[\frac{E^{\rm tot}}{n} + 10.7 C_3\right]^3 }
           \right\} \ ,
  \end{equation}
  where $n$ is the number of excitons, $C_1$, $C_2$, and $C_3$ are the
  adjustable parameters included. Since the two-component exciton
  model in TALYS distinguishes an exciton to be a neutron or a proton,
  the interaction strength between neutron ($\nu$) and proton $(\pi)$
  can be adjusted by introducing the scaling factors $M^2_{\pi\pi} =
  R_{\pi\pi}M^2$, $M^2_{\nu\nu} = R_{\nu\nu}M^2$, $M^2_{\pi\nu} =
  R_{\pi\nu}M^2$, and $M^2_{\nu\pi} = R_{\nu\pi}M^2$. These $R$
  factors are also adjusted.

\end{itemize}

The model parameter tuning processes were automatically carried out
using the specially developed tuning system, as shown in
Fig.~\ref{fig:KAERIprocedure}. The parameter tuning system consists of a
preprocessor and a tuning tool. The preprocessor automatically
generates the necessary input files for the tuning tool by reading the
output file of a default TALYS run and EXFOR files. The tuning tool
then repeatedly adjusts the model parameters, creates an input file
for TALYS, runs TALYS, compares the calculated results with the
experimental data, and computes $\chi^2$. It searches for the optimum
model parameters with minimum $\chi^2$ using gradient search, grid
search and/or random search. To reduce the computation time, multiple
CPUs are employed simultaneously using MPI. The whole process ends
after a user-specified number of loops and a user may monitor the
current status by looking at the plots which are automatically
generated after each calculation. In the following we compare the KAERI
evaluations with available experimental data for selected cases.

\subsubsection{$^{94}$Zr}

\begin{figure}[!htb]
 \begin{center}
   \resizebox{0.9\columnwidth}{!}{\includegraphics{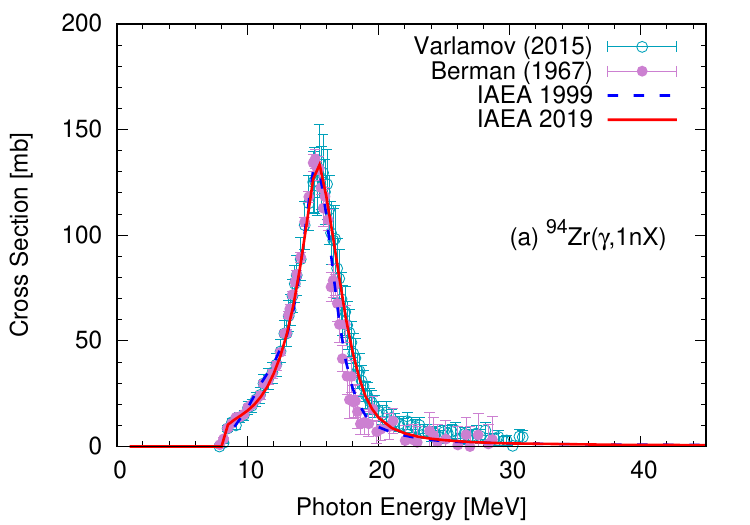}}\\
   \resizebox{0.9\columnwidth}{!}{\includegraphics{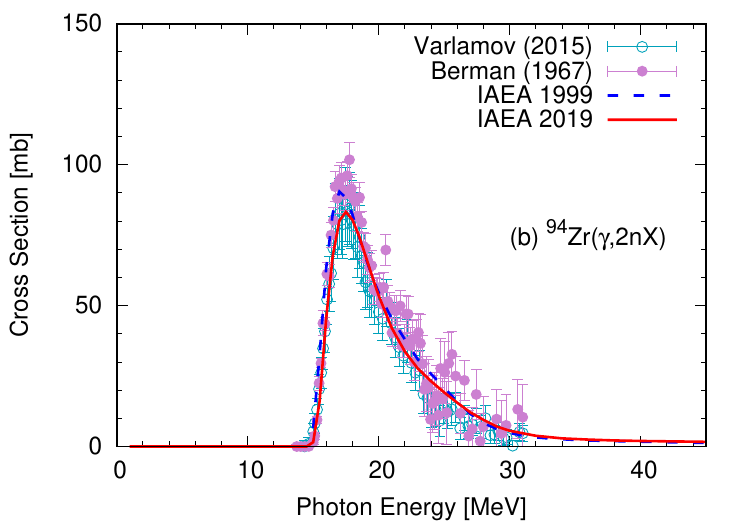}}\\
   \resizebox{0.9\columnwidth}{!}{\includegraphics{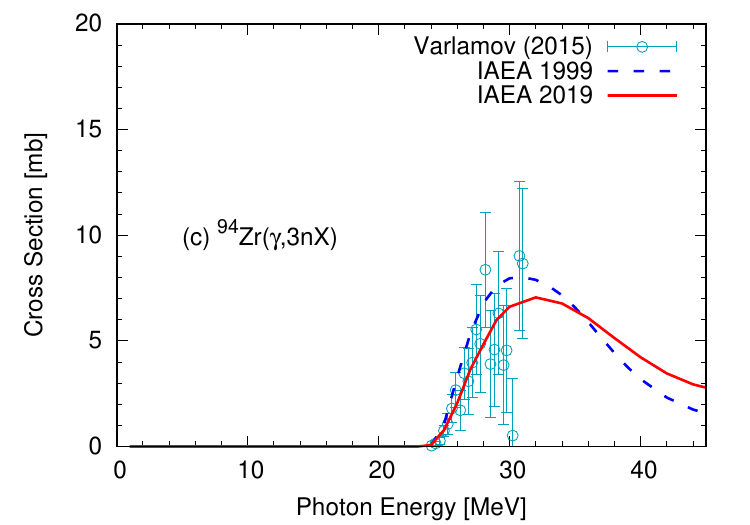}}\\
   \resizebox{0.9\columnwidth}{!}{\includegraphics{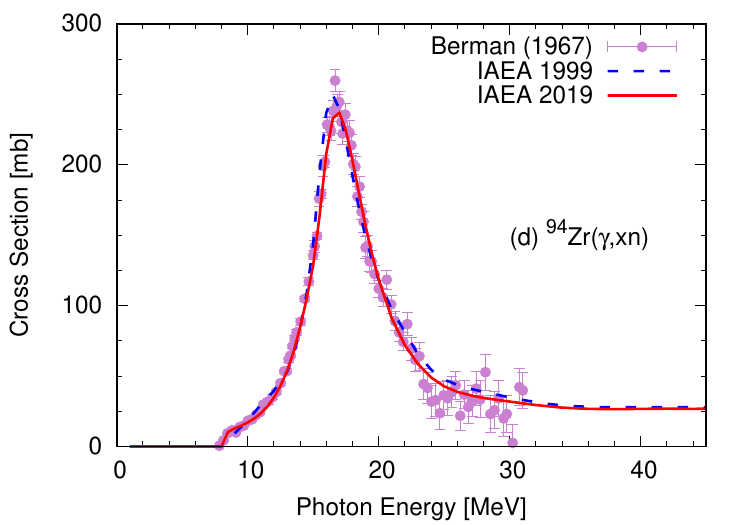}}
 \end{center}
 \caption{(Color online) Comparison of evaluated and experimental cross sections for
   the $\gamma$-ray induced reaction on $^{94}$Zr.  (a)
   $\sigma(\gamma,1nX)$, (b) $\sigma(\gamma,2nX)$, (c)
   $\sigma(\gamma,3nX)$, and (d) $\sigma_{xn}$.}
 \label{fig:KAERIZr94}
\end{figure}

The photonuclear data for $^{94}$Zr in the previous IAEA 1999 library were
evaluated by KAERI using the GUNF and GNASH code. The previous evaluation
was done based on the $(\gamma,1nX)$, $(\gamma,2nX)$, $\sigma_{Sn}$ and
$\sigma_{xn}$ reaction cross sections of Berman {\it et al.}~\cite{Berman1967}.
The present evaluation was performed to reproduce the new $(\gamma,1nX)$, 
$(\gamma,2nX)$, $(\gamma,3nX)$ and $\sigma_{Sn}$ cross sections of 
Varlamov {\it et al.}~\cite{Varlamov2015a} which are $F_i$-corrected ones.
The $(\gamma,1nX)$ cross sections of Varlamov are higher than those of Berman
by 10\% and  the new $(\gamma,2nX)$ cross sections of Varlamov are lower
than those of Berman by 15\% in the peak region.
The calculated photonuclear cross sections
for $^{94}$Zr are compared with the experimental and evaluated
data~\cite{Berman1967, Varlamov2015a} in
Fig.~\ref{fig:KAERIZr94}. As described above, the model parameters were adjusted to
reproduce the evaluated experimental data of Varlamov {\it et al.}

\subsubsection{$^{133}$Cs}

\begin{figure}[!htb]
 \begin{center}
   \resizebox{0.9\columnwidth}{!}{\includegraphics{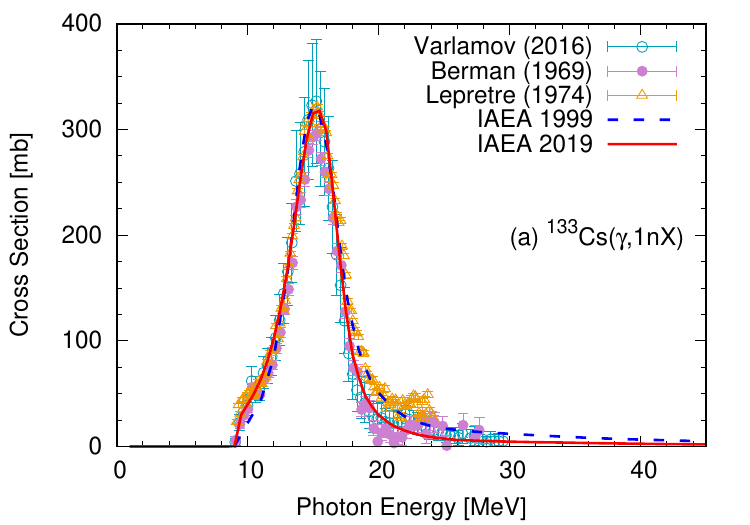}}\\
   \resizebox{0.9\columnwidth}{!}{\includegraphics{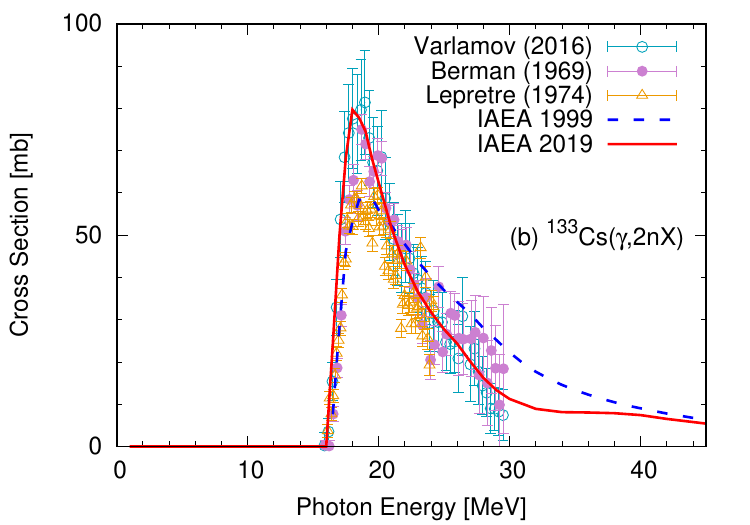}}\\
   \resizebox{0.9\columnwidth}{!}{\includegraphics{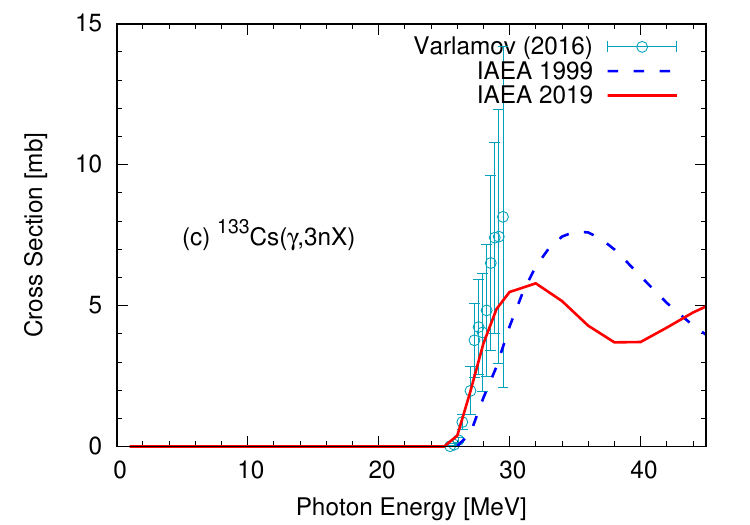}}\\
   \resizebox{0.9\columnwidth}{!}{\includegraphics{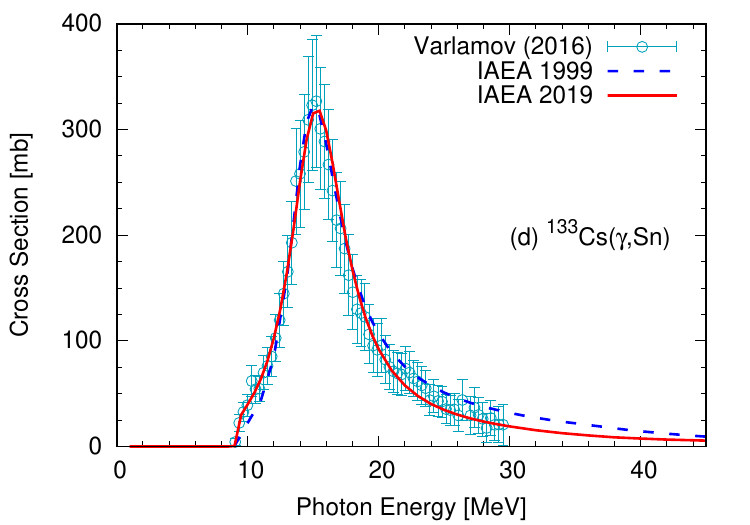}}
 \end{center}
 \caption{(Color online) Comparison of evaluated and experimental cross sections for
   the $\gamma$-ray induced reaction on $^{133}$Cs.  (a)
   $\sigma(\gamma,1nX)$, (b) $\sigma(\gamma,2nX)$, (c)
   $\sigma(\gamma,3nX)$, and (d) $\sigma_{Sn}$.}
 \label{fig:KAERICs133}
\end{figure}

The photonuclear data for $^{133}$Cs in the previous IAEA 1999 library were
evaluated by KAERI using the GUNF and GNASH code. The previous evaluation
was done based on the $(\gamma,1nX)$, $(\gamma,2nX)$, $\sigma_{Sn}$ and
$\sigma_{xn}$ reaction cross sections of Lepr\^{e}tre {\it et al.}~\cite{Lepretre1974}.
The present evaluation was performed to reproduce the new $(\gamma,1nX)$,
$(\gamma,2nX)$, $(\gamma,3nX)$ and $\sigma_{Sn}$ cross sections of
Varlamov {\it et al.}~\cite{Varlamov2016b} which are $F_i$-corrected ones.
The $(\gamma,1nX)$ cross sections of Varlamov are similar to those of Lepr\^{e}tre
and  the $(\gamma,2nX)$ cross sections of Varlamov are much higher 
than those of Lepr\^{e}tre by 30\% in the peak region.
The calculated photonuclear cross sections for $^{133}$Cs are compared
with the evaluated and experimental data~\cite{Varlamov2016b, Berman1969,
Lepretre1974} in Fig.~\ref{fig:KAERICs133}.

\subsubsection{$^{138}$Ba}

\begin{figure}[!htb]
 \begin{center}
   \resizebox{0.9\columnwidth}{!}{\includegraphics{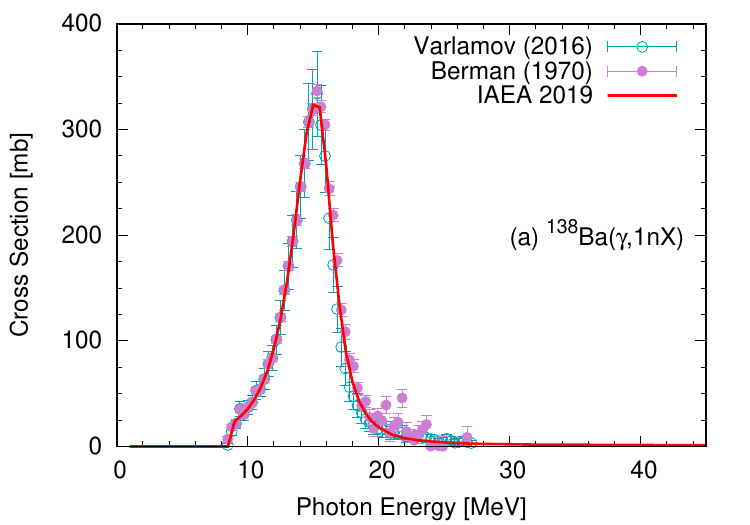}}\\
   \resizebox{0.9\columnwidth}{!}{\includegraphics{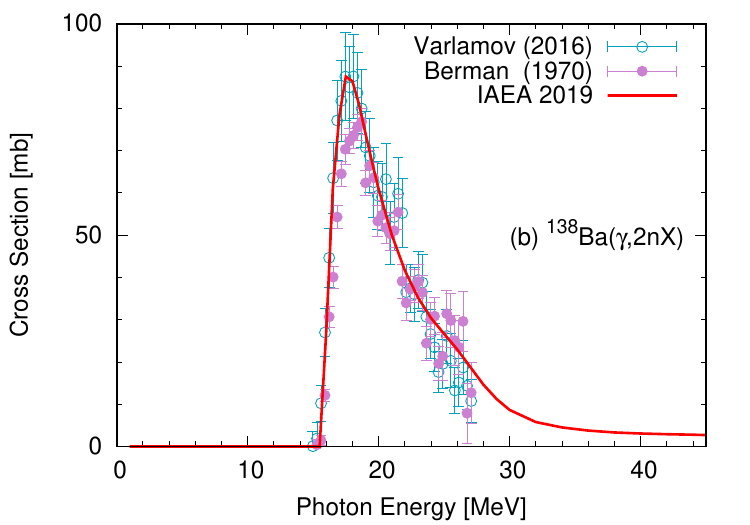}}\\
   \resizebox{0.9\columnwidth}{!}{\includegraphics{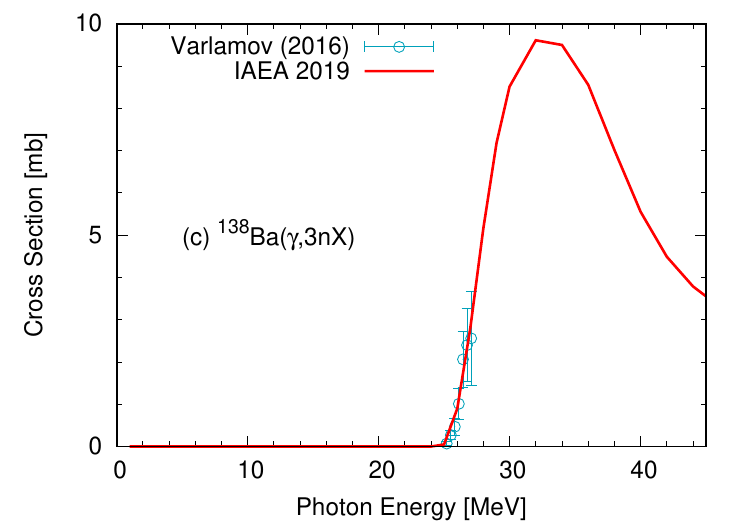}}
 \end{center}
 \caption{(Color online) Comparison of evaluated and experimental cross sections for
   the $\gamma$-ray induced reaction on $^{138}$Ba.  (a)
   $\sigma(\gamma,1nX)$, (b) $\sigma(\gamma,2nX)$, and (c)
   $\sigma(\gamma,3nX)$.}
 \label{fig:KAERIBa138}
\end{figure}

The photonuclear data for $^{138}$Ba were not included in the previous IAEA 1999 library.
The present evaluation was performed to reproduce the new $(\gamma,1nX)$,
$(\gamma,2nX)$, $(\gamma,3nX)$ and $\sigma_{Sn}$ cross sections of
Varlamov {\it et al.}~\cite{Varlamov2016b} which are $F_i$-corrected ones.
The calculated photonuclear cross sections for $^{138}$Ba are compared
with the experimental and evaluated data~\cite{Berman1970, Varlamov2016b} in
Fig.~\ref{fig:KAERIBa138}.

\subsubsection{$^{142}$Ce}

\begin{figure}[!htb]
 \begin{center}
   \resizebox{0.9\columnwidth}{!}{\includegraphics{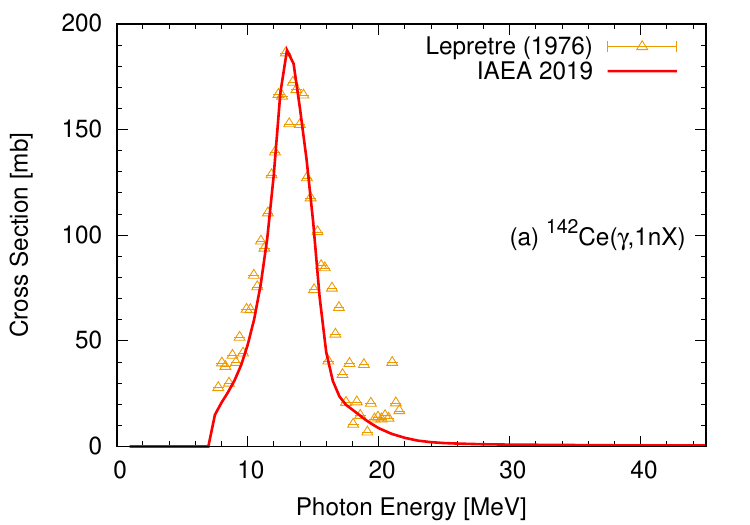}}\\
   \resizebox{0.9\columnwidth}{!}{\includegraphics{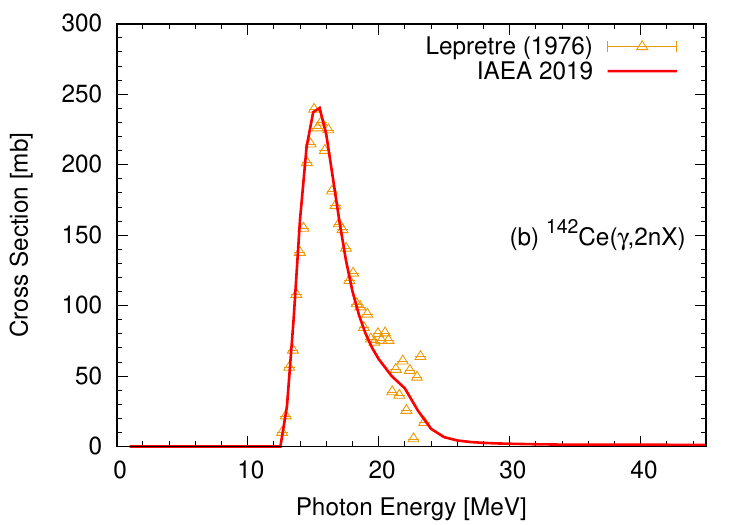}}\\
   \resizebox{0.9\columnwidth}{!}{\includegraphics{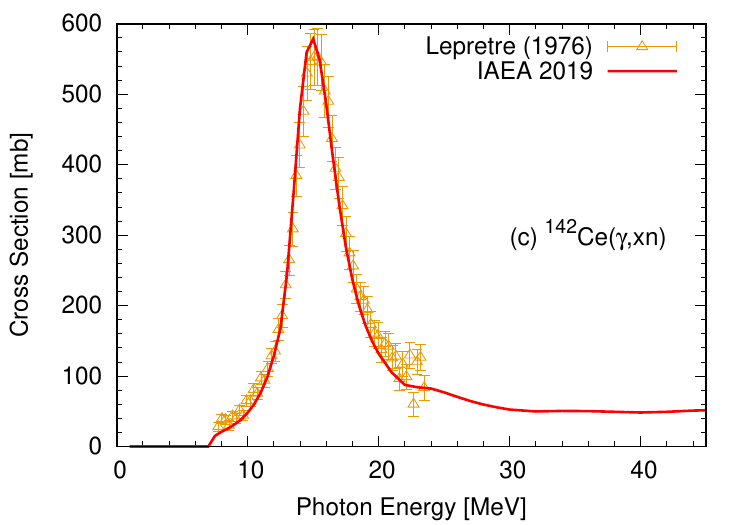}}
 \end{center}
 \caption{(Color online) Comparison of evaluated and experimental cross sections for
   the $\gamma$-ray induced reaction on $^{142}$Ce.  (a)
   $\sigma(\gamma,1nX)$, (b) $\sigma(\gamma,2nX)$, and (c)
   $\sigma_{xn}$.}
 \label{fig:KAERICe142}
\end{figure}

The photonuclear data for $^{142}$Ce were not included in the previous IAEA 1999 library.
The present evaluation was performed to reproduce the new $(\gamma,1nX)$,
$(\gamma,2nX)$ and $\sigma_{xn}$ cross sections of
Lepr\^{e}tre {\it et al.}~\cite{Lepretre1976}.
The calculated photonuclear cross sections for $^{142}$Ce are compared
with the experimental data~\cite{Lepretre1976} in
Fig.~\ref{fig:KAERICe142}.

  \clearpage

  \subsection{Data Evaluation at IFIN-HH}
  \label{subsec:evalIFINHH}
Photo-neutron reaction cross sections for $^{59}$Co, $^{89}$Y,
$^{103}$Rh, $^{159}$Tb, $^{165}$Ho, $^{169}$Tm and $^{181}$Ta have
been recently measured in the GDR region using quasi-monochromatic LCS
$\gamma$-ray beams at the NewSUBARU facility. The partial
photo-neutron cross sections, $\sigma_{inX}$ with $i = 1$ to 4, were
determined with the direct neutron multiplicity sorting method
described in Sec.~\ref{subsubsec:DNMSFED}, which is based on a slow
response flat efficiency neutron detector. The average neutron
emission energy corresponding to the total neutron-yield cross section
was also provided using the ring ratio method, as described in
Sec.~\ref{subsubsec:QAP}. The total $\sigma_{Sn}$ and partial
$\sigma_{inX}$ ($i=1$ -- 4) cross sections have been communicated
among the CRP members for evaluations, leaving possible small
modifications to be made on the final results which will be published
separately. We evaluated the newly measured NewSUBARU data for the
seven nuclei listed above in an attempt to resolve the long-standing
discrepancies between the Livermore and Saclay measurements detailed
in Sec.~\ref{subsec:evaldata}.

The EMPIRE statistical model code~\cite{INDC0603} has been employed
for the evaluations. For each evaluated nucleus, we have investigated
which of the MLO1, MLO2, SLO and SMLO closed-forms for E1 $\gamma$-ray
strength functions~\cite{RIPL3} implemented in EMPIRE reproduces best
the experimental data. We also implemented in EMPIRE a third SLO
component, which was used if necessary. EMPIRE also includes the
quasi-deuteron contribution to describe the photo-absorption cross
section above the GDR region, up to 200~MeV.

When evaluating the NewSUBARU experimental data, we took into account
that, as also in the case of Saclay and Livermore data, the detection
system was not sensitive to the charged particle emission from the
target. The NewSUBARU experiments provided the sum of cross sections
with $i$ neutrons in the final state $(\gamma,inX)$, where $i=1$ to
4. The total photo-neutron cross section $\sigma_{Sn}$ was obtained as
the sum $\sum_i \sigma_{inX}$. For each nucleus, we adjusted the
photo-absorption to reproduce the total photo-neutron cross section
$\sigma_{Sn}$. Finally we tuned the level density parameters, and the
single particle level density parameters from the PCROSS
pre-equilibrium model in order to reproduce the experimental
$\sigma_{inX}$ data with $i=1$ -- 4. The global Koning-Delaroche
potentials~\cite{Koning2003} have been used to obtain the transmission
coefficients for proton and neutron emission.

In order to compute complex $\sigma_{inX}$ cross sections, one has to
determine cross sections for different reaction chains which populate
the same residual nucleus. For example, let us consider the $_Z^A$X
photo-disintegration to $_{Z-1}^{A-3}$X nucleus, which can be produced
by six different reaction chains $(\gamma,nnp)$, $(\gamma,npn)$,
$(\gamma,pnn)$, $(\gamma,nd)$, $(\gamma,dn)$ and $(\gamma,t)$. We
compute the sum cross sections $(\gamma,nnp) + (\gamma,npn) +
(\gamma,pnn)$ and $(\gamma,nd) + (\gamma,dn)$ separately, which
contribute to the $(\gamma,2nX)$ and $(\gamma,1nX)$ reactions,
respectively.

Because of limitations imposed by the available computing power and
huge memory requirement, the EMPIRE statistical model code uses only
one memory buffer per compound nucleus (CN) which is incremented by
all reactions chains that populate the same CN. Thus, the code
provides the population cross sections of each accessible CN
disregarding the way it was produced and the emission cross section of
each particle that can be emitted from the given CN. To obtain the
cross sections for different reaction chains which populate the same
residual nucleus, we developed a method of reconstructing all possible
pathways by processing the statistical model code results. This
treatment is particularly important for low mass nuclei which have a
large contribution of charged particle emission.

The method uses the additional information provided by EMPIRE
concerning exclusive cross sections, which are defined as the double
differential cross section / occupation for obtaining a nucleus in a
certain state after having emitted a certain particle (proton,
neutron, $\alpha$-particle, $\ldots$). This cross section takes into
account all emissions of that particular particle that lead to the
given final state.  However, the method does not provide exact
results, because the information concerning spin and parity population
distribution of the CN is taken into account only as an
average. EMPIRE could easily avoid this approximation, but the
calculation would grow by many orders of magnitude, an increase which
would not be justified by an advantage of more precise splitting of
inclusive spectra into their exclusive components.  Also, obtaining
cross sections for all pathways is possible up to a certain energy
over which, due to the complexity of all the possible pathways, the
information concerning exclusive population cross section alone is not
anymore sufficient to solve very complex equation systems. The
NewSUBARU measurements were performed up to energies located under
this energy limit to takes all possible pathways into account
properly.

\subsubsection{$^{59}$Co}
\label{subsubsec:EVALIFIN_CO}

\begin{figure}[!htb]
 \begin{center}
  \resizebox{0.9\columnwidth}{!}{\includegraphics{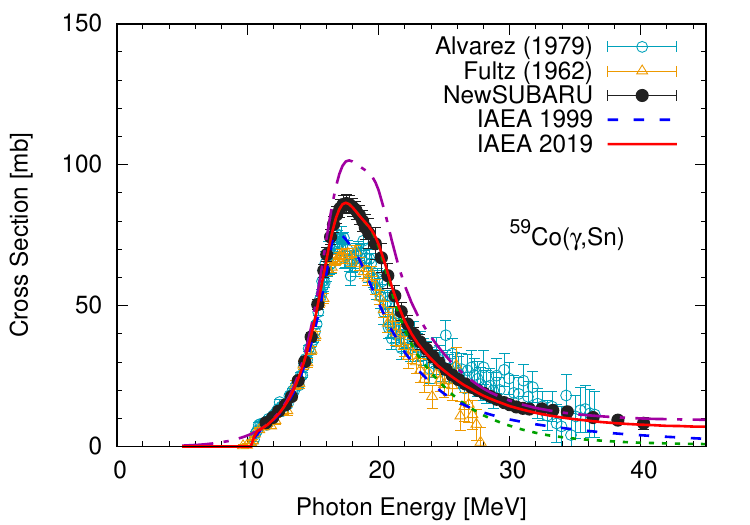}} 
 \end{center}
 \caption{(Color online) Comparison between experimental and evaluated (solid line) 
 total cross sections $\sigma_{Sn}$ for $^{59}$Co. 
 Evaluated photo-absorption cross section including charged-particle
 channels is also shown by the dot-dashed line. The evaluated sum
 cross section $\sum_i\sigma_{in}$ ($i=1$ -- 3) of all  channels
 without charged-particle emission is also shown by the dotted line.}
 \label{fig:IFINHHCo59sn}
\end{figure}

\begin{figure}
 \begin{center}
  \resizebox{0.9\columnwidth}{!}{\includegraphics{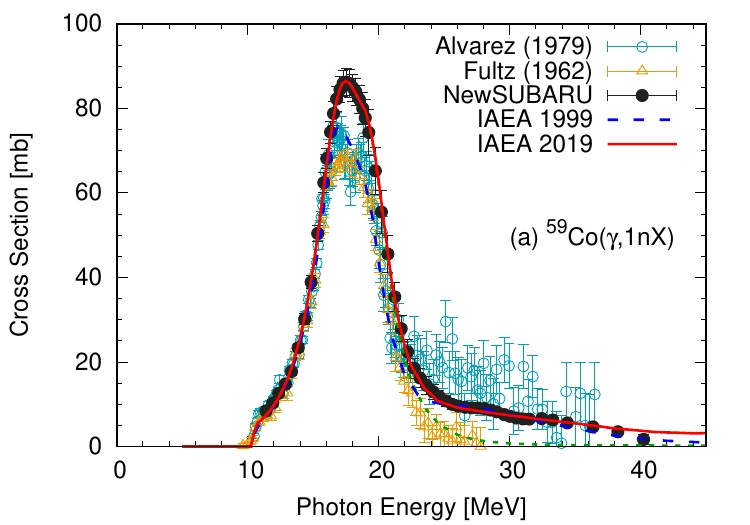}} 
  \resizebox{0.9\columnwidth}{!}{\includegraphics{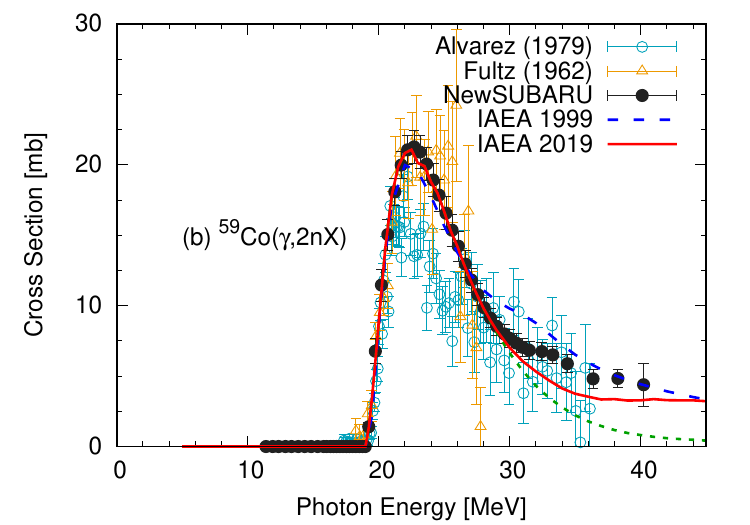}} 
  \resizebox{0.9\columnwidth}{!}{\includegraphics{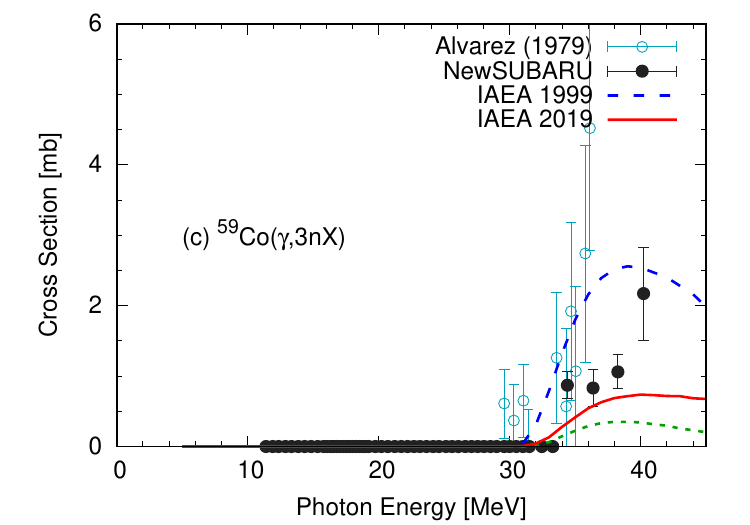}} 
 \end{center}
 \caption{(Color online) Comparison between experimental and evaluated (solid lines) 
 partial cross sections for $^{59}$Co: (a)~$\sigma_{1nX}$, 
 (b)~$\sigma_{2nX}$, and (c)~$\sigma_{3nX}$. See Fig.~\ref{fig:IFINHHCo59sn}
 for the dotted line.}
 \label{fig:IFINHHCo59}
\end{figure}

The previous $^{59}$Co evaluation has been performed by KAERI using
the GNASH code~\cite{LA6947, LA12343}. The KAERI evaluation modeled the
$(\gamma,1nX)$ and $(\gamma,2nX)$ reaction cross sections of Alvarez
{\it et al.}~\cite{Alvarez1979} and Fultz {\it et
al.}~\cite{Fultz1962}, the $(\gamma,3nX)$ reaction cross sections of
Alvarez and the $\sigma_{xn}$ data of Alvarez and Bazhanov {\it et
al.}~\cite{Bazhanov1964}. New $\sigma_{Sn}$, $\sigma_{1nX}$,
$\sigma_{2nX}$ and $\sigma_{3nX}$ have been measured at NewSUBARU for
$^{59}$Co in the energy region between 10 and 40~MeV. The NewSUBARU
data are shown by the filled circles in Figs.~\ref{fig:IFINHHCo59sn}
and \ref{fig:IFINHHCo59}.  The NewSUBARU $(\gamma,Sn)$ cross section
is about 20\% higher than the Livermore data of Fultz and Alvarez and
the peaks of the two GDR Lorentzians are not well separated. The MLO1
$\gamma$-ray strength function with two centroids reproduced best the
$(\gamma,Sn)$ cross sections.  A third SLO has been introduced to
reproduce the large width of the GDR region.

For this particular nucleus newly measured at NewSUBARU, the
charged-particle emission contribution is stronger, as can be seen in
Figs~\ref{fig:IFINHHCo59sn} and \ref{fig:IFINHHCo59}. The EMPIRE
photo-absorption and the photo-neutron cross sections are displayed in
Fig.~\ref{fig:IFINHHCo59sn} by the dot-dashed and continuous lines,
respectively, where the difference between the two cross sections is
attributed to the charged-particle emission channel which is
unaccompanied by neutron emission.  The sum cross section of all
neutron emission only channels is shown with the dotted line.  We can
observe that, because of the strong charged particle contribution, the
GDR parameters, namely the peak cross section and width, cannot be
obtained by directly fitting the measured photo-neutron cross section
with a set of Lorentzians. Figure~\ref{fig:IFINHHCo59} shows the
comparison between $(\gamma,inX)$ (solid lines) and $(\gamma,in)$
(dotted lines) cross sections for $i = 1$ -- 3, where the emission of
accompanying charged particles increases with increasing energy.

\clearpage

\subsubsection{$^{89}$Y}
\label{subsubsec:EVALIFIN_Y}

\begin{figure}
 \begin{center}
  \resizebox{0.9\columnwidth}{!}{\includegraphics{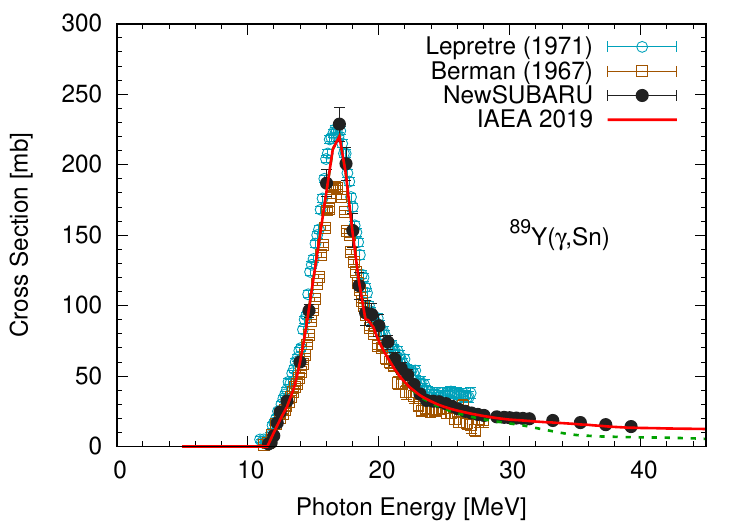}} 
 \end{center}
 \caption{(Color online) Comparison between evaluated (solid line) and experimental 
 total cross sections $\sigma_{Sn}$ for $^{89}$Y. The evaluated 
 sum cross section $\sum_i\sigma_{in}$ ($i=1$ -- 3) of all 
 channels without charged-particle emission is also shown by the dotted line.}
 \label{fig:IFINHHY89sn}       
\end{figure}

\begin{figure}
 \begin{center}
  \resizebox{0.9\columnwidth}{!}{\includegraphics{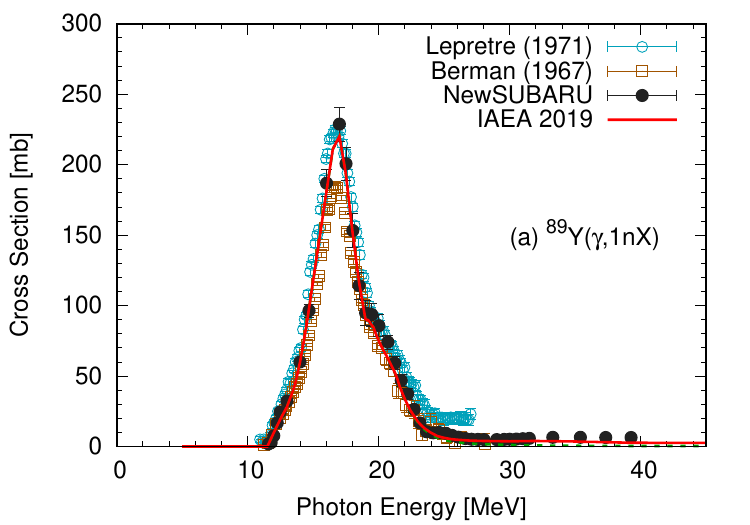}} 
  \resizebox{0.9\columnwidth}{!}{\includegraphics{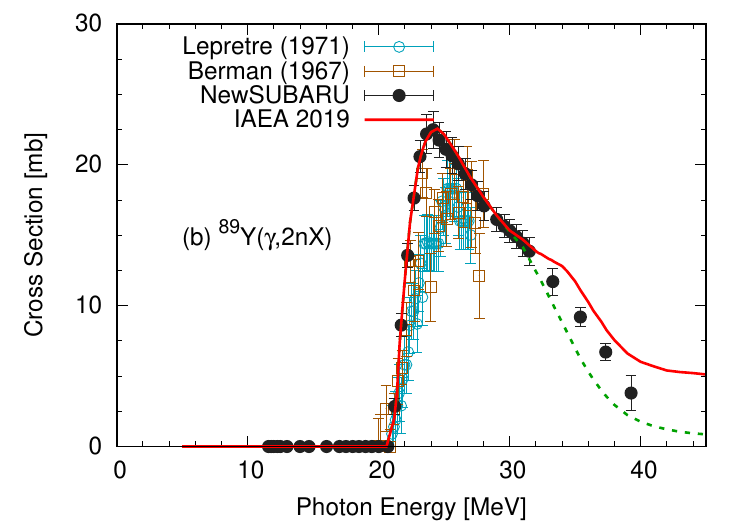}} 
  \resizebox{0.9\columnwidth}{!}{\includegraphics{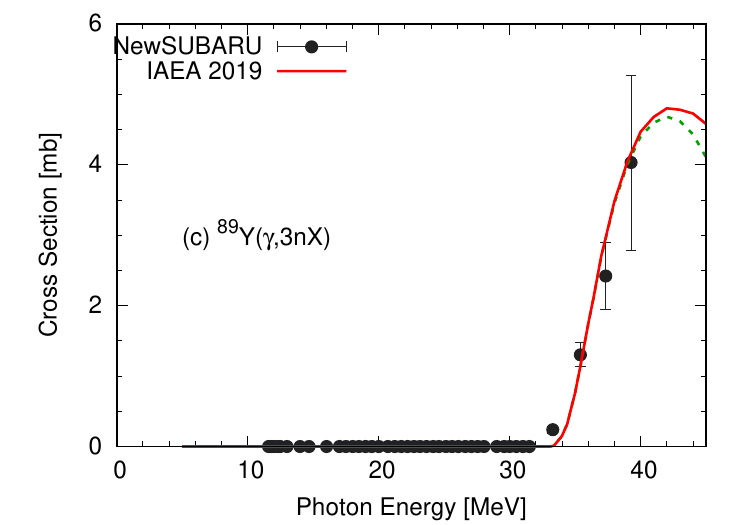}} 
 \end{center}
 \caption{(Color online) Comparison between experimental and evaluated (solid lines) 
 partial cross sections for $^{89}$Y: (a)~$\sigma_{1nX}$, 
 (b)~$\sigma_{2nX}$, and (c)~$\sigma_{3nX}$.
 See Fig.~\ref{fig:IFINHHY89sn} for the dotted lines.}
 \label{fig:IFINHHY89}
\end{figure}

$^{89}$Y was not included in the previous IAEA 1999 database. The
$(\gamma,1nX)$, $(\gamma,2nX)$, $(\gamma,Sn)$ reaction cross sections
and the neutron yield $\sigma_{xn}$ have been measured by Berman {\it
et al.}~\cite{Berman1967} and by Lepr\^{e}tre {\it et
al.}~\cite{Lepretre1971} between 11 and 28~MeV excitation energy. New
$(\gamma,Sn)$, $(\gamma,1nX)$, $(\gamma,2nX)$ and $(\gamma,3nX)$ cross
sections have been measured at NewSUBARU for $^{89}$Y in the energy
region between 11 and 40~MeV. The NewSUBARU and the Saclay cross
sections are in good agreement in the GDR peak energy region. The
$(\gamma,Sn)$ reaction cross section was best reproduced by using the
SMLO model for the $\gamma$-ray strength function with two centroids.
The evaluation is compared to the experimental data in
Figs.~\ref{fig:IFINHHY89sn} and \ref{fig:IFINHHY89}.

\subsubsection{$^{103}$Rh}
\label{subsubsec:EVALIFIN_RH}

\begin{figure}
 \begin{center}
  \resizebox{0.9\columnwidth}{!}{\includegraphics{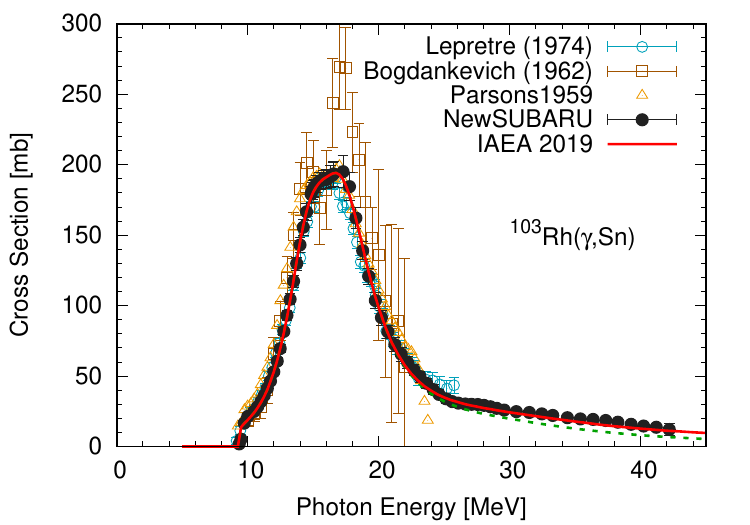}} 
 \end{center}
 \caption{(Color online) Comparison between experimental and evaluated (solid line) 
 total cross sections $\sigma_{Sn}$ for $^{103}$Rh.
 The evaluated sum cross section $\sum_i\sigma_{in}$ ($i=$1 -- 3) 
 of all channels without charged-particle emission is also 
 shown by the dotted line.}
 \label{fig:IFINHHRh103sn}
\end{figure}

\begin{figure}
 \begin{center}
  \resizebox{0.9\columnwidth}{!}{\includegraphics{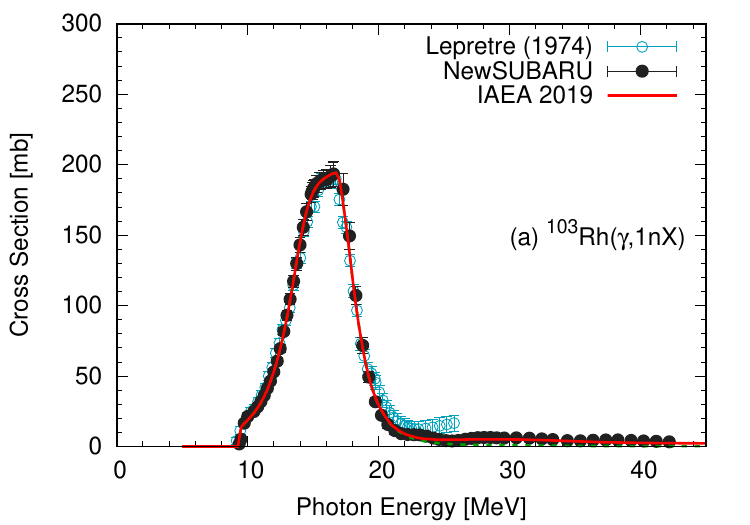}} 
  \resizebox{0.9\columnwidth}{!}{\includegraphics{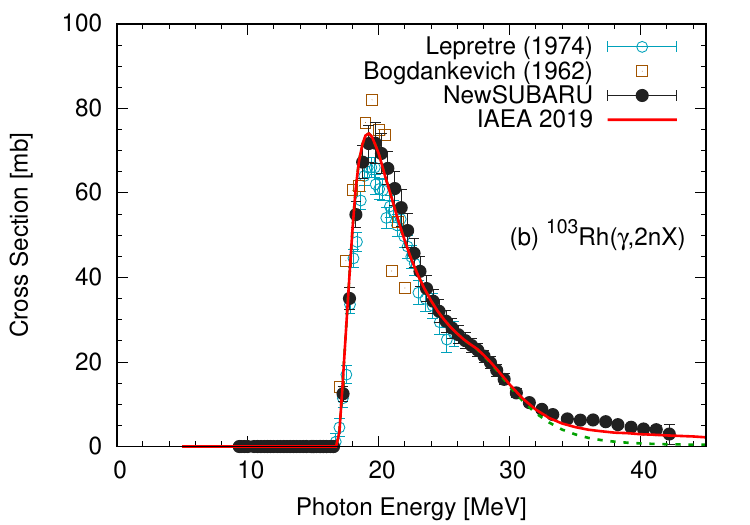}} 
  \resizebox{0.9\columnwidth}{!}{\includegraphics{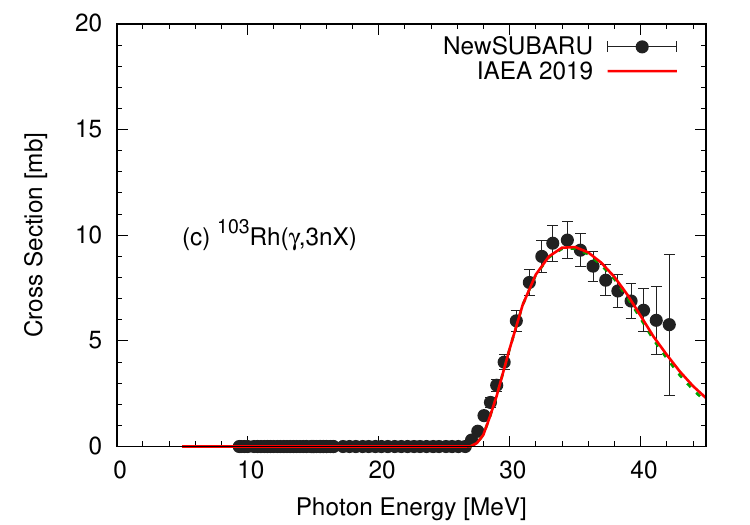}} 
 \end{center}
 \caption{(Color online) Comparison between experimental and evaluated (solid lines) 
 partial cross sections for $^{103}$Rh: (a)~$\sigma_{1nX}$, 
 (b)~$\sigma_{2nX}$, and (c)~$\sigma_{3nX}$.
 See Fig.~\ref{fig:IFINHHRh103sn} for the dotted lines.}
 \label{fig:IFINHHRh103}
\end{figure}

$^{103}$Rh was not included in the previous IAEA 1999 database. The
$(\gamma,1nX)$, $(\gamma,2nX)$, $(\gamma,Sn)$ reaction cross sections
and the neutron yield $\sigma_{xn}$ have been measured by Lepr\^{e}tre
{\it et al.}~\cite{Lepretre1974} between 9 and~26 MeV excitation
energy. The $(\gamma,Sn)$ cross section has been measured using
bremsstrahlung beams by Parsons~\cite{Parsons1959} and by Bogdankevich
{\it et al.}~\cite{Bogdankevich1962}, where Bogdankevich reported also
the $(\gamma,2nX)$ cross section. New $(\gamma,Sn)$, $(\gamma,1nX)$,
$(\gamma,2nX)$ and $(\gamma,3nX)$ cross sections have been measured at
NewSUBARU for $^{103}$Rh in the energy region between 9 and 42~MeV.
The NewSUBARU and the Saclay cross sections are in overall good
agreement in the entire investigated energy region. The MLO2 with two
centroids reproduced best the $(\gamma,Sn)$ cross sections. A third
SLO has been introduced to reproduce the large width of the GDR peak
region. The evaluation is compared to the experimental data in
Figs.~\ref{fig:IFINHHRh103sn} and \ref{fig:IFINHHRh103}.  The
comparison between the experimental and calculated average neutron
emission energy corresponding to the neutron-yield cross section is
displayed in Fig.~\ref{fig:IFINHHRh103}. We underline that the
model parameter was tuned for the experimental cross sections not for
the average neutron energies.

\begin{figure}
 \begin{center}
  \resizebox{0.9\columnwidth}{!}{\includegraphics{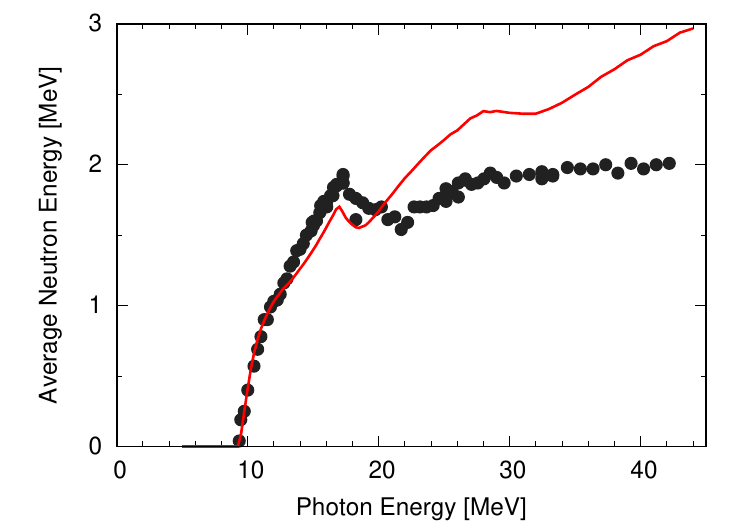}} 
 \end{center}
 \caption{(Color online) Comparison between evaluated and experimental values for
   average energy of neutrons emitted in photo-induced
   reactions on $^{103}$Rh.}
 \label{fig:IFINHHRh103eave}
\end{figure}

\subsubsection{$^{159}$Tb}
\label{subsubsec:EVALIFIN_TB}

\begin{figure}
 \begin{center}
  \resizebox{0.9\columnwidth}{!}{\includegraphics{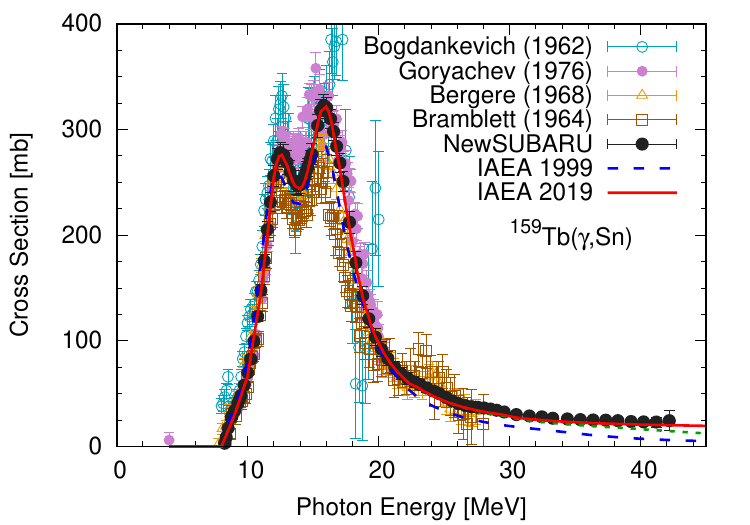}} 
 \end{center}
 \caption{(Color online) Comparison between experimental and evaluated 
 total cross sections $\sigma_{Sn}$ for $^{159}$Tb. The solid line is the current evaluation,
 and the dashed line is IAEA 1999 library~\cite{IAEAPhoto1999}. The evaluated sum cross section 
 $\sum_i\sigma_{in}$ ($i=1$ -- 4) of all channels without 
 charged-particle emission is also shown by the dotted line.}
 \label{fig:IFINHHTb159sn}       
\end{figure}

\begin{figure}
 \begin{center}
  \resizebox{0.9\columnwidth}{!}{\includegraphics{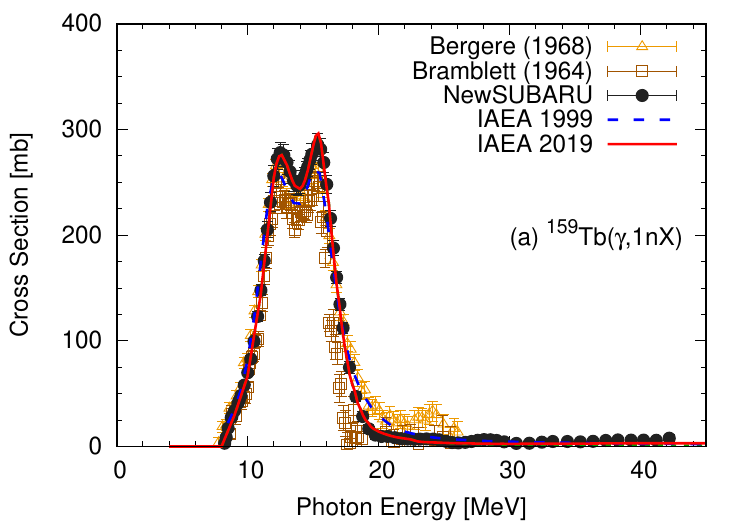}} 
  \resizebox{0.9\columnwidth}{!}{\includegraphics{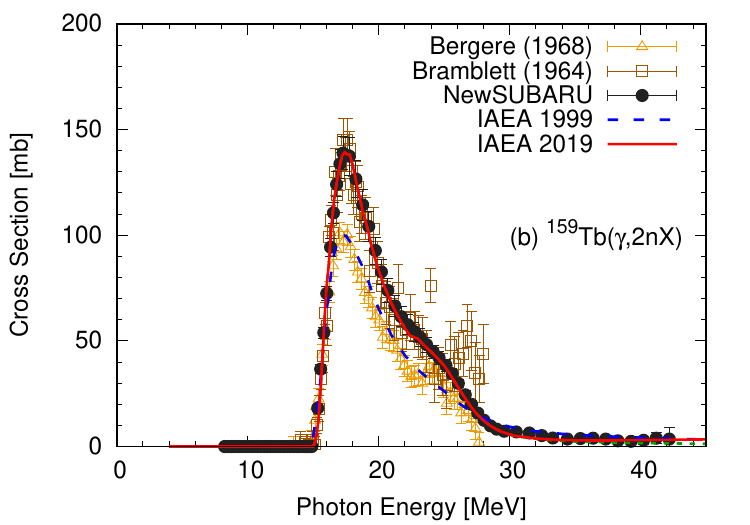}} 
  \resizebox{0.9\columnwidth}{!}{\includegraphics{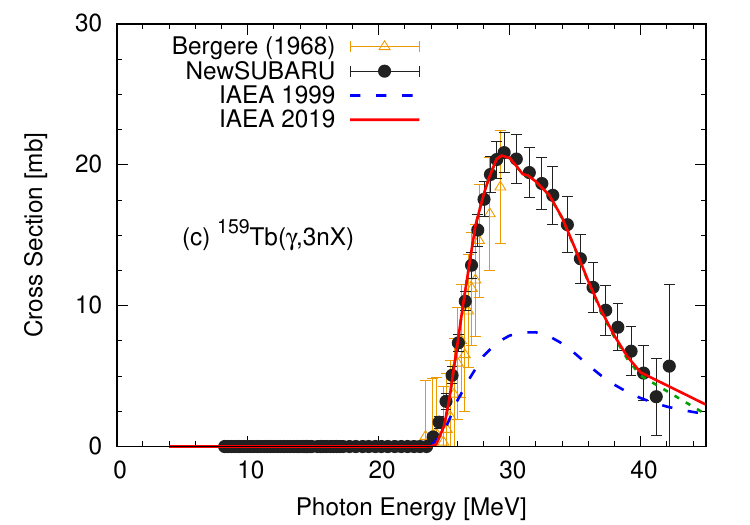}} 
  \resizebox{0.9\columnwidth}{!}{\includegraphics{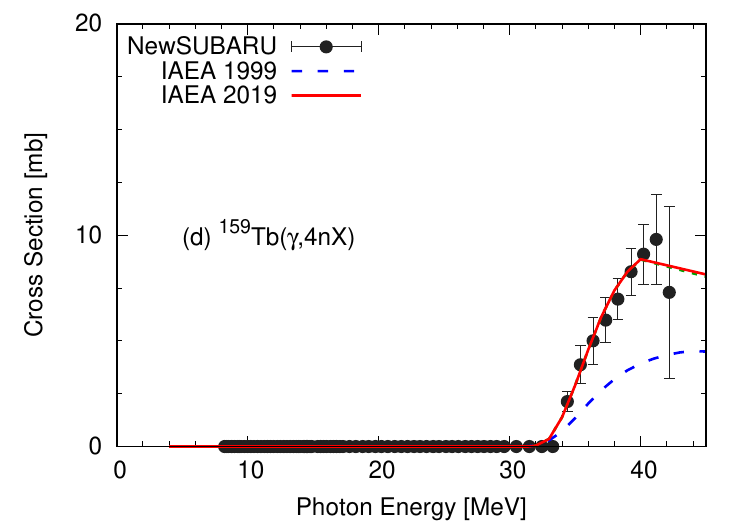}} 
 \end{center}
 \caption{(Color online) Comparison between experimental and evaluated (solid and dashed lines) 
 partial cross sections for $^{159}$Tb: (a)~$\sigma_{1nX}$, 
 (b)~$\sigma_{2nX}$, (c)~$\sigma_{3nX}$, and (d)~$\sigma_{4nX}$.
 See Fig.~\ref{fig:IFINHHTb159sn} for the dotted lines.}
 \label{fig:IFINHHTb159}
\end{figure}

The previous $^{159}$Tb evaluation has been performed by KAERI using
the GNASH code. The $(\gamma,1nX)$, $(\gamma,2nX)$ and $(\gamma,xn)$
cross sections have been measured both by Bramblett {\it et
al.}~\cite{Bramblett1964} and by Berg\`{e}re {\it et
al.}~\cite{Bergere1968}. Berg\`{e}re also reported the $(\gamma,3nX)$
and $(\gamma,Sn)$ cross sections.  The $\sigma_{Sn}$ have been
measured using bremsstrahlung beams by Bogdankevich {\it et
al.}~\cite{Bogdankevich1962} and by Goryachev {\it et
al.}~\cite{Goryachev1976}.  The KAERI evaluation follows the
Berg\`{e}re $(\gamma,Sn)$, $(\gamma,1 \sim 3nX)$ cross sections. New
$(\gamma,Sn)$, $(\gamma,1nX)$, $(\gamma,2nX)$, $(\gamma,3nX)$ and
$(\gamma,4nX)$ cross sections have been measured at NewSUBARU for
$^{159}$Tb in the energy region between 8 and 42~MeV. The NewSUBARU
$(\gamma,1nX)$ and $(\gamma,3nX)$ cross sections are in agreement with
the Saclay data, while the new $(\gamma,2nX)$ cross sections are in
agreement with the Livermore data. For the evaluation, the NewSUBARU
$(\gamma,Sn)$ reaction cross section was best reproduced by using the
SMLO model for the $\gamma$-ray strength function with two
centroids. The evaluation is compared to the experimental data in
Figs.~\ref{fig:IFINHHTb159sn} and \ref{fig:IFINHHTb159}.

\subsubsection{$^{165}$Ho}
\label{subsubsec:EVALIFIN_HO}

\begin{figure}
 \begin{center}
  \resizebox{0.9\columnwidth}{!}{\includegraphics{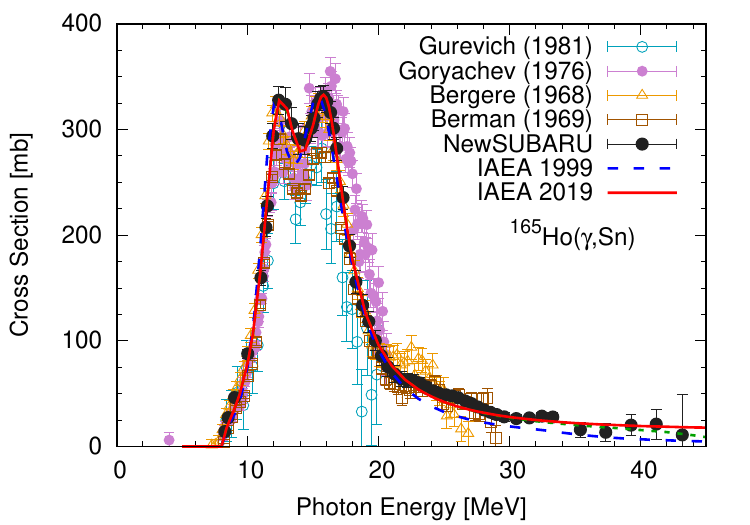}} 
 \end{center}
 \caption{(Color online) Comparison between experimental and evaluated (solid line) 
 $\sigma_{Sn}$ for $^{165}$Ho. The evaluated sum cross section 
 $\sum_i\sigma_{in}$ ($i=1$ -- 4) of all channels without 
 charged-particle emission is also shown by the dotted lines.}
 \label{fig:IFINHHHo165sn}
\end{figure}

\begin{figure}[!h]
 \begin{center}
   \resizebox{0.9\columnwidth}{!}{\includegraphics{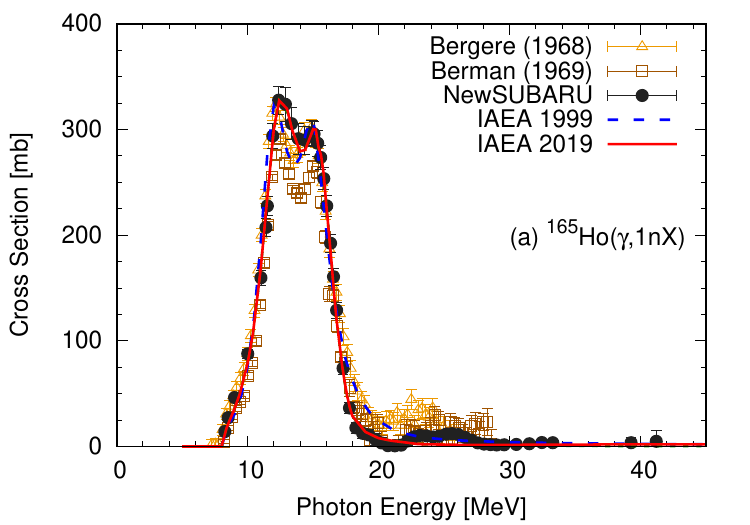}}
   \resizebox{0.9\columnwidth}{!}{\includegraphics{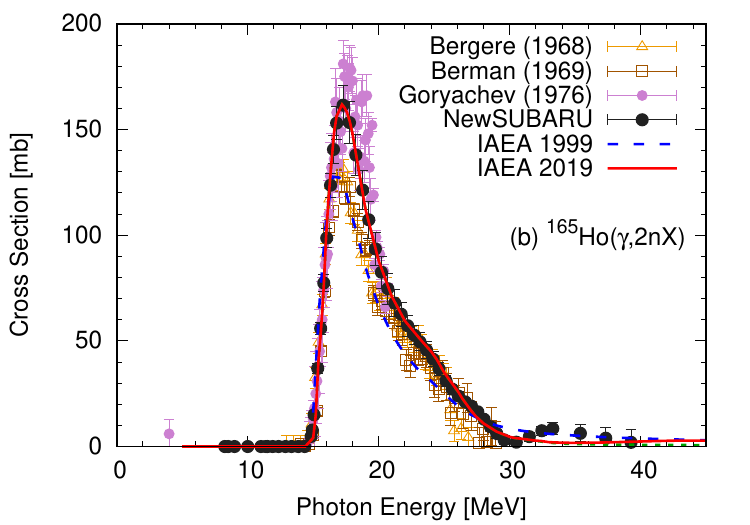}}
   \resizebox{0.9\columnwidth}{!}{\includegraphics{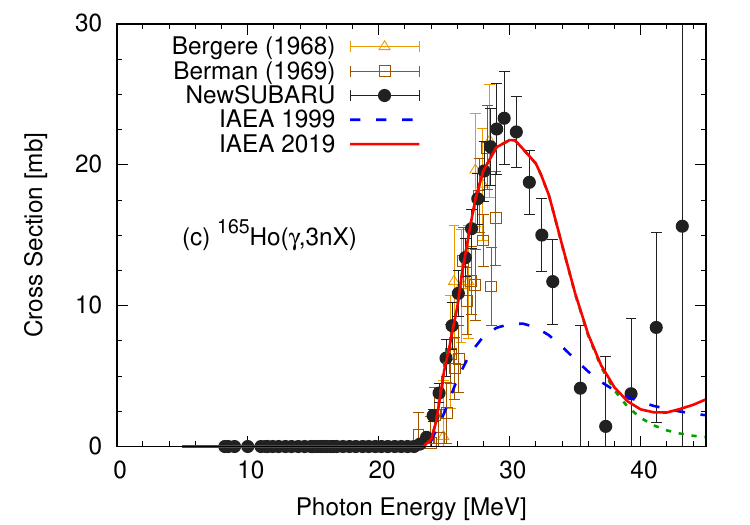}}
   \resizebox{0.9\columnwidth}{!}{\includegraphics{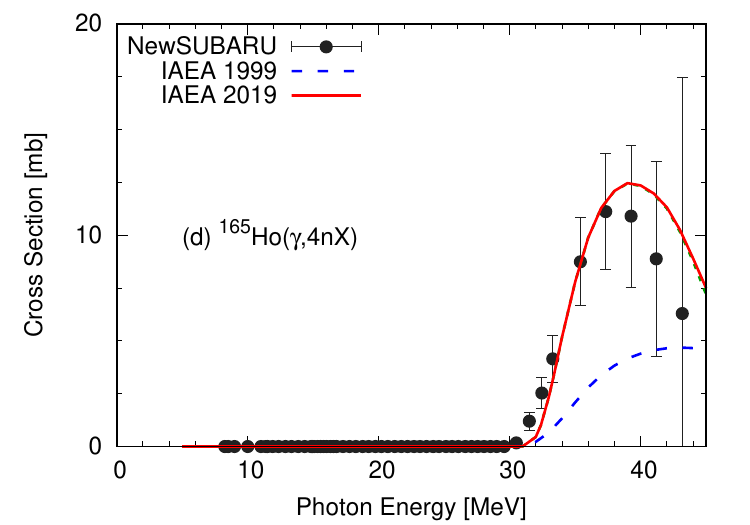}}
 \end{center}
 \caption{(Color online) Comparison between experimental and evaluated (solid lines) 
 partial cross sections for $^{165}$Ho: (a)~$\sigma_{1nX}$, 
 (b)~$\sigma_{2nX}$, (c)~$\sigma_{3nX}$, and (d)~$\sigma_{4nX}$.
 See Fig.~\ref{fig:IFINHHHo165sn} for the dotted lines.}
 \label{fig:IFINHHHo165}
\end{figure}

The previous $^{165}$Ho evaluation has been performed by KAERI using
the GNASH code. The $(\gamma,Sn)$, $(\gamma,1nX)$, $(\gamma,2nX)$,
$(\gamma,3nX)$ cross sections have been measured both by Berman {\it
et al.}~\cite{Berman1969b} and by Berg\`{e}re {\it et al.}~\cite{Bergere1968}. 
The total nuclear photo-absorption cross section $\sigma_{abs}$ has been 
measured by Gurevich {\it et al.}~\cite{Gurevich1981}. The $\sigma_{Sn}$ 
and $\sigma_{2nx}$ have also been reported by Goryachev 
{\it et al.}~\cite{Goryachev1976}. 
The KAERI evaluation followed the Berg\`{e}re cross sections. 
New $(\gamma,Sn)$, $(\gamma,1nX)$, $(\gamma,2nX)$,
$(\gamma,3nX)$ and $(\gamma,4nX)$ cross sections have been measured at
NewSUBARU for $^{165}$Ho in the energy region between 8 and
43~MeV. The NewSUBARU $(\gamma,1nX)$ cross sections are in better
agreement with the Saclay measurement while the new $(\gamma,2nX)$
cross sections are higher than both the Saclay and Livermore data by
$\sim$20\% in the peak region. For the evaluation, the NewSUBARU
$(\gamma,Sn)$ reaction cross section was best reproduced by using the
SMLO model for the $\gamma$-ray strength function with two centroids.
The evaluation is compared to the experimental data in 
Figs.~\ref{fig:IFINHHHo165sn} and \ref{fig:IFINHHHo165}.

\subsubsection{$^{169}$Tm}
\label{subsubsec:EVALIFIN_TM}

\begin{figure}
 \begin{center}
  \resizebox{0.9\columnwidth}{!}{\includegraphics{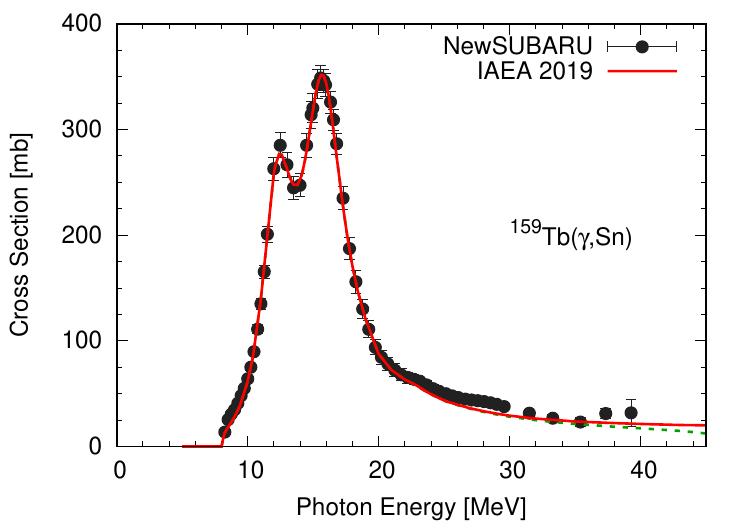}} 
 \end{center}
 \caption{(Color online) Comparison between evaluated and NewSUBARU
 $\sigma_{Sn}$ for $^{169}$Tm. The evaluated sum cross section 
 $\sum_i\sigma_{in}$ ($i=1$ -- 4) of all channels without 
 charged-particle emission is also shown by the dotted lines.}
 \label{fig:IFINHHTm169sn}
\end{figure}

\begin{figure}
 \begin{center}
  \resizebox{0.9\columnwidth}{!}{\includegraphics{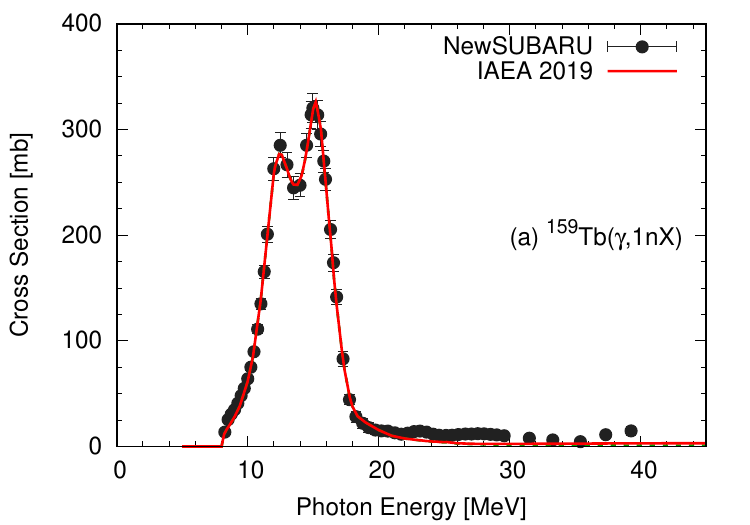}} 
  \resizebox{0.9\columnwidth}{!}{\includegraphics{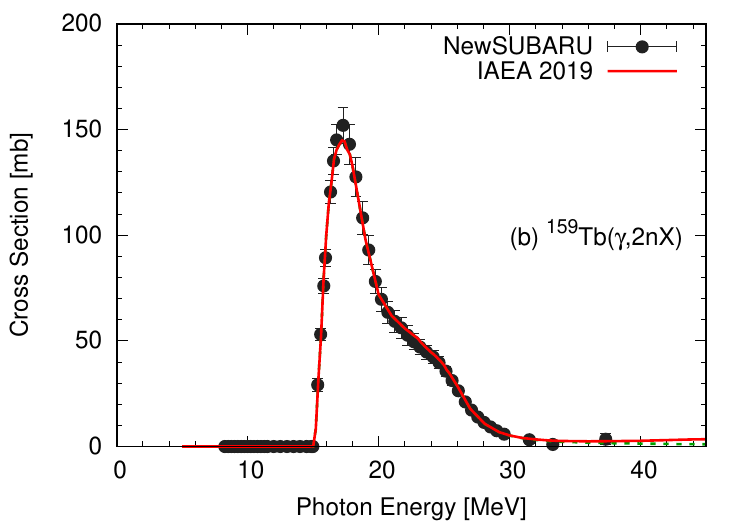}} 
  \resizebox{0.9\columnwidth}{!}{\includegraphics{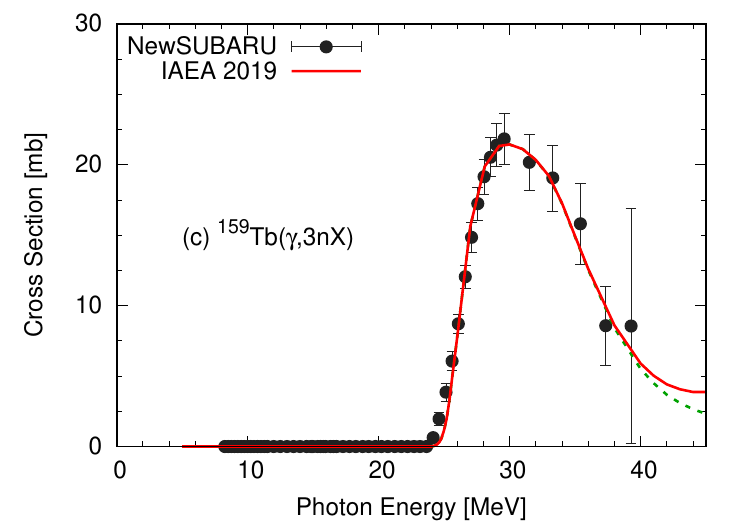}} 
  \resizebox{0.9\columnwidth}{!}{\includegraphics{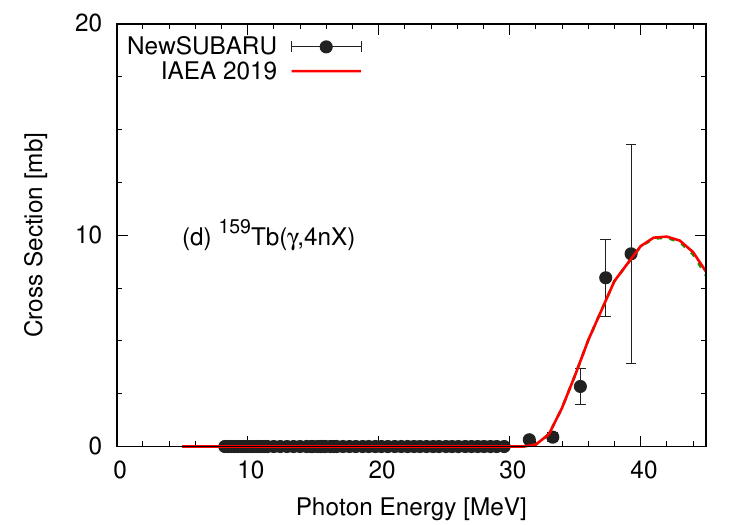}} 
 \end{center}
 \caption{(Color online) Comparison between evaluated and NewSUBARU 
 partial cross sections for $^{169}$Tm: (a)~$\sigma_{1nX}$, 
 (b)~$\sigma_{2nX}$, (c)~$\sigma_{3nX}$, and (d)~$\sigma_{4nX}$. 
 See Fig.~\ref{fig:IFINHHTm169sn} for the dotted lines.}
 \label{fig:IFINHHTm169}
\end{figure}

The $(\gamma,Sn)$, $(\gamma,1nX)$, $(\gamma,2nX)$, $(\gamma,3nX)$ and
$(\gamma,4nX)$ cross sections have been measured at NewSUBARU for
$^{169}$Tm in the energy region between 8 and 40~MeV. No other
data exist. For the evaluation, the NewSUBARU $(\gamma,Sn)$
reaction cross section was best reproduced by using the SMLO model for
the gamma ray strength function with two centroids.
The evaluation is compared to the experimental data in 
Figs.~\ref{fig:IFINHHTm169sn} and \ref{fig:IFINHHTm169}.

\subsubsection{$^{181}$Ta}
\label{subsubsec:EVALIFIN_TA}

\begin{figure}
 \begin{center}
  \resizebox{0.9\columnwidth}{!}{\includegraphics{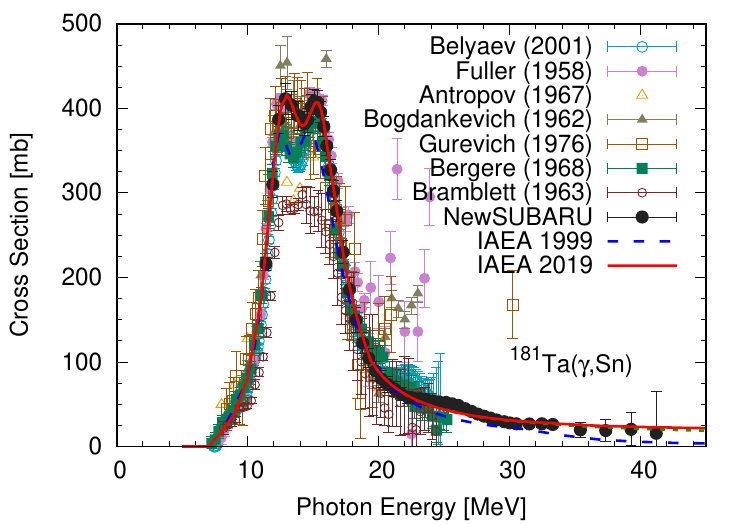}} 
 \end{center}
 \caption{(Color online) Comparison between experimental and evaluated 
 $\sigma_{Sn}$ for $^{181}$Ta. The solid line is the current evaluation,
 and the dashed line is IAEA 1999 library~\cite{IAEAPhoto1999}. The evaluated sum cross section 
 $\sum_i\sigma_{in}$ ($i=1$ -- 4) of all channels without 
 charged-particle emission is also shown by the dotted line.}
 \label{fig:IFINHHTa181sn}
\end{figure}

\begin{figure}
 \begin{center}
  \resizebox{0.9\columnwidth}{!}{\includegraphics{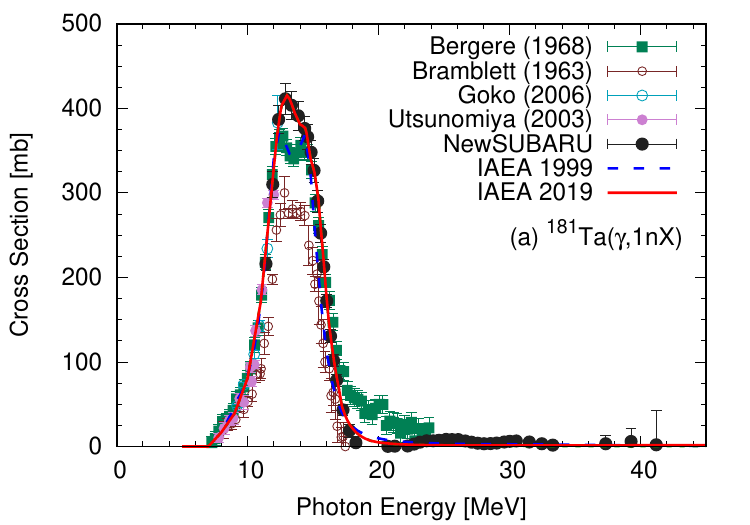}} 
  \resizebox{0.9\columnwidth}{!}{\includegraphics{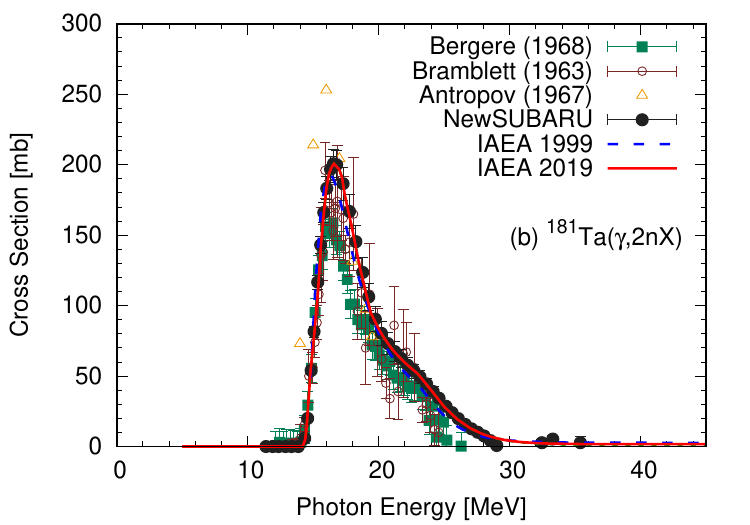}} 
  \resizebox{0.9\columnwidth}{!}{\includegraphics{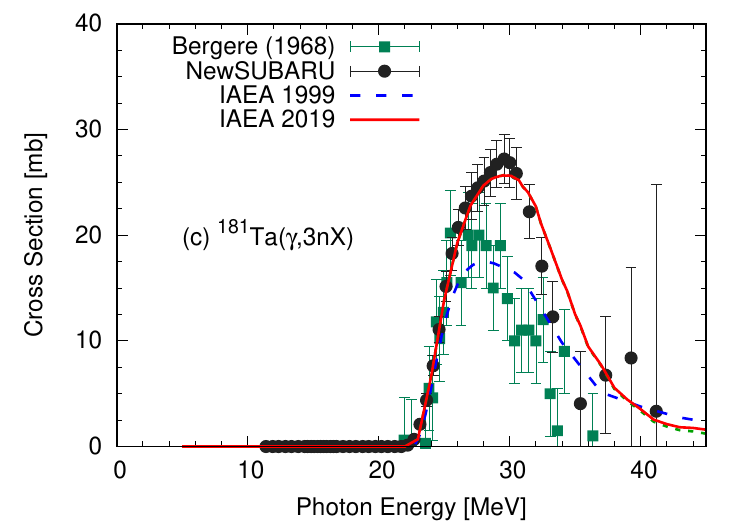}} 
  \resizebox{0.9\columnwidth}{!}{\includegraphics{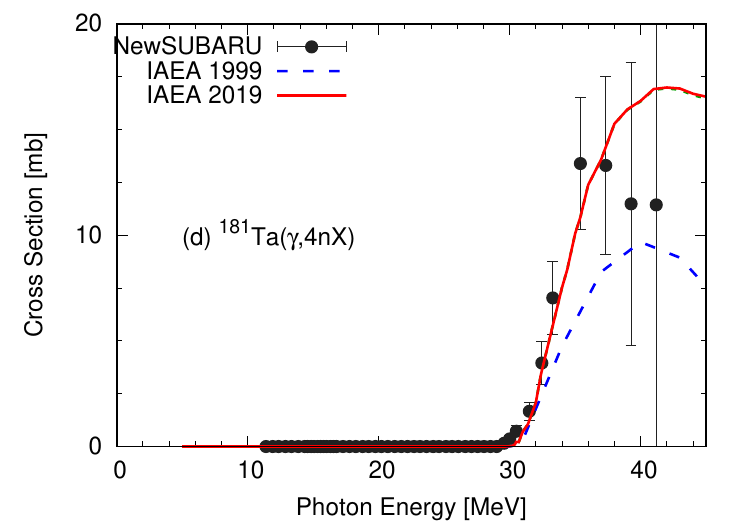}} 
 \end{center}
 \caption{(Color online) Comparison between evaluated and NewSUBARU
 partial cross sections for $^{181}$Ta: (a)~$\sigma_{1nX}$, 
 (b)~$\sigma_{2nX}$, (c)~$\sigma_{3nX}$, and (d)~$\sigma_{4nX}$.
 See Fig.~\ref{fig:IFINHHTa181sn} for the dotted line.}
 \label{fig:IFINHHTa181}
\end{figure}

The previous $^{181}$Ta evaluation has been taken from the JENDL
Photonuclear Data Library~\cite{Kishida2004} constructed with the
ALICE-F~\cite{Fukahori1992} code. The JAERI evaluation was performed
on existing $(\gamma,Sn)$, $(\gamma,1nX)$ and $(\gamma,2nX)$ cross
sections measured by Berg\`{e}re {\it et al.}~\cite{Bergere1968} and
by Bramblett {\it et al.}~\cite{Bramblett1963}, where the Livermore
cross sections have been reconstructed based on a study performed at
the Sao Paulo laboratory~\cite{Wolynec1984, INDC0364}. The
$\sigma_{Sn}$ has been measured using bremsstrahlung beams by Belyaev
{\it et al.}~\cite{Belyaev2001}, Fuller {\it et
al.}~\cite{Fuller1958}, Antropov {\it et al.}~\cite{Antropov1967},
Bogdankevich {\it et al.}~\cite{Bogdankevich1962} and Gurevich {\it et
al.}~\cite{Gurevich1976}. Bogdankevich \cite{Bogdankevich1962}
reported also the $(\gamma,2nX)$.  Additional $(\gamma,1nX)$
measurements have been performed using LCS $\gamma$-ray beams by
Utsunomiya {\it et al.}~\cite{Utsunomiya2003} and Goko {\it et
al.}~\cite{Goko2006}. New $(\gamma,Sn)$, $(\gamma,1nX)$,
$(\gamma,2nX)$, $(\gamma,3nX)$ and $(\gamma,4nX)$ cross sections have
been measured at NewSUBARU for $^{181}$Ta in the energy region between
12 and 41~MeV. The NewSUBARU $(\gamma,1nX)$ cross sections are in
better agreement with the Saclay measurement, but have higher cross
sections in the GDR peak region. The NewSUBARU results do not separate
well the peaks of the two GDR Lorentzians, resembling better the
Livermore data. The new $(\gamma,2nX)$ cross sections are in better
agreement with the Livermore data. For the evaluation, the NewSUBARU
$(\gamma,Sn)$ reaction cross section was best reproduced by using the
SMLO model for the $\gamma$-ray strength function with two centroids.
The evaluation is compared to the experimental data in
Figs.~\ref{fig:IFINHHTa181sn} and \ref{fig:IFINHHTa181}.

  \clearpage

  \subsection{Data Evaluation at JAEA}
  \label{subsec:evalJAEA}
For quite a few nuclides, the photonuclear data in the IAEA 2019
library are adopted from the JENDL photonuclear data file 2016
(JENDL/PD-2016), which was released in December, 2017. These nuclides
are listed in Appendix~\ref{appendix:libindex}. JENDL/PD-2016 includes
photo-induced reaction data (photo-absorption cross section,
photo-fission cross section, the average number of prompt and delayed
fission neutrons, particle and photon emission cross sections and
energy-angle distribution, residual cross sections, {\it etc.})  up to
140 MeV energy. The evaluation was made by the CCONE
code~\cite{Iwamoto2016} for medium to heavy nuclides.

Some of the light nuclides with atomic number $\le 20$ were also taken
from JENDL/PD-2016; $^3$He, $^{6,7}$Li, and $^{19}$F.  Those nuclear
data were evaluated by the method described by Murata {\it et
al.}~\cite{Murata2011} with the ALICE-F code~\cite{Fukahori1992},
except for $^{35,37}$Cl, $^{36,38,40}$Ar and $^{39,40,41}$K, which
were evaluated with CCONE~\cite{Iwamoto2016}.  In Murata's method,
measured photo-absorption and $(\gamma,1nX)$ reaction cross sections
were reproduced by resonances and quasi-deuteron disintegration
models. The other cross sections were calculated with ALICE-F, which
provided cross section ratios of particle emission cross sections to
the photo-absorption one. The energy-angle distributions calculated
with ALICE-F were replaced by the results of CCONE calculation. The
nuclear data of light nuclides evaluated by this method have an upper
energy limit of 140~MeV, which is the same as JENDL/PD-2016.

The CCONE code consists of the Hauser-Feshbach model for statistical
decay and the two-component exciton model for pre-equilibrium decay.
The deexcitation by $\gamma$-rays is expressed by the E1, M1 and E2
transitions.  For the E1 radiation, the MLO1 model is basically
utilized for the photon strength function. The default GDR parameters
are taken from the RIPL-2 systematics~\cite{RIPL2}. Those parameters
are modified so as to reproduce measured photo-absorption and/or
photoneutron cross sections. For the M1 and E2 radiations the
formulations of Kopecky and Uhl~\cite{Kopecky1990} are adopted. The
discrete level information is taken from the RIPL-3
database~\cite{RIPL3}. The Gilbert-Cameron formula~\cite{Gilbert1965}
is employed as a model describing the level density above the discrete
levels, where the Fermi-gas model is replaced by the revised model of
Mengoni and Nakajima~\cite{Mengoni1994}. In the statistical decay, we
consider six particle emission channels, namely neutron, proton,
deuteron, triton, $^{3}$He, and $\alpha$-particle, as well as the
$\gamma$-ray emission. For the particle emission, the optical model
potential parameters of Koning-Delaroche~\cite{Koning2003} for
neutrons and protons, Han {\it et al.}~\cite{Han2006} for deuteron,
folding potential generated from the neutron and proton potentials of
Koning-Delaroche for triton, Xu {\it et al.}~\cite{Xu2011} for
$^{3}$He and Avrigeanu-Avrigeanu~\cite{Avrigeanu2010} for
$\alpha$-particles are adopted.  The photon incident energy region is
extended up to 200~MeV. The lowest energy of 1~MeV is unchanged in the
evaluations by CCONE. The quasi-deuteron disintegration model, which
has an important contribution above the photon energy of 40~MeV, is
also implemented.  The Pauli-blocking function $P_b(E_\gamma)$ in the
model is originally derived in the energy range of 20 to 140~MeV. The
extension to lower and higher energies is necessary to produce
nuclear data from 1 to 200~MeV. The function is extrapolated by
Eq.~(\ref{eq:pbextra}).

The inclusive data form (MF/MT=6/5) in the ENDF-6 format is adopted in
the JAEA evaluations. The experimental $(\gamma,1nX)$ reaction cross
section, which is the sum of $(\gamma,1n)$, $(\gamma,1np)$ and
$(\gamma,1n\alpha)$ reaction ones and so on, for example, cannot be
separated into individual reaction. Thus, in the figures shown below,
the present results for $(\gamma,1n)$ and $(\gamma,2n)$ reactions are
used for comparisons with measured $(\gamma,1nX)$ and $(\gamma,2nX)$
reaction cross sections. Such a comparison is justified for nuclides
with high atomic number, since reaction cross sections accompanying
charged particle emissions such as $(\gamma,1np)$ and
$(\gamma,1n\alpha)$ reactions are highly suppressed due to the large
Coulomb barrier.  Note that the $(\gamma,1n)$ and $(\gamma,2n)$
reaction cross sections are given as the production cross sections of
residual nuclides with the mass numbers of $A-1$ and $A-2$ for a
target with $A$ in the ENDF-6 format.

In the case of nuclides with fission reaction, the average number of
prompt neutrons per fission $\overline{\nu}_p$ is calculated with the
linear functions with respect to photon energy fitted by Berman {\it
et al.}~\cite{Berman1986} and Caldwell {\it et
al.}~\cite{caldwell1980}, complemented with an exponential
function~\cite{Ethvignot2005} at the higher energies. The average
number of delayed neutrons per fission $\overline{\nu}_d$ and the
delayed neutron spectra, if available, are adopted from the neutron
nuclear data of JENDL-4.0, if a compound nuclide is the same as the
target in the photon-incident case.

\subsubsection{$^{40}$Ca}

The IAEA 1999 library~\cite{IAEAPhoto1999} has the nuclear data of
$^{40}$Ca performed with the GNASH code at LANL. There are a few
experimental data available for the evaluation: photo-absorption,
$(\gamma,1nX)$, $(\gamma,1pX)$, $(\gamma,Sn)$ and $(\gamma,xn)$
reactions. The present evaluation was carried out by applying Murata's
method.  The photo-absorption cross section measured by Ahrens {\it et
al.}~\cite{Ahrens1975} was reconstructed by 10 resonances, and this
was also estimated from the $(\gamma,1nX)$ reaction cross section of
Veyssi\`{e}re {\it et al.}~\cite{Veyssiere1974}. The other particle
emissions and residual production cross sections were calculated with
ALICE-F.

\begin{figure}
 \begin{center}
  \resizebox{0.9\columnwidth}{!}{\includegraphics{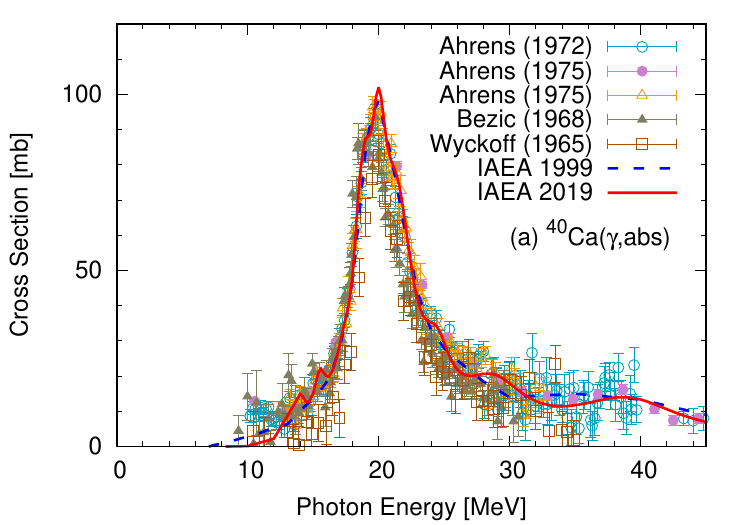}}
  \resizebox{0.9\columnwidth}{!}{\includegraphics{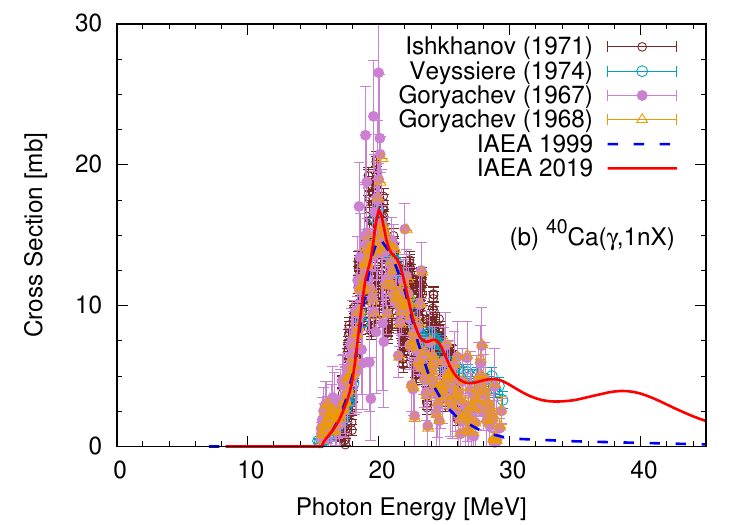}}
  \resizebox{0.9\columnwidth}{!}{\includegraphics{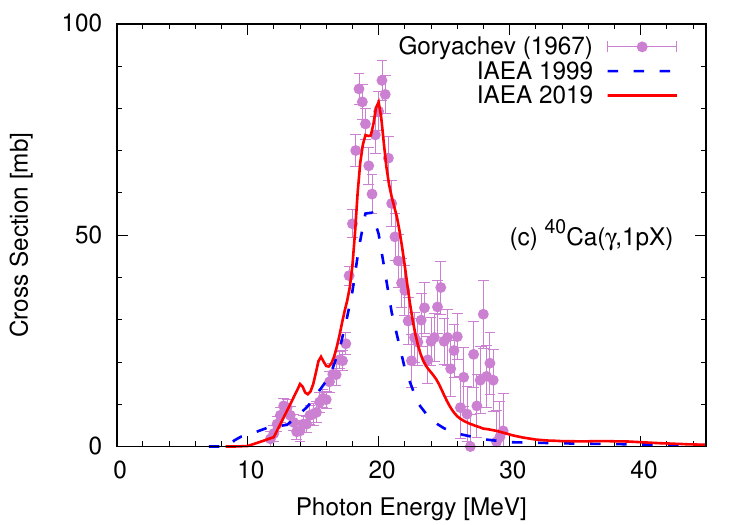}}
 \end{center}
 \caption{(Color online) Comparisons of the present results with
 the evaluated (IAEA 1999) and experimental data for $^{40}$Ca. The
 cross sections of (a) photo-absorption, (b) $(\gamma,1nX)$ and (c)
 $(\gamma,1pX)$ reactions are illustrated in each panel.}
 \label{fig:JAEACa40}
\end{figure}

Figure~\ref{fig:JAEACa40} shows the comparison of the present
evaluation with previous evaluation~\cite{IAEAPhoto1999} and measured
data for photo-absorption, $(\gamma,1nX)$ and $(\gamma,1pX)$ reaction
cross sections. The measurements of photo-absorption cross sections
are for the natural Ca target. The comparison of the present
evaluation of $^{40}$Ca with the measured data is physically
meaningful, since the abundance of $^{40}$Ca accounts for 96.9\% of
natural Ca. The present evaluation gives similar photo-absorption
cross section as the IAEA 1999 library. The evaluated $(\gamma,1p)$
reaction cross section is in good agreement with experimental data of
Goryachev {\it et al.}~\cite{Goryachev1967} and higher than the
previous evaluation in the IAEA 1999 library by 45\% at the peak cross
section.

\subsubsection{$^{139}$La}

\begin{figure}[!h]
 \begin{center}
   \resizebox{0.9\columnwidth}{!}{\includegraphics{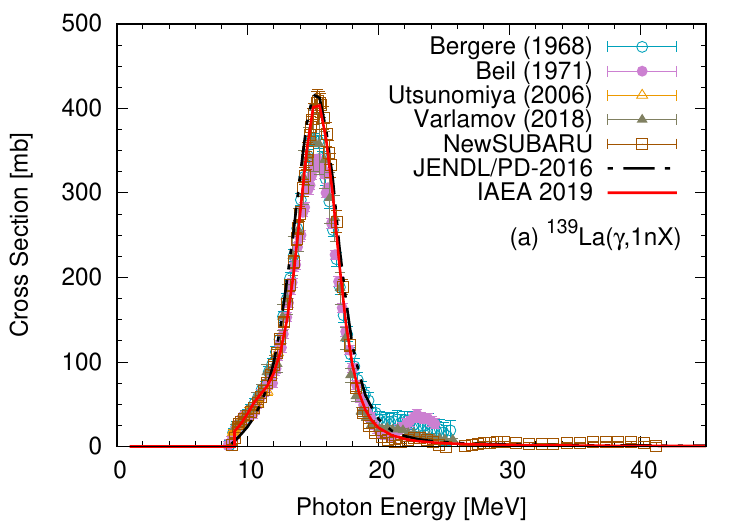}}
   \resizebox{0.9\columnwidth}{!}{\includegraphics{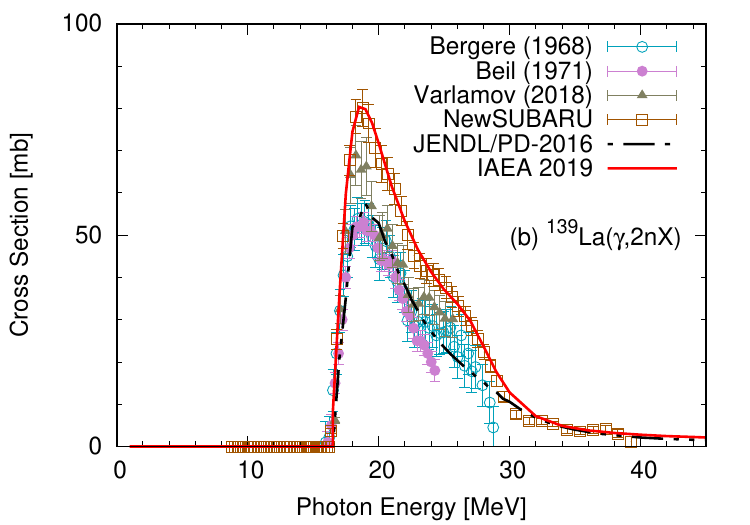}}
   \resizebox{0.9\columnwidth}{!}{\includegraphics{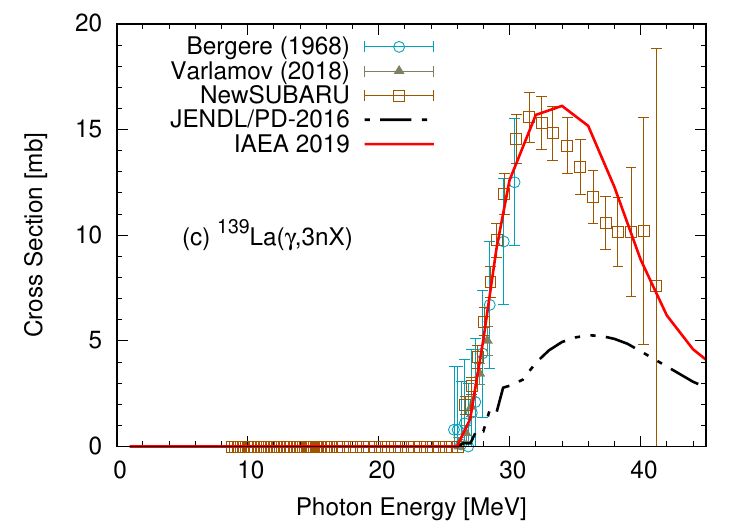}}
   \resizebox{0.9\columnwidth}{!}{\includegraphics{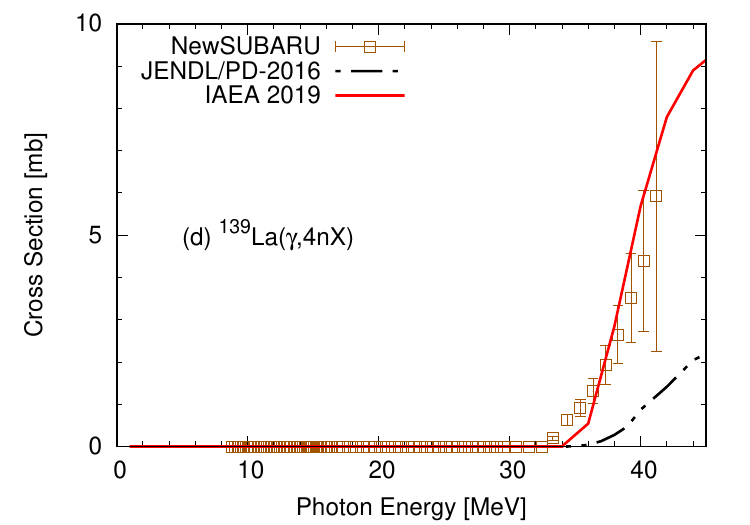}}
 \end{center}
 \caption{(Color online) Comparisons of the present photoneutron cross sections with
 the evaluated (JENDL/PD-2016) and experimental data for
 $^{139}$La; (a) $(\gamma,1nX)$, (b) $(\gamma,2nX)$, (c) $(\gamma,3nX)$ and (d)
 $(\gamma,4nX)$.} 
 \label{fig:JAEALa139} 
\end{figure}

The photonuclear data of $^{139}$La were not included in the IAEA 1999
library. New measurement was carried out at the NewSUBARU facility and
provided the cross sections of $(\gamma,1nX)$, $(\gamma,2nX)$,
$(\gamma,3nX)$ and $(\gamma,4nX)$ reactions. The measured data are
compared with previous ones in Fig.~\ref{fig:JAEALa139}. It is found
that the $(\gamma,1nX)$ and $(\gamma,2nX)$ reaction cross sections of
NewSUBARU are much higher than the previous ones. The present
evaluation with the SMLO model for E1 photon strength function was made
on the basis of the data measured at NewSUBARU.

The new evaluated cross sections for the $(\gamma,1n)$, $(\gamma,2n)$,
$(\gamma,3n)$ and $(\gamma,4n)$ reactions are compared with
JENDL/PD-2016~\cite{JENDL-PD2016} in Fig.~\ref{fig:JAEALa139}. The
present evaluations are larger than JENDL/PD-2016, except for the
$(\gamma,1n)$ reaction cross section, where both evaluations agree
since there is only one neutron emission below the threshold energy of
$(\gamma,2n)$ at 16.2~MeV. So, there is no ambiguity in detecting 1n
or 2n neutrons. The recent data of Varlamov {\it et al.} have been
evaluated on the basis of the old data of Beil {\it et
al.}~\cite{Beil1971} from Saclay and are also smaller than those of
NewSUBARU.

\subsubsection{$^{239}$Pu}

The nuclear data of $^{239}$Pu in the previous IAEA 1999 library was
evaluated by IPPE. The energy region is limited to 20~MeV.  The
average number of prompt neutrons per fission $\overline{\nu}_p$ is
given in the MF/MT=1/456 section (MF/MT combination defined in the
ENDF-6 format), based on an estimation from neutron nuclear data of
$^{238}$Pu. The MF/MT=3/5 format is especially used as the sum of
$(\gamma,1n)$ and $(\gamma,1np)$ reaction cross sections. The current
new library adopts the data from JENDL/PD-2016, but with the incident
energy range extended to 200~MeV. The evaluation was made with
CCONE. $\overline{\nu}_p$ and $\overline{\nu}_d$ were included in the
evaluation.

\begin{figure}
 \begin{center}
   \resizebox{0.9\columnwidth}{!}{\includegraphics{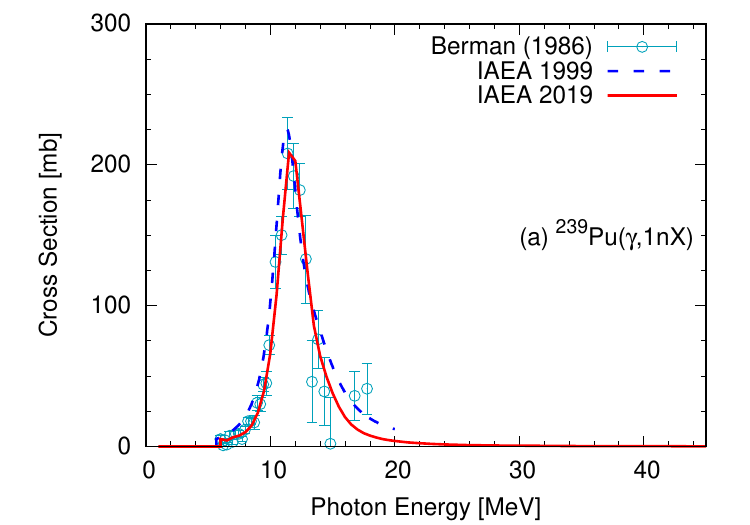}}
   \resizebox{0.9\columnwidth}{!}{\includegraphics{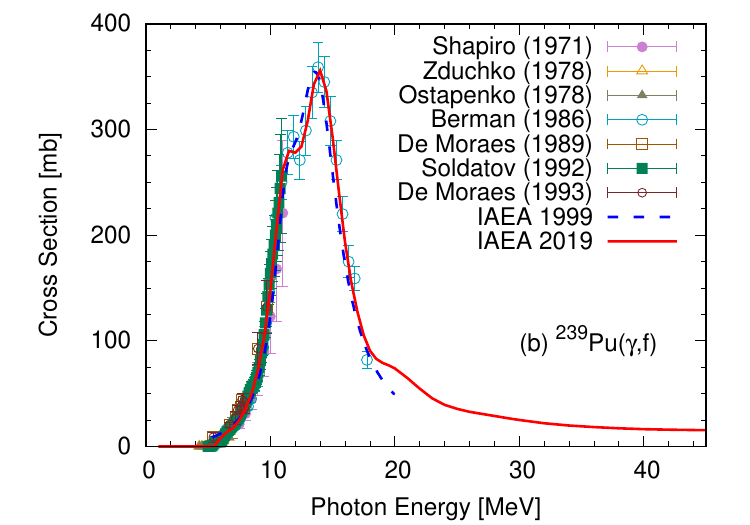}}
   \resizebox{0.9\columnwidth}{!}{\includegraphics{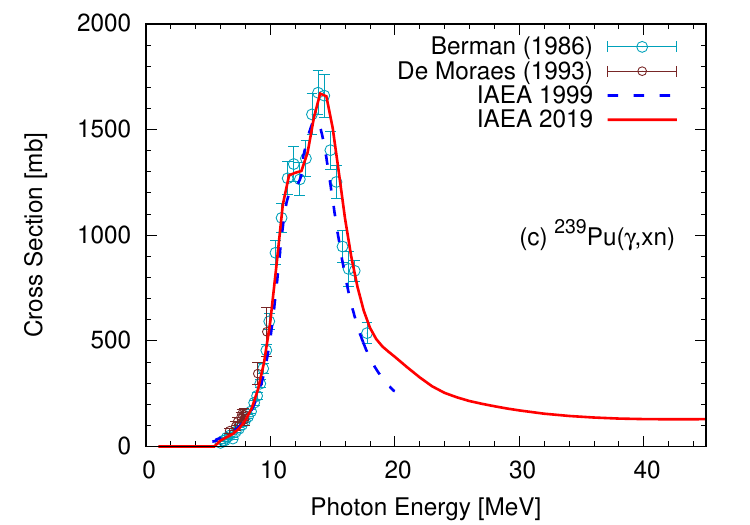}}
 \end{center}
 \caption{(Color online) Comparisons of the present results with
 the evaluated (IAEA 1999) and experimental data for $^{239}$Pu;
 (a) $(\gamma,1nx)$, (b) photo-fission, and (c) $(\gamma,xn)$.}
 \label{fig:JAEAPu239} 
\end{figure}

Figure~\ref{fig:JAEAPu239} shows the comparison of the new evaluations
with the previous IAEA 1999 library and the measured data for
$(\gamma,1nX)$, photo-fission and $(\gamma,xn)$ reactions. The
$(\gamma,1nX)$ reaction cross section in the previous library shifts
slightly to the lower energy region, relative to the measured
data. The photo-fission cross sections are almost consistent with
each other. The accuracy of this reaction cross section may be
important for the development of inspection technology for nuclear
fuel material. The small bump seen around 20~MeV comes from the
contribution of third chance fission. This is not seen in the previous
evaluation. The $(\gamma,xn)$ reaction cross section has a
contribution of fission neutrons as well as neutron emission cross
sections multiplied by neutron multiplicity. The present $(\gamma,xn)$
reaction cross section reproduces well the data of Berman {\it et
al.}~\cite{Berman1986}. In contrast, the data in the previous library
are lower than the measured data. This is attributed to the smaller
$(\gamma,2nX)$ reaction cross section.

  \clearpage

  \subsection{Data Evaluation at CIAE}
  \label{subsec:evalCIAE}
\begin{figure}[!hbt]
 \begin{center}
   \resizebox{0.9\columnwidth}{!}{\includegraphics{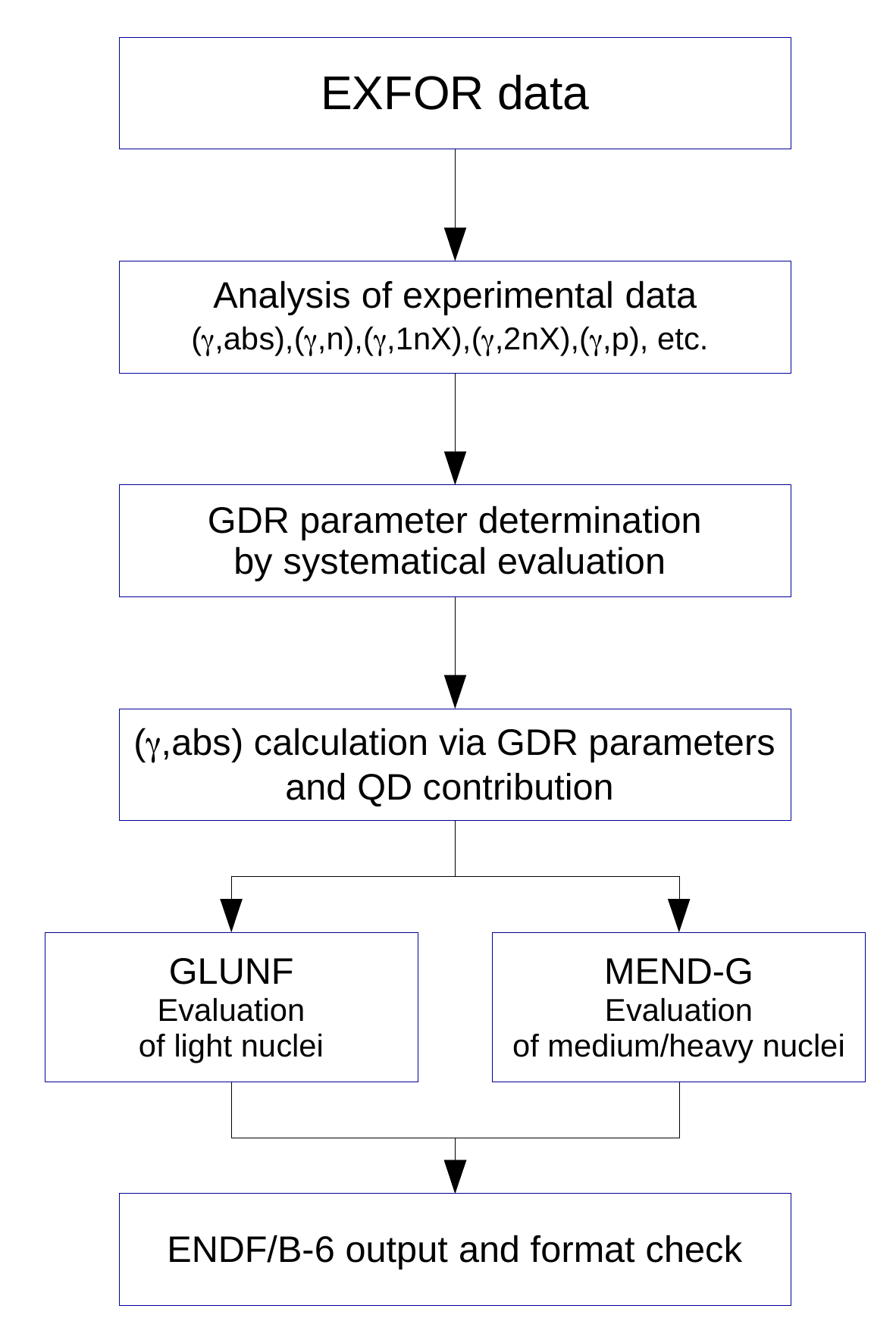}}
 \end{center}
 \caption{Scheme of photonuclear data evaluation at CIAE/CNDC.}
 \label{fig:CIAEflow}
\end{figure}

The evaluation scheme of photonuclear data at CIAE/CNDC is shown in
Fig.~\ref{fig:CIAEflow}. By collecting experimental data given in the
EXFOR library, the following photonuclear reaction cross sections ---
$(\gamma,{\rm abs})$, $(\gamma,xn)$, $(\gamma,1nX)$, $(\gamma,2nX)$,
$(\gamma,3nX)$, $(\gamma,1p)$, {\it etc.} --- are recommended to guide
our theoretical calculation. The new photonuclear reaction calculation
systems, GLUNF for the light nuclei ($^9$Be, $^{10,11}$B, $^{12}$C,
$^{14}$N, and $^{16}$O) and MEND-G for the medium-heavy nuclei, are
specially built for the incident $\gamma$-ray energy $E_\gamma$ below
200~MeV under the corporation between CNDC and Nankai University.
Reactions where up to eighteen particles are allowed to be emitted are
considered in MEND-G, and the spherical optical model, equilibrium and
pre-equilibrium statistical models are included in the code. Since
determination of the scheme for competing particle emission is the
most important in the light element case, a different reaction scheme
in GLUNF is constructed for an individual photonuclear reaction chain
for each of the light nuclei. Once the reaction chain was determined,
the statistical theory of light-nucleus reaction~\cite{Sun2015,
  Sun2016} is applied to calculate all the reactions in the GLUNF
code.

The photon strength function (PSF) used in deriving the
photo-absorption cross sections is also systematically studied. For
this purpose the code CPSF was developed at CNDC to produce the
various PSFs, which includes several empirical Lorentzian
functions~\cite{Goriely2019}, known as SLO, MLO1, MLO2, MLO3, EGLO,
GFL, and SMLO.  In addition, a semi-microscopic model, RQRPA, is also
implemented in CPSF. In this calculation, we combine the GDR process
below about 30~MeV and the phenomenological QD model up to 200 MeV.

\subsubsection{$^{9}$Be}

$^9$Be is a natural stable isotope of beryllium with the abundance of
100\%. The evaluation in the previous IAEA 1999
library~\cite{IAEAPhoto1999} was performed with the GLUNF code at
CNDC.  In the GLUNF reaction scheme used in our present evaluation,
the photo-absorption cross section $\sigma_{\rm abs}$ is split into
the reaction channels of $(\gamma,2n)$, $(\gamma,2n\alpha)$,
$(\gamma,np)$, $(\gamma,n2\alpha)$, $(\gamma,nd)$, $(\gamma,nt)$,
$(\gamma,2np)$, $(\gamma,pd)$, $(\gamma,d\alpha)$, $(\gamma,p_0)$,
$(\gamma,p_1)$, $(\gamma,d_0)$, $(\gamma,d_1)$, $(\gamma,t_0)$,
$(\gamma,t_1)$, $(\gamma,{}^3{\rm He}_0)$, and $(\gamma,{}^3{\rm
  He}_1)$, where $(\gamma,p_0)$ stands for the proton production by
leaving the residual in its ground state.

\begin{figure}[!h]
 \begin{center}
   \resizebox{0.9\columnwidth}{!}{\includegraphics{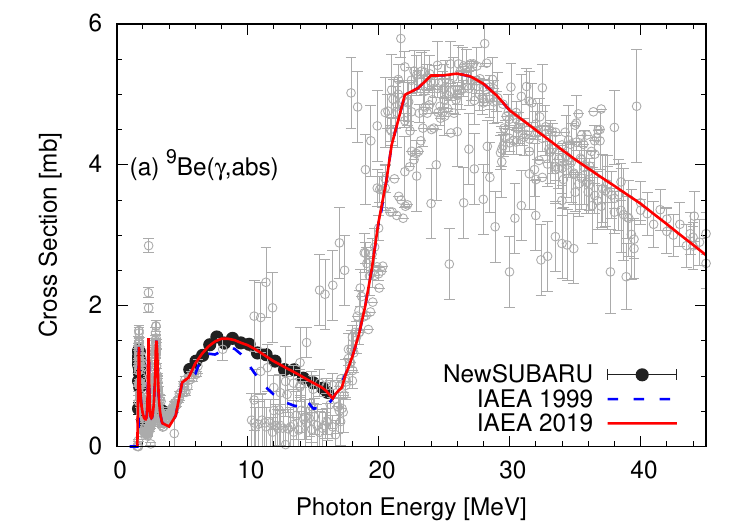}}
   \resizebox{0.9\columnwidth}{!}{\includegraphics{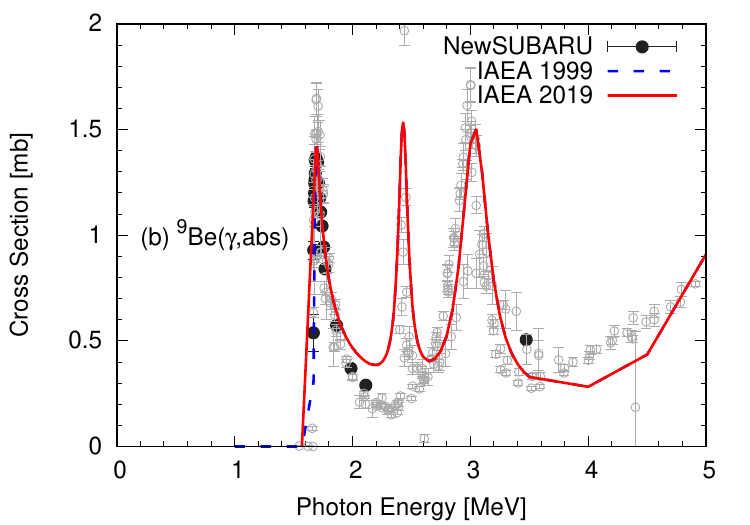}}
 \end{center}
 \caption{(Color online) The evaluated photo-absorption cross section for $^9$Be.
  The bottom panel (b) is the zoom in the 1 -- 5~MeV region.
  All the data are taken from EXFOR database, except for the NewSUBARU data
  shown by the filled circles.}
 \label{fig:CIAEBe9abs}
\end{figure}

The experimental data in EXFOR cover the data of $\sigma_{\rm abs}$,
$(\gamma,n)$, $(\gamma,2n)$, and the charged particle emissions. In
addition, the new NewSUBARU data for $(\gamma,n)+(\gamma,np)$
presented at this CRP are the main new experimental data used in our
current update. Figure~\ref{fig:CIAEBe9abs} compares our evaluated
photo-absorption cross section with these experimental data.  In the
energy range 30--150~MeV, the evaluation of absorption cross section
is based on the measurement of Ahrens {\it et al.}~\cite{Ahrens1975}.

\begin{figure}[!htb]
 \begin{center}
   \resizebox{0.9\columnwidth}{!}{\includegraphics{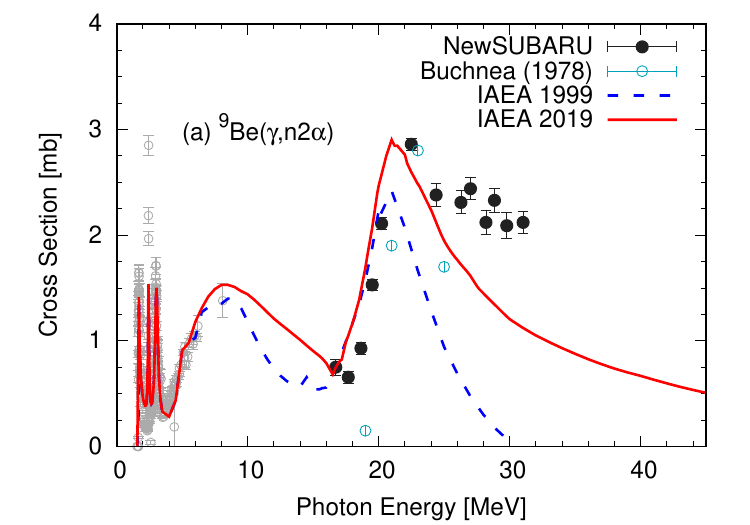}}
   \resizebox{0.9\columnwidth}{!}{\includegraphics{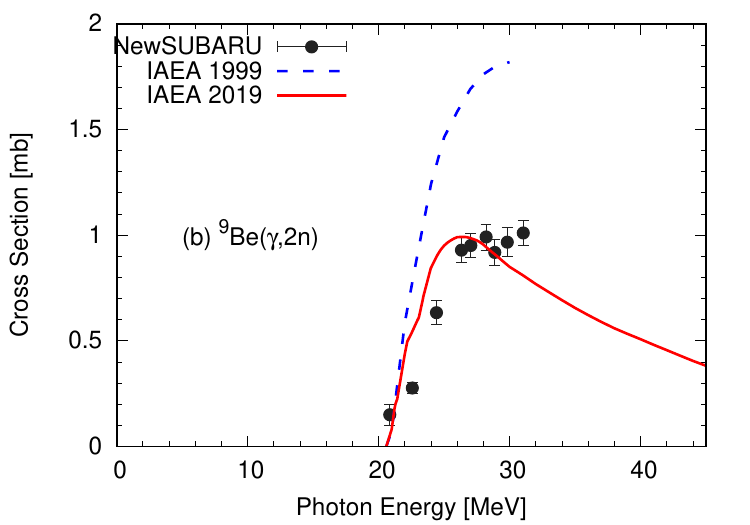}}
 \end{center}
 \caption{(Color online) Comparison of the evaluated and experimental
  (a) $^9$Be$(\gamma,n2\alpha)$ and (b) $(\gamma,2n)$ cross sections.}
 \label{fig:CIAEBe9}
\end{figure}

The new reaction chain scheme in GLUNF for $^9$Be was developed at
CNDC to evaluate the photonuclear data below 150~MeV.  The optical
model parameters were adjusted to the experimental data adopted in our
evaluation. In Fig.~\ref{fig:CIAEBe9}, the calculated
$(\gamma,n2\alpha)$ is compared with the available experimental
data. The $(\gamma,n)$ experimental data below 15~MeV were taken from
EXFOR. Above 15~MeV, the measurement of Buchnea {\it et
  al.}~\cite{Buchnea1978} is compiled as the $^{9}{\rm
  Be}(\gamma,n\alpha){}^4{\rm He}$ reaction in EXFOR.  The data shown
by the filled circles are from the new measurement of Utsunomiya {\it
  et al.} reported in this CRP. The measurement of $(\gamma,n)$ is
represented by $(\gamma,n2\alpha)$. The results of $(\gamma,2n)$ in
our work are in good agreement with the experimental data of
Utsunomiya as shown in Fig.~\ref{fig:CIAEBe9}. Also it is seen that
the new evaluation is naturally extrapolated to the higher energy
region oppose to the previous GLUNF evaluation, which predicts almost
zero cross section of $(\gamma,n2\alpha)$ at 30~MeV.

\subsubsection{$^{14}$N}

$^{14}$N is a natural stable isotope of nitrogen, which has the
largest abundance of 99.64\%. A new evaluation for the photo-induced
reaction on $^{14}$N has been performed at CNDC.  By setting up a
reaction scheme in GLUNF, we produced $\sigma_{\rm abs}$,
$(\gamma,2n)$, $(\gamma,n3\alpha)$, $(\gamma,np)$, $(\gamma,nd)$,
$(\gamma,n {}^3{\rm He})$, $(\gamma,nd2\alpha)$, $(\gamma,nt2\alpha)$,
$(\gamma,n2p)$, $(\gamma,n_0)$, $(\gamma,2\alpha)$,
$(\gamma,3\alpha)$, $(\gamma,2p)$, $(\gamma,p\alpha)$,
$(\gamma,t2\alpha)$, $(\gamma,d2\alpha)$, $(\gamma,pd)$,
$(\gamma,pt)$, $(\gamma,d\alpha)$, as well as the proton, deuteron,
triton, $^3$He, and $\alpha$-particle emissions leaving the residuals in their
discrete states.

\begin{figure}[!h]
 \begin{center}
   \resizebox{0.9\columnwidth}{!}{\includegraphics{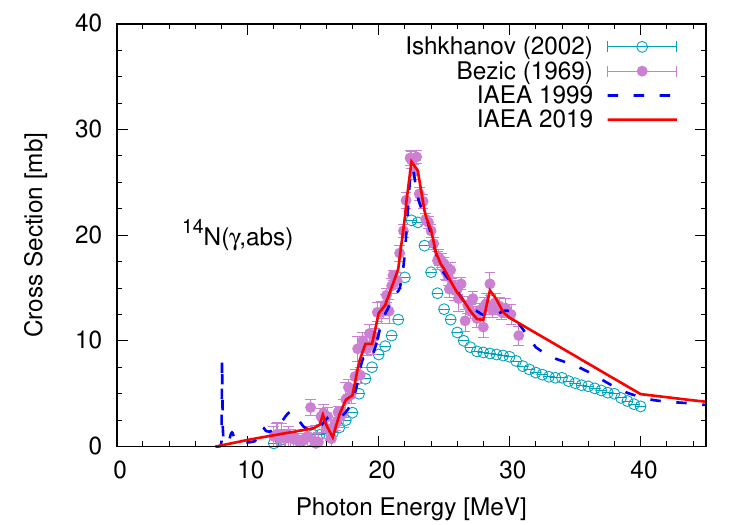}}
 \end{center}
 \caption{(Color online) Comparison of evaluated and experimental data of
   the photo-absorption cross section for $^{14}$N.}
 \label{fig:CIAEN14abs}
\end{figure}

\begin{figure}[!h]
 \begin{center}
   \resizebox{0.9\columnwidth}{!}{\includegraphics{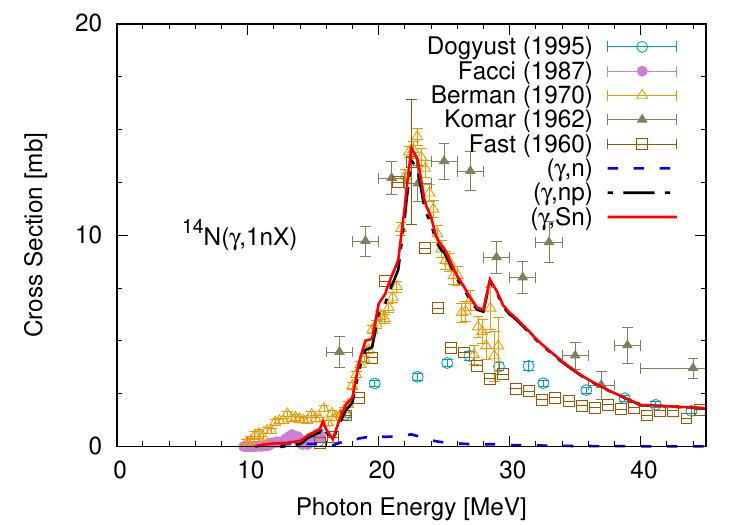}}
 \end{center}
 \caption{(Color online) Comparison of evaluated and experimental data of
   neutron-production reaction cross sections, $(\gamma,n)$ and $(\gamma,np)$
   for the photo-induced reaction on $^{14}$N.}
 \label{fig:CIAEN14n}
\end{figure}

The photo-absorption cross section is evaluated based on the
measurement of Bezic {\it et al.}~\cite{Bezic1969}, which is shown in
Fig.~\ref{fig:CIAEN14abs}. Below 30~MeV, the $(\gamma,1nX)$ cross
section consists of the $(\gamma,n)$, $(\gamma,np)$ and
$(\gamma,np3\alpha)$ reactions in our estimation, and the comparison
of evaluated and experimental data of $(\gamma,1nX)$ is shown in
Fig.~\ref{fig:CIAEN14n}.  In Table I of Ref.~\cite{Komar1962}, a
relative yield of the $(\gamma,np)$ reaction is reported to be 35\%,
while there are two channels given, 16\% of $(\gamma,n)$ and 11\% of
$(\gamma,3\alpha pn)$, which produce one neutron. In addition, the
residual of $(\gamma,p)$ reaction at higher energies, $^{13}$C$^*$
undergoes neutron-decay, with a branching ratio estimated to be 28\%.
Since the data of Komar are for $(\gamma,np)$ only, we multiplied their
data by $(28+35+16+11)/35=2.57$. The re-scaled data better agree with
other experimental data as well as our model calculation, as shown in
Fig.~\ref{fig:CIAEN14n}.

The excited $^{14}$N$^*$ emits a neutron and the residual nucleus
$^{13}$N continues decaying by emitting a proton. The first excited
state in $^{13}$N would be 100\% proton-decay and the probability of
13N ground state production would be very small. This is predicted by
our code as we get a very small $(\gamma,n_0)$ cross section. In
addition, $(\gamma,2n)$ is also included in GLUNF, although it has very
small values, typically less than 0.01~mb.

\subsubsection{$^{16}$O}

\begin{figure}[!htb]
 \begin{center}
   \resizebox{0.9\columnwidth}{!}{\includegraphics{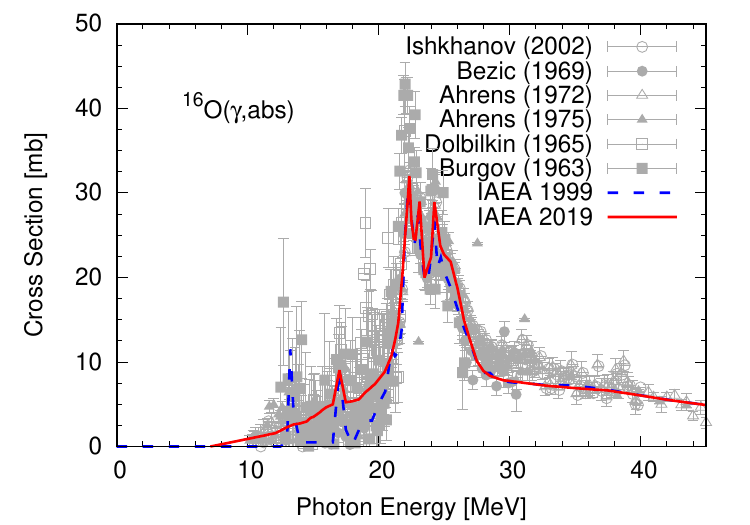}}
 \end{center}
 \caption{(Color online) Comparison of theoretical results and experimental data
   for the photo-absorption cross section of $^{16}$O.}
 \label{fig:CIAEO16abs}
\end{figure}

\begin{figure}[!htb]
 \begin{center}
   \resizebox{0.9\columnwidth}{!}{\includegraphics{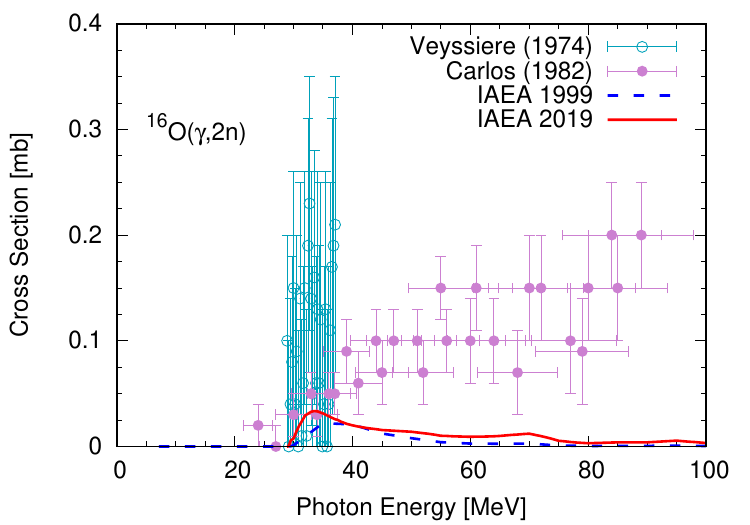}}
 \end{center}
 \caption{(Color online) Comparison of evaluation and experimental data for the cross sections
    of $^{16}{\rm O}(\gamma,2n){}^{14}{\rm O}$.}
 \label{fig:CIAEO162n}
\end{figure}

$^{16}$O is a natural stable isotope for oxygen with the abundance of
99.76\%.  We produced evaluated $\sigma_{\rm abs}$ and all of the
exclusive cross sections for the $(\gamma,2n)$, $(\gamma,n\alpha)$,
$(\gamma,n3\alpha)$, $(\gamma,np)$, $(\gamma,n2\alpha)$,
$(\gamma,nd)$, $(\gamma,n {}^3{\rm He})$, $(\gamma,2np)$,
$(\gamma,n_0)$ -- $(\gamma,n_6)$, $(\gamma,3\alpha)$, $(\gamma,2p)$,
$(\gamma,p\alpha)$, $(\gamma,t2\alpha)$, $(\gamma,d2\alpha)$,
$(\gamma,pd)$, $(\gamma,pt)$, and $(\gamma,d\alpha)$ reactions, as
well as the discrete transitions for the proton, deuteron, triton,
$^3$He, and $\alpha$-particle emission reactions. As examples, the evaluated
photo-absorption cross section and $(\gamma,2n)$ reaction are shown in
Figs.~\ref{fig:CIAEO16abs} and \ref{fig:CIAEO162n}, respectively.  A
notable difference in the evaluated photo-absorption cross sections
between the previous IAEA 1999 data file and the current evaluation is seen
in the energy range below 20~MeV. While the older IAEA data have a
resonance near 13~MeV, the new data exclude this peak.  The new
evaluation for the $(\gamma,2n)$ reaction in Fig.~\ref{fig:CIAEO162n}
is significantly lower than the experimental data in the energy range
above 40~MeV. We are unable to reproduce the data within our
theoretical model framework, and we have no clear explanation for
this.  However, it should be noted that the previous evaluation, which
employed a different statistical model, also shows the same tendency.

\subsubsection{$^{27}$Al}

\begin{figure*}[!htb]
 \begin{center}
   \resizebox{0.9\columnwidth}{!}{\includegraphics{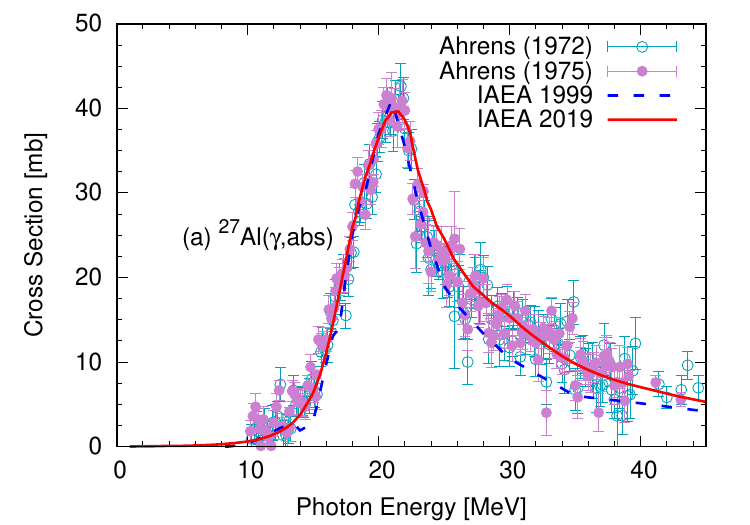}}
   \resizebox{0.9\columnwidth}{!}{\includegraphics{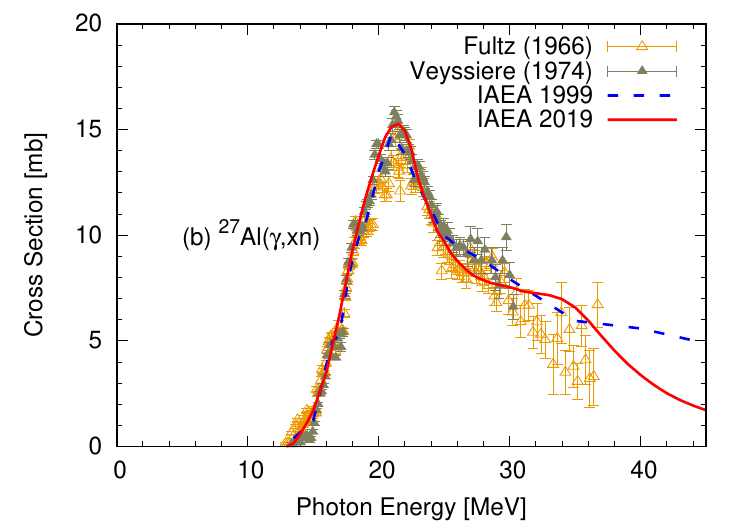}}
   \resizebox{0.9\columnwidth}{!}{\includegraphics{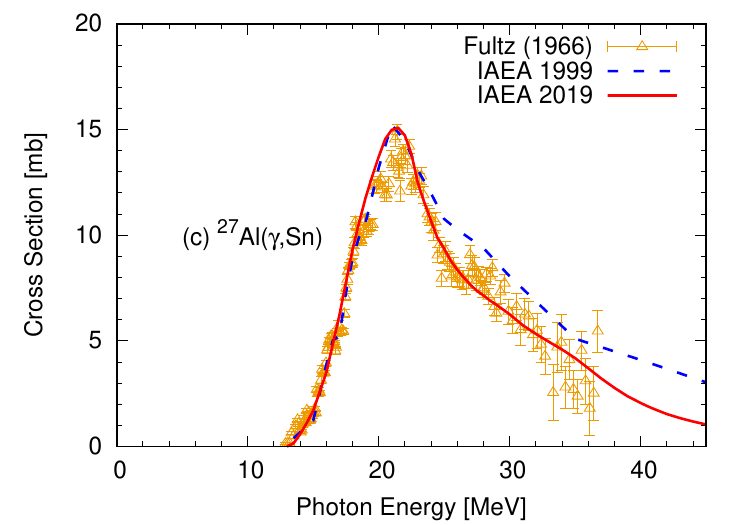}}
   \resizebox{0.9\columnwidth}{!}{\includegraphics{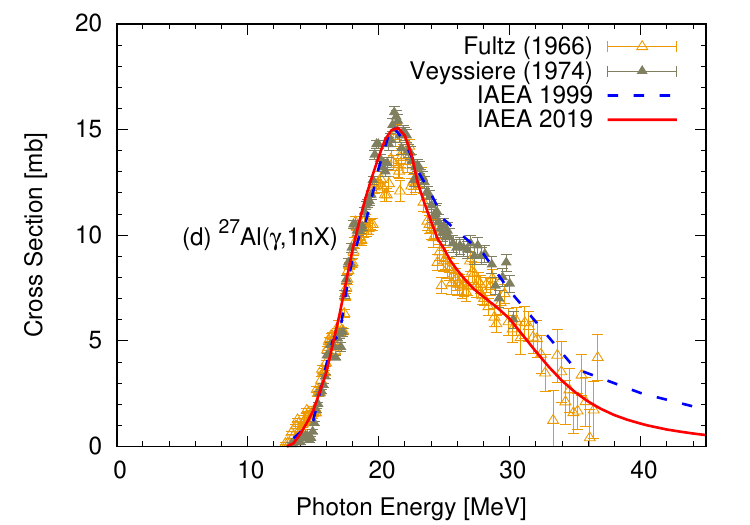}}
   \resizebox{0.9\columnwidth}{!}{\includegraphics{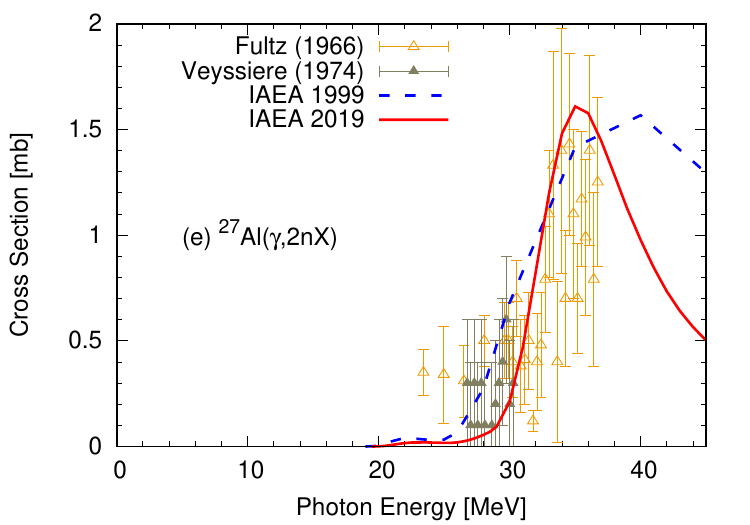}}
   \resizebox{0.9\columnwidth}{!}{\includegraphics{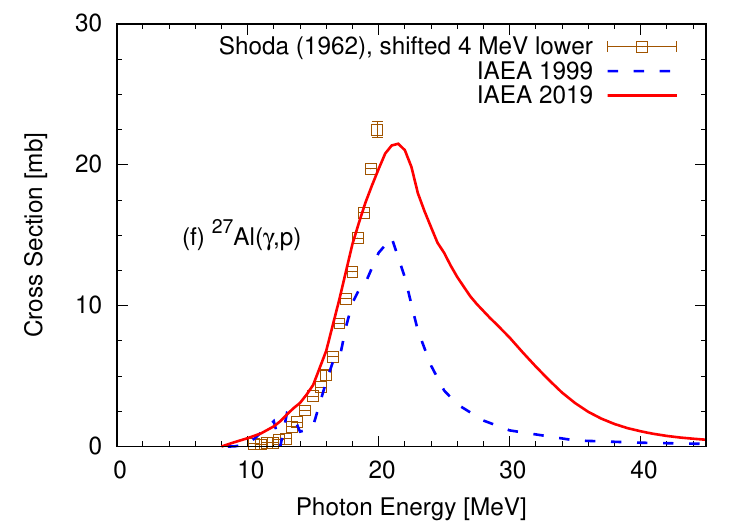}}
 \end{center}
 \caption{(Color online) Comparison of evaluated and experimental data for $^{27}$Al;
    (a) photo-absorption, (b) $\sigma_{xn}$, (c) $\sigma_{Sn}$, 
    (d) $(\gamma,1n) + (\gamma,1n+p)$, (e) $(\gamma,2n) + (\gamma,2n+p)$,
    and (f) $(\gamma,p)$ reactions.
    In the panel (f), the data of Shoda {\it et al.}~\cite{Shoda1962}
    are lower shifted by 4~MeV.}
 \label{fig:CIAEAl27}
\end{figure*}

A new evaluation for $^{27}$Al based on the MEND-G code calculation
was adopted in the new photonuclear data library.  The
photo-absorption cross section is evaluated based on the experimental
data of Ahrens {\it et al.}~\cite{Ahrens1972, Ahrens1975}, and the PSF
calculation using the SMLO model. The evaluated photo-absorption cross
section, shown in Fig.~\ref{fig:CIAEAl27} (a), is very similar to the
older evaluation up to the GDR peak of 20~MeV, then tends to be higher
at higher energies and agrees better with the Ahrens data.

The evaluated reaction cross sections are compared with measurements
of Fultz {\it et al.}~\cite{Fultz1966}, Veyssi\`{e}re {\it et
  al.}~\cite{Veyssiere1974}, and Shoda {\it et al.}~\cite{Shoda1962}
in Fig.~\ref{fig:CIAEAl27} (b)--(f).  The new evaluations agree
better with Fultz rather than Veyssi\`{e}re, although these data sets
do not reveal critical discrepancies. In the case of the $(\gamma,p)$
reaction shown in Fig.~\ref{fig:CIAEAl27} (f), the experimental data
of Shoda {\it et al.}  are shifted to lower energies by 4~MeV to match
the realistic reaction threshold energy.

\subsubsection{$^{50}$Cr}

There are four natural stable isotopes for chromium, and all of them
were evaluated by CIAE in the former IAEA Photonuclear data library.
The new library includes an updated evaluation of $^{50}$Cr whose
abundance is 4.35\%. On the other hand $^{53,54}$Cr data were left
unchanged, and the $^{52}$Cr evaluation was replaced by a new JAEA
evaluation.

\begin{figure}[!htb]
 \begin{center}
   \resizebox{0.9\columnwidth}{!}{\includegraphics{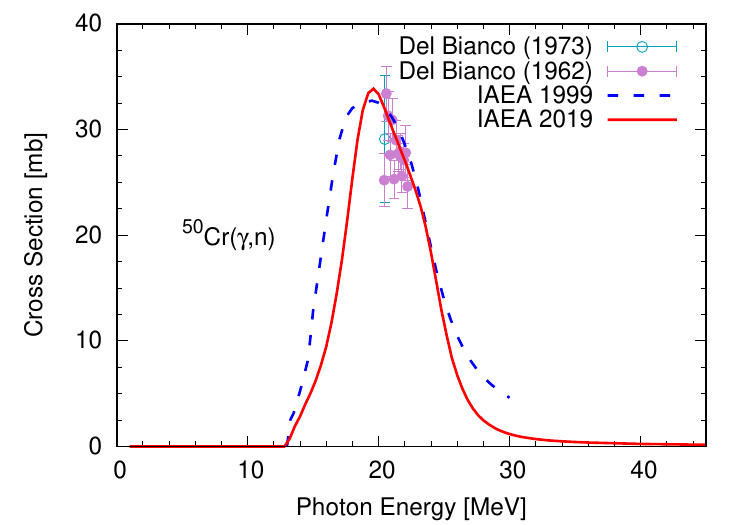}}
 \end{center}
 \caption{(Color online) Comparison of evaluation and experimental data for $^{50}$Cr$(\gamma,n)$.}
 \label{fig:CIAECr50n}
\end{figure}

The evaluation of the new $^{50}$Cr data was performed with the MEND-G
code. We employed the EGLO model for producing the photo-absorption
cross section of $^{50}$Cr.  Figure~\ref{fig:CIAECr50n} compares the
current evaluation with some available experimental data for the
$(\gamma,n)$ reaction and the previous evaluation. The peak location
of the new evaluation is slightly shifted toward the higher energy
side, and the cross section drops rapidly above 25~MeV. This is more
physical as the number of open channels increases at higher energies.

\subsubsection{$^{90}$Zr}

\begin{figure}[!htb]
 \begin{center}
   \resizebox{0.9\columnwidth}{!}{\includegraphics{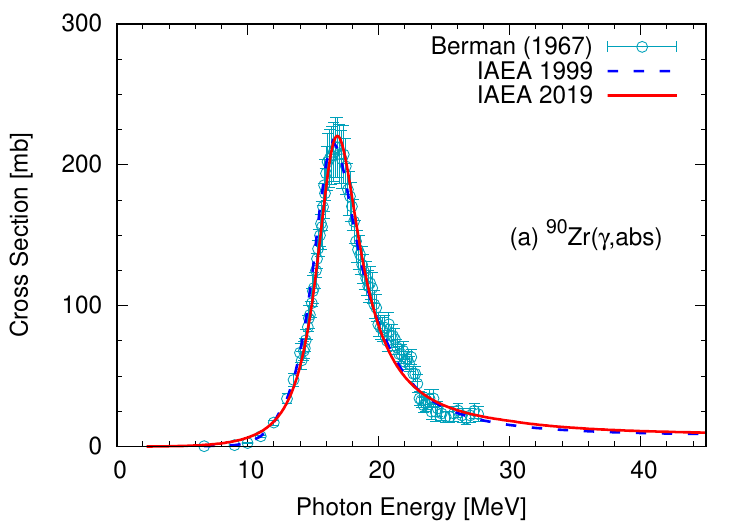}}
   \resizebox{0.9\columnwidth}{!}{\includegraphics{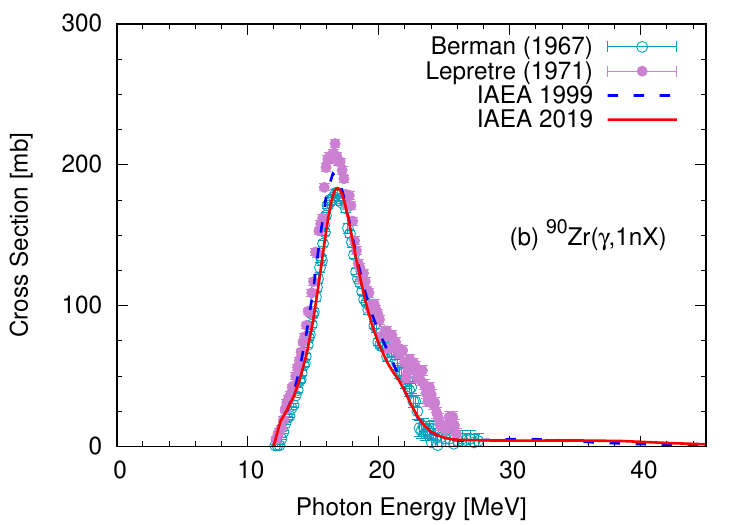}}
   \resizebox{0.9\columnwidth}{!}{\includegraphics{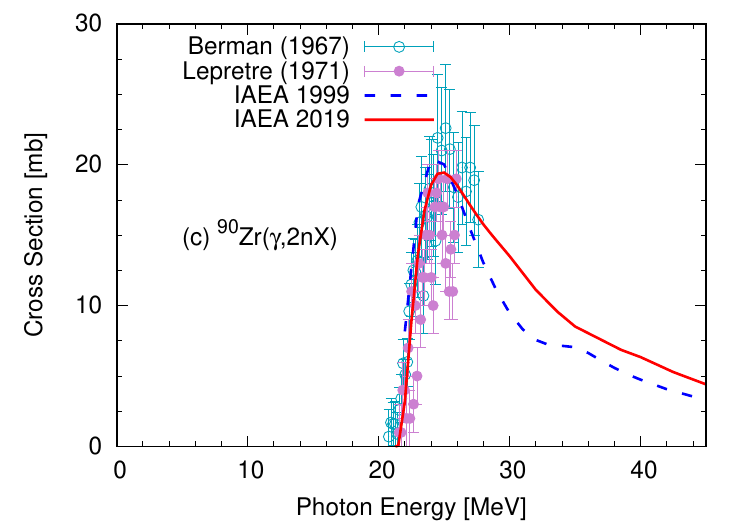}}
 \end{center}
 \caption{(Color online) Comparison of evaluated and experimental data for $^{90}$Zr;
    (a) photo-absorption, (b) $(\gamma,n)+(\gamma,n+p)$, and
    (c) $(\gamma,2n)+(\gamma,2n+p)$ cross sections.}
 \label{fig:CIAEZr90}
\end{figure}

The natural zirconium includes five stable isotopes, and CIAE
evaluated $^{90}$Zr that has the largest abundance of 51.45\%.  The
new evaluation was performed with the MEND-G code.  The experimental
data of Berman {\it et al.}~\cite{Berman1967} and Lepr\^{e}tre {\it et
  al.}~\cite{Lepretre1971} are reported for $^{90}$Zr. Varlamov {\it
  et al.} also published their evaluations of the partial neutron
production cross sections~\cite{Varlamov2018} in 2018. As has been
mentioned in previous sections, the data of Berman (Livermore) and
Lepr\^{e}tre (Saclay) show some inconsistencies. In this case, after
considering all the available data, we decided to adopt Berman's data
for our evaluation.  SMLO is adopted to produce the photo-absorption
cross sections. The evaluated photo-absorption cross section and the
related neutron emission cross sections are in good agreement with the
experimental data of Berman as shown in Fig.~\ref{fig:CIAEZr90}. We
concluded that the Lepr\^{e}tre data for $\sigma_{1nX}$ and
$\sigma_{2nX}$ could have a neutron mis-counting issue, hence the
current evaluation does not follow the Lepr\^{e}tre data.

\subsubsection{$^{118}$Sn}

\begin{figure}[!htb]
 \begin{center}
   \resizebox{0.9\columnwidth}{!}{\includegraphics{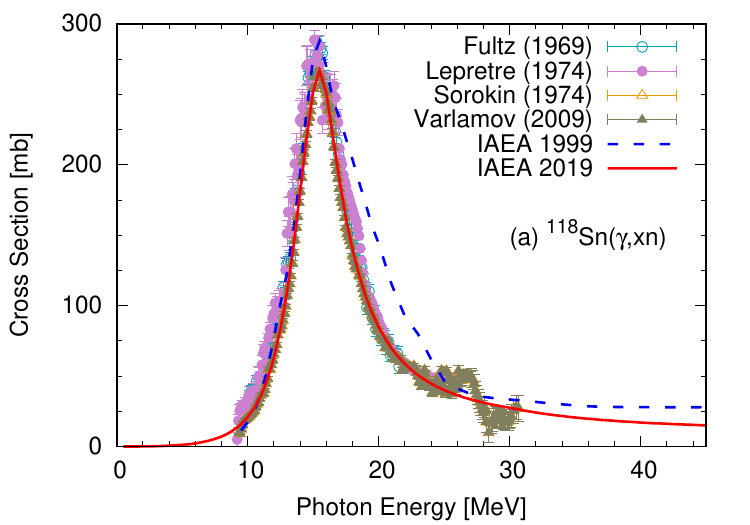}}
   \resizebox{0.9\columnwidth}{!}{\includegraphics{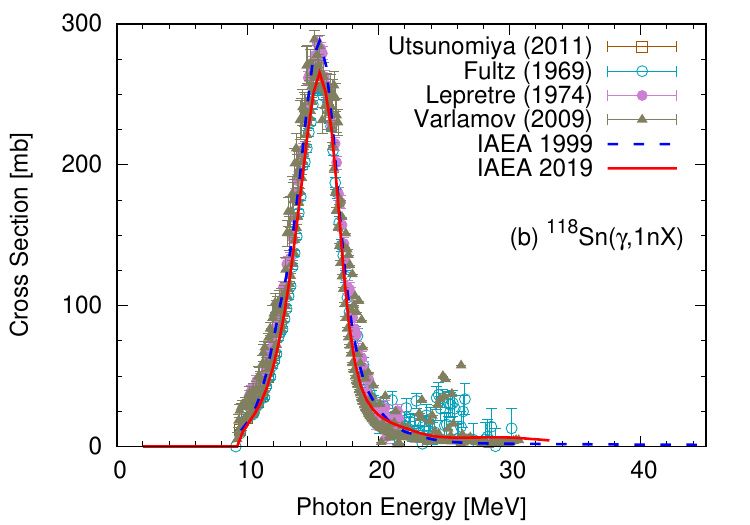}}
   \resizebox{0.9\columnwidth}{!}{\includegraphics{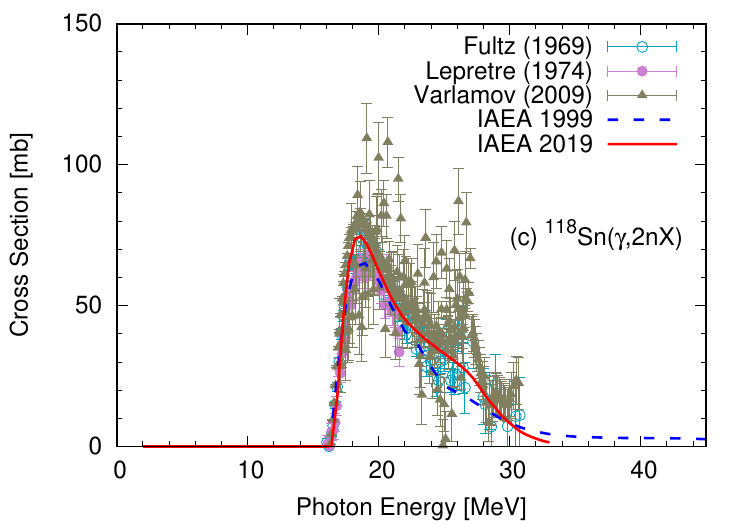}}
   \resizebox{0.9\columnwidth}{!}{\includegraphics{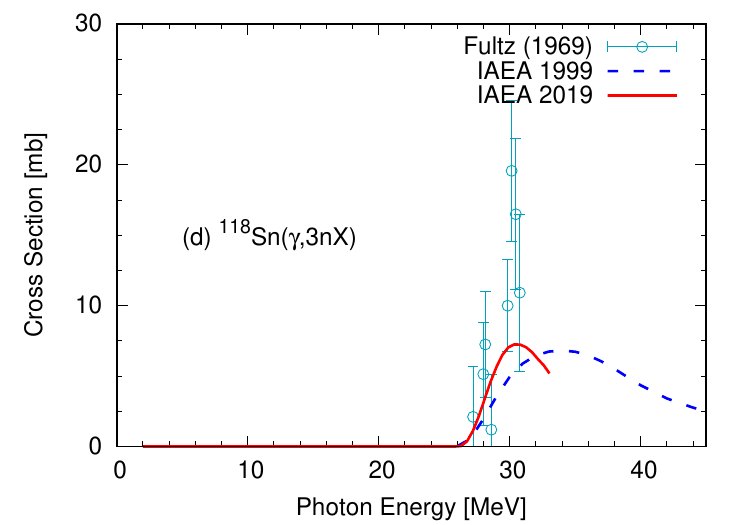}}
 \end{center}
 \caption{(Color online) Comparison of evaluated and experimental data for $^{118}$Sn;
  (a) $(\gamma,xn)$, (b) $(\gamma,1nX)$, (c)  $(\gamma,2nX)$, and
  (d) $(\gamma,3nX)$}
 \label{fig:CIAESn118}
\end{figure}

There are 10 stable isotopes in natural tin. The IAEA photonuclear
data library adopted $^{120}$Sn from
JENDL/PD-2016~\cite{JENDL-PD2016}, which has the largest abundance of
32.6\%. CIAE undertook the evaluation of $^{118}$Sn, whose abundance
is the second largest of 24.22\%.  The experimental data of Fultz {\it
  et al.}~\cite{Fultz1969} and Lepr\^{e}tre~\cite{Lepretre1974}
reported in 1960--1970s provide overall excitation functions of
$\sigma_{xn}$ and $\sigma_{inX}$ for $i=1,2$ and 3. On the other hand,
new neutron emission data by Utsunomiya {\it et
  al.}~\cite{Utsunomiya2011} were measured in 2011. Our evaluation
strictly adopts the new data by Utsunomiya, which are compare with the
other experimental data as well as the corrected data by Varlamov {\it
  et al.}  in the four panels of Fig.~\ref{fig:CIAESn118}.

\clearpage
\section{CONTENTS OF THE LIBRARY}
\label{sec:libcontent}
%input{libcontent}
The new IAEA photonuclear data library contains photo-induced reaction
data for 220 nuclides for photon energies up to 200~MeV. The energy
and mass ranges are expanded compared to the previous 1999
version~\cite{IAEAPhoto1999} which included 164 nuclides with the
energies mostly up to 140~MeV. Here we call new library IAEA 2019. In
Appendix~\ref{appendix:libindex} we summarize the source of evaluated
data for each isotope in both IAEA 1999 and IAEA 2019.

Many of the isotopes evaluations included in IAEA 1999 originated from
different evaluated nuclear data libraries maintained by different institutes. For
example, the actinides data were taken from BOFOD~\cite{Blokhin1992},
the JAERI evaluations were from JENDL photonuclear data file
2004~\cite{Kishida2004}. KAERI had significant contributions to both
JENDL 2004 and IAEA 1999.

The sources of the IAEA 2019 library are classified into;
\begin{itemize}
\item New photonuclear data evaluations for IAEA 2019 at CNDC/CIAE,
 as well as special upgrades of their new upcoming CENDL (Chinese
 Evaluated Nuclear Data Library);

\item JENDL Photonuclear Data File~\cite{JENDL-PD2016}, and
 upgrades and high energy extension, as well as new evaluations
 specialized for IAEA 2019 at NDC/JAEA;

\item New evaluations at IFIN-HH;

\item New evaluations at NDC/KAERI; and

\item Carry-over from IAEA 1999 photonuclear data library.
\end{itemize}

In some cases, the same isotopes were evaluated by more than one
institutes. The inter-comparison of the different evaluations led to
improvements in the evaluations performed by the individual
institutes. They also triggered discussions on potential issues in the
experimental data themselves, as most often, the differences in the
evaluations were a result of adopting different experimental data
sets. In the end, one set of evaluations was selected for these
isotopes. The criteria for selection were based mainly on how well
these evaluations reproduce the recommended available experimental
data. An external review team that consisted of nuclear data
evaluation experts, some of whom contributed to the previous IAEA 1999
library, reviewed the evaluations and following discussions with the
evaluators arrived at a consensus on a unique set of the data
library. It should be emphasized that all the evaluated data files that
were produced within this CRP, including those that were not included
in the new IAEA library, are of extremely good quality and could be
available as part of other libraries such as CENDL, JENDL, {\it etc.}
in near future.

The new IAEA Photonuclear Data Library is produced in the ENDF-6
format. The ENDF-6 format adopted for the stored cross sections and
energy spectra depends on the isotopes, due to a diffculty in applying
a common evaluation technique and format for all the isotopes. For
some nuclides, typically for light elements, exclusive nuclear
reaction channels are explicitly given in MF=3 (MF: the file number
defined in ENDF-6). While in many cases the MF=3 contains the total
photo-absorption cross section only, and all individual information,
such as the particle multiplicities and the residual nucleus
production probabilities, is stored in MF=6. The MF=6 representation is the
most common method since the library contains the high energy data
part where the number of exclusive reaction channels exceeds the
limitation of the ENDF-6 rule.

The new IAEA Photonuclear Data Library is available through the IAEA
ENDF web interface ({\tt http://www-nds.iaea.org/endf/}). In addition,
users will be able to access all the graphical comparisons of the new
evaluated data with experimental data and previous evaluations from a
dedicated web interface that is under construction on the IAEA web
server. The evaluated data will be downloadable as simple asci files
from this web interface as well.

\section{CONCLUSIONS}
\label{sec:conclusions}
he new 2019 IAEA Photonuclear Data Library contains 220 isotopes,
including 56 newly added data files. This library is one of the major
final products of the IAEA CRP ``Updating the Photonuclear Data Library
and Generating a Reference Database for Photon Strength Functions.''
The photon energy range was extended to 200~MeV to cover many
applications of photonuclear data for radiation transport calculations
as well as isotope production. In this paper we reviewed the
experimental techniques used to measure photonuclear reaction cross
sections, and summarized possible issues in the reported experimental
data due to deficiencies in the techniques. We discussed the method of
assessing published partial photoneutron cross section data developed
by Varlamov {\it et al.}, to identify neutron mis-counting issues that
lurk in the neutron-multiplicity sorting technique. The method which
is based on the $F_i$ correction factors, has been applied to several of
the new evaluations, however, the need to consider the model-dependent
uncertainties when applying it to correct the experimental data was
highlighted in this review. One important accomplishment of this CRP
was the completion of the new direct multiplicity counting technique
at the NewSUBARU facility that led to new measurements of $(\gamma,
inX)$ cross sections for seven isotopes. These new data were
considered in the new evaluations performed for the updated library.

The statistical Hauser-Feshbach model calculation was heavily involved
in the data evaluation; GLUNF and MEND-G codes at CIAE, CCONE at JAEA,
TALYS at KAERI, and EMPIRE at IFIN-HH.  Since understanding
characteristics of each Hauser-Feshbach code is crucial to estimate
uncertainties coming from the limitation in the evaluation technique,
a code inter-comparison was performed.  The comparison also included
the CoH$_3$ code at LANL, as well as the CPNRM code at SINP/MSU, which
was used for the $F_i$ value correction to experimental data. We
reported that the largest difference amongst the model codes is seen in
the pre-equilibrium modeling in each code.

We assembled the new data library by collecting five main sources; new
evaluations performed at CIAE along with some special upgrades of the
new CENDL photodata library for the IAEA, JENDL/PD-2016 evaluations
and new evaluations at JAEA, new evaluations at KAERI, and new
evaluations at IFIN-HH based on the experimental data from NewSUBARU
Laser Compton Scattering Facility, and some data files carried-over
from the 1999 IAEA Photonuclear Data Library. Finally some selected
comparisons of the newly evaluated data with available experimental
data were given to demonstrate the improved evaluations over the
previous IAEA data library. The new 2019 IAEA Photonuclear Data
Library will be available on the IAEA web server ({\tt
http://www-nds.iaea.org/endf}).

\acknowledgments

This work was performed within the IAEA CRP on ``Updating the
Photonuclear Data Library and Generating a Reference Database for
Photon Strength Functions'' (F41032). Contributors from the following
institutes are grateful to the IAEA for financial support within the
CRP: CIAE, KAERI, IFIN-HH, Taras Shevchenko National University,
Lomonosov Moscow State University, Centre for Energy Research (HAS),
Charles University (Prague), and iThemba LABS.
The work described in this paper would not have been
possible without contributions by the IAEA member states.  HU
acknowledges support extended to the PHOENIX collaboration for the
IAEA-CRP by the Premier Project of Konan University.  YC was supported
by the R\&D program of the Korea Atomic Energy Research Institute
funded by the Korean Ministry of Science and ICT.  TK carried out this
work under the auspices of the National Nuclear Security
Administration of the U.S. Department of Energy at Los Alamos National
Laboratory under Contract No. 89233218CNA000001.

%\clearpage
%\bibliography{reference}
%\input{reference}

\clearpage
\appendix

\section{Contents of IAEA 1999 and 2019 Photonuclear Data Libraries}
\label{appendix:libindex}
In the following table, we summarize the source of evaluated file for
each isotope. In the IAEA 1999 column, an institute name that
undertook the evaluation is given.  When this column is empty, no
evaluation is given in IAEA 1999.  In the IAEA 2019 column,
JENDL/PD-2016 means these files were taken from JENDL Photonuclear Data
File 2016~\cite{JENDL-PD2016}, the $=$ sign indicates the IAEA 1999 data
were carried over, and the institution names stand for the new
evaluations performed at each institute.

\begin{table}[!h]
\caption{Contents of IAEA 1999 and 2019 photonuclear data libraries. The source of the data or leading evaluation institute is given. The $=$ sign means the 2019 library is the same as 1999.}
\label{table:libcont}
\begin{center}
\begin{tabular}{c|l|l}
\hline\hline
 Nuclide  & IAEA 1999& IAEA 2019 \\
\hline
 $^{ 2}$H  & JAERI   & = \\
 $^{ 3}$He &         & JENDL/PD-2016 \\
 $^{ 6}$Li &         & JENDL/PD-2016 \\
 $^{ 7}$Li &         & JENDL/PD-2016 \\
 $^{ 9}$Be & CIAE    & CIAE \\
 $^{12}$C  & LANL    & CIAE \\
 $^{13}$C  & KAERI   & = \\
 $^{14}$C  &         & KAERI \\
 $^{14}$N  & JAERI   & CIAE \\
 $^{15}$N  & KAERI   & = \\
 $^{16}$O  & LANL    & = \\
 $^{17}$O  & KAERI   & = \\
 $^{18}$O  & KAERI   & = \\
 $^{19}$F  &         & JENDL/PD-2016 \\
 $^{23}$Na & KAERI   & = \\
 $^{24}$Mg & KAERI   & = \\
 $^{25}$Mg & KAERI   & = \\
 $^{26}$Mg & KAERI   & = \\
 $^{27}$Al & LANL    & CIAE \\
 $^{27}$Si & KAERI   & = \\
 $^{28}$Si & KAERI   & = \\
 $^{29}$Si & KAERI   & = \\
 $^{30}$Si & KAERI   & = \\
 $^{32}$S  & KAERI   & = \\
 $^{33}$S  & KAERI   & = \\
 $^{34}$S  & KAERI   & = \\
 $^{36}$S  & KAERI   & = \\
 $^{35}$Cl & KAERI   & JAEA \\
 $^{37}$Cl & KAERI   & JAEA \\
 $^{36}$Ar & KAERI   & JAEA \\
 $^{38}$Ar & KAERI   & JAEA \\
 $^{40}$Ar & KAERI   & JAEA \\
 $^{39}$K  & KAERI   & JAEA \\
 $^{40}$K  & KAERI   & JAEA \\
 $^{41}$K  & KAERI   & JAEA \\
 $^{40}$Ca  & LANL   & JENDL/PD-2016 \\
 $^{42}$Ca  & KAERI  & = \\
 $^{43}$Ca  & KAERI  & = \\
 $^{44}$Ca  & KAERI  & JENDL/PD-2016 \\
 $^{46}$Ca  & KAERI  & = \\
\hline
\end{tabular}
\end{center}
\end{table}

\noindent TABLE~\ref{table:libcont} continued.
\begin{center}
\begin{tabular}{c|l|l}
\hline
 Nuclide  & IAEA 1999& IAEA 2019 \\
\hline
 $^{48}$Ca  & KAERI  & JENDL/PD-2016 \\
 $^{45}$Sc  &        & JAEA \\
 $^{46}$Ti  & KAERI  & JAEA \\
 $^{47}$Ti  & KAERI  & JAEA \\
 $^{48}$Ti  & KAERI  & JAEA \\
 $^{49}$Ti  & KAERI  & JAEA \\
 $^{50}$Ti  & KAERI  & JAEA \\
 $^{50}$V   &        & JAEA \\
 $^{51}$V   & CIAE   & JAEA \\
 $^{50}$Cr  & CIAE   & CIAE \\
 $^{52}$Cr  & CIAE   & JAEA \\
 $^{53}$Cr  & CIAE   & = \\
 $^{54}$Cr  & CIAE   & = \\
 $^{55}$Mn  & KAERI  & JAEA \\
 $^{54}$Fe  & JAERI  & JAEA \\
 $^{56}$Fe  & JAERI  & JAEA \\
 $^{57}$Fe  & KAERI  & JAEA \\
 $^{58}$Fe  & KAERI  & JAEA \\
 $^{59}$Co  & KAERI  & IFIN-HH \\
 $^{58}$Ni  & JAERI  & JAEA \\
 $^{60}$Ni  & KAERI  & JAEA \\
 $^{61}$Ni  & KAERI  & JAEA \\
 $^{62}$Ni  & KAERI  & JAEA \\
 $^{64}$Ni  & KAERI  & JAEA \\
 $^{63}$Cu  & LANL   & JAEA \\
 $^{65}$Cu  & JAERI  & JAEA \\
 $^{64}$Zn  & JAERI  & KAERI \\
 $^{66}$Zn  & KAERI  & JENDL/PD-2016 \\
 $^{67}$Zn  & KAERI  & JENDL/PD-2016 \\
 $^{68}$Zn  & KAERI  & JENDL/PD-2016 \\
 $^{70}$Zn  & KAERI  & JENDL/PD-2016 \\
 $^{70}$Ge  & KAERI  & JENDL/PD-2016 \\
 $^{72}$Ge  & KAERI  & JENDL/PD-2016 \\
 $^{73}$Ge  & KAERI  & JENDL/PD-2016 \\
 $^{74}$Ge  & KAERI  & JENDL/PD-2016 \\
 $^{76}$Ge  & KAERI  & JENDL/PD-2016 \\
 $^{75}$As  &        & KAERI \\
 $^{76}$Se  &        & KAERI \\
 $^{78}$Se  &        & KAERI \\
 $^{80}$Se  &        & KAERI \\
 $^{84}$Sr  & KAERI  & JENDL/PD-2016 \\
 $^{86}$Sr  & KAERI  & JENDL/PD-2016 \\
 $^{87}$Sr  & KAERI  & JENDL/PD-2016 \\
 $^{88}$Sr  & KAERI  & JENDL/PD-2016 \\
 $^{90}$Sr  & KAERI  & JENDL/PD-2016 \\
 $^{89}$Y   &        & IFIN-HH \\
 $^{90}$Zr  & KAERI  & CIAE \\
 $^{91}$Zr  & CIAE   & KAERI \\
 $^{92}$Zr  & CIAE   & JENDL/PD-2016 \\
 $^{93}$Zr  & KAERI  & JENDL/PD-2016 \\
 $^{94}$Zr  & KAERI  & KAERI \\
 $^{96}$Zr  & CIAE   & JENDL/PD-2016 \\
 $^{93}$Nb  & KAERI  & JENDL/PD-2016 \\
 $^{94}$Nb  & KAERI  & JENDL/PD-2016 \\
 $^{82}$Se  &        & KAERI \\
\hline
\end{tabular}
\end{center}

\noindent TABLE~\ref{table:libcont} continued.
\begin{center}
\begin{tabular}{c|l|l}
\hline
 Nuclide  & IAEA 1999& IAEA 2019 \\
\hline
 $^{ 92}$Mo & KAERI  & JENDL/PD-2016 \\
 $^{ 94}$Mo & KAERI  & JENDL/PD-2016 \\
 $^{ 95}$Mo & KAERI  & JENDL/PD-2016 \\
 $^{ 96}$Mo & KAERI  & JENDL/PD-2016 \\
 $^{ 97}$Mo & KAERI  & JENDL/PD-2016 \\
 $^{ 98}$Mo & KAERI  & JENDL/PD-2016 \\
 $^{100}$Mo & KAERI  & JENDL/PD-2016 \\
 $^{ 98}$Ru &        & JAEA \\
 $^{103}$Rh &        & IFIN-HH \\
 $^{102}$Pd & KAERI  & JENDL/PD-2016 \\
 $^{104}$Pd & KAERI  & JENDL/PD-2016 \\
 $^{105}$Pd & KAERI  & JENDL/PD-2016 \\
 $^{106}$Pd & KAERI  & JENDL/PD-2016 \\
 $^{107}$Pd & KAERI  & JENDL/PD-2016 \\
 $^{108}$Pd & KAERI  & JENDL/PD-2016 \\
 $^{110}$Pd & KAERI  & JENDL/PD-2016 \\
 $^{107}$Ag & KAERI  & JENDL/PD-2016 \\
 $^{108}$Ag & KAERI  & JAEA \\
 $^{109}$Ag & KAERI  & JENDL/PD-2016 \\
 $^{106}$Cd & KAERI  & JENDL/PD-2016 \\
 $^{108}$Cd & KAERI  & JENDL/PD-2016 \\
 $^{110}$Cd & KAERI  & JENDL/PD-2016 \\
 $^{111}$Cd & KAERI  & JENDL/PD-2016 \\
 $^{112}$Cd & KAERI  & JENDL/PD-2016 \\
 $^{113}$Cd & KAERI  & JENDL/PD-2016 \\
 $^{114}$Cd & KAERI  & JENDL/PD-2016 \\
 $^{116}$Cd & KAERI  & JENDL/PD-2016 \\
 $^{115}$In &        & KAERI \\
 $^{112}$Sn & KAERI  & JENDL/PD-2016 \\
 $^{114}$Sn & KAERI  & JENDL/PD-2016 \\
 $^{115}$Sn & KAERI  & JENDL/PD-2016 \\
 $^{116}$Sn & KAERI  & KAERI \\
 $^{117}$Sn & KAERI  & JENDL/PD-2016 \\
 $^{118}$Sn & KAERI  & CIAE \\
 $^{119}$Sn & KAERI  & JENDL/PD-2016 \\
 $^{120}$Sn & KAERI  & JENDL/PD-2016 \\
 $^{122}$Sn & KAERI  & JENDL/PD-2016 \\
 $^{124}$Sn & KAERI  & JENDL/PD-2016 \\
 $^{121}$Sb & KAERI  & JENDL/PD-2016 \\
 $^{123}$Sb & KAERI  & JENDL/PD-2016 \\
 $^{120}$Te & KAERI  & JENDL/PD-2016 \\
 $^{122}$Te & KAERI  & JENDL/PD-2016 \\
 $^{123}$Te & KAERI  & JENDL/PD-2016 \\
 $^{124}$Te & KAERI  & JENDL/PD-2016 \\
 $^{125}$Te & KAERI  & JENDL/PD-2016 \\
 $^{126}$Te & KAERI  & JENDL/PD-2016 \\
 $^{128}$Te & KAERI  & JENDL/PD-2016 \\
 $^{130}$Te & KAERI  & JENDL/PD-2016 \\
 $^{127}$I  & KAERI  & JENDL/PD-2016 \\
 $^{129}$I  & KAERI  & JENDL/PD-2016 \\
 $^{132}$Xe &        & JAEA \\
 $^{133}$Cs & KAERI  & KAERI \\
 $^{135}$Cs & KAERI  & JENDL/PD-2016 \\
 $^{137}$Cs & KAERI  & JENDL/PD-2016 \\
 $^{138}$Ba &        & KAERI \\
\hline
\end{tabular}
\end{center}

\noindent TABLE~\ref{table:libcont} continued.
\begin{center}
\begin{tabular}{c|l|l}
\hline
 Nuclide  & IAEA 1999& IAEA 2019 \\
\hline
 $^{139}$La &        & JAEA \\
 $^{140}$Ce &        & KAERI \\
 $^{142}$Ce &        & KAERI \\
 $^{141}$Pr & KAERI  & JENDL/PD-2016 \\
 $^{142}$Nd &        & KAERI \\
 $^{143}$Nd &        & KAERI \\
 $^{144}$Nd &        & KAERI \\
 $^{145}$Nd &        & KAERI \\
 $^{146}$Nd &        & KAERI \\
 $^{148}$Nd &        & KAERI \\
 $^{150}$Nd &        & KAERI \\
 $^{144}$Sm & KAERI  & JAEA \\
 $^{147}$Sm & KAERI  & JAEA \\
 $^{148}$Sm & KAERI  & JAEA \\
 $^{149}$Sm & KAERI  & JAEA \\
 $^{150}$Sm & KAERI  & JAEA \\
 $^{151}$Sm & KAERI  & JAEA \\
 $^{152}$Sm & KAERI  & JAEA \\
 $^{154}$Sm & KAERI  & JAEA \\
 $^{153}$Eu &        & KAERI \\
 $^{156}$Gd &        & JAEA \\
 $^{157}$Gd &        & JAEA \\
 $^{158}$Gd &        & JAEA \\
 $^{160}$Gd &        & JENDL/PD-2016 \\
 $^{158}$Tb & KAERI  & JENDL/PD-2016 \\
 $^{159}$Tb & KAERI  & IFIN-HH \\
 $^{162}$Dy &        & JAEA \\
 $^{163}$Dy &        & JAEA \\
 $^{165}$Ho & KAERI  & IFIN-HH \\
 $^{166}$Er &        & JAEA \\
 $^{170}$Er &        & JAEA \\
 $^{169}$Tm &        & IFIN-HH \\
 $^{175}$Lu &        & KAERI \\
 $^{174}$Hf &        & JAEA \\
 $^{176}$Hf &        & JAEA \\
 $^{177}$Hf &        & JAEA \\
 $^{178}$Hf &        & JAEA \\
 $^{179}$Hf &        & JAEA \\
 $^{180}$Hf &        & JAEA \\
 $^{181}$Ta & JAERI  & IFIN-HH \\
 $^{180}$W  & CIAE   & CIAE \\
 $^{182}$W  & JAERI  & CIAE \\
 $^{183}$W  & CIAE   & CIAE \\
 $^{184}$W  & LANL   & CIAE \\
 $^{186}$W  & JAERI  & CIAE \\
 $^{185}$Re &        & JAEA \\
 $^{187}$Re &        & JAEA \\
 $^{186}$Os &        & KAERI \\
 $^{188}$Os &        & KAERI \\
 $^{189}$Os &        & KAERI \\
 $^{190}$Os &        & KAERI \\
 $^{192}$Os &        & KAERI \\
 $^{194}$Pt &        & JAEA \\
 $^{197}$Au & KAERI  & JAEA \\
 $^{206}$Pb & LANL   & JENDL/PD-2016 \\
\hline
\end{tabular}
\end{center}

\noindent TABLE~\ref{table:libcont} continued.
\begin{center}
\begin{tabular}{c|l|l}
\hline
 Nuclide  & IAEA 1999& IAEA 2019 \\
\hline
 $^{207}$Pb & LANL   & JENDL/PD-2016 \\
 $^{208}$Pb & LANL   & JENDL/PD-2016 \\
 $^{209}$Bi & CIAE   & KAERI \\
 $^{226}$Ra &        & JAEA \\
 $^{232}$Th & IPPE   & JENDL/PD-2016 \\
 $^{233}$U  & IPPE   & JENDL/PD-2016 \\
 $^{234}$U  & IPPE   & JENDL/PD-2016 \\
 $^{235}$U  & IPPE   & JENDL/PD-2016 \\
 $^{236}$U  & IPPE   & JENDL/PD-2016 \\
 $^{238}$U  & IPPE   & JENDL/PD-2016 \\
 $^{237}$Np &        & JENDL/PD-2016 \\
 $^{238}$Pu & IPPE   & JENDL/PD-2016 \\
 $^{239}$Pu & IPPE   & JENDL/PD-2016 \\
 $^{241}$Pu & IPPE   & JENDL/PD-2016 \\
\hline\hline
\end{tabular}
\end{center}

\vfill\null
\break

%\clearpage

\section{GDR ATLAS}
\label{appendix:GDR}
Here we summarize the updated tables for the recommended experimental
GDR parameters within the Standard Lorentzian (SLO) and the Simplified
version of the modified Lorentzian (SMLO) approaches. In the case of
double humped shape, two sets of GDR parameters are given. A complete
list of all the references is given at {\tt
https://www-nds.iaea.org/CRP-photonuclear/}, and only abbreviations
are given for the sake of simplicity.

\begin{table}[!h]
\caption{Recommended experimental GDR parameters within the Standard Lorentzian
(SLO) approach.}
\label{table:GDRAtlas1}
\begin{center}
\begin{tabular}{p{1cm}|p{1cm}p{1cm}p{1cm}|c|c}
\hline\hline
           & $E$ & $\Gamma$ & $\sigma$ & Range & Ref. \\
           & [MeV] & [MeV]      & [mb]      & [MeV] & \\
\hline
$^{  6}$Li &  23.69 &  5.26 &    3.71 &21.5 -- 27.0 &  1986Var \\
$^{  7}$Li &  18.59 & 16.28 &    3.65 &13.2 -- 25.6 &  1985Ahr \\
$^{  9}$Be &  23.75 &  9.47 &    5.17 &17.5 -- 26.0 &  1975Ahr \\
$^{ 10}$B  &  21.72 &  9.08 &    4.64 & 8.5 -- 24.9 &  1987Ahs \\
$^{ 12}$C  &  22.86 &  3.61 &   21.30 &20.1 -- 25.0 &  1969Bez \\
$^{ 13}$C  &  24.60 &  8.43 &   12.71 &14.5 -- 29.0 &  2002Ish \\
$^{ 14}$C  &  15.41 &  5.82 &    7.49 &14.5 -- 30.0 &  2002Ish \\
           &  26.13 &  7.78 &    8.13 &             & \\
$^{\rm nat}$C  &  23.12 &  4.19 &   19.28 &19.5 -- 25.6 &  1985Ahr \\
$^{ 14}$N  &  23.05 &  6.95 &   22.95 &18.2 -- 28.0 &  1969Bez \\
$^{ 15}$N  &  24.78 & 12.82 &   13.82 &14.5 -- 28.0 &  2002Ish \\
$^{ 16}$O  &  23.70 &  5.36 &   27.96 &18.1 -- 26.0 &  1975Ahr \\
$^{ 17}$O  &  23.40 &  5.48 &   21.82 &18.5 -- 26.5 &  2002Ish \\
$^{ 18}$O  &  19.08 &  2.12 &    5.13 &18.5 -- 26.0 &  2002Ish \\
           &  24.10 &  5.25 &   13.48 &             & \\
$^{\rm nat}$O  &  23.60 &  5.82 &   27.07 &18.9 -- 27.9 &  1985Ahr \\
$^{ 19}$F  &  21.61 & 12.57 &   16.58 &10.0 -- 24.0 &  2002Ish \\
$^{ 23}$Na &  17.43 &  3.10 &   12.38 &14.2 -- 23.0 &  1981Ish \\
           &  21.13 &  4.51 &   26.98 &             & \\
$^{ 24}$Mg &  19.51 &  2.71 &   21.40 &16.5 -- 27.0 &  1966Dol \\
           &  23.88 &  8.86 &   25.20 &             & \\
$^{ 25}$Mg &  22.06 &  6.09 &   34.99 & 9.0 -- 24.2 &  2002Ish \\
$^{ 26}$Mg &  17.38 &  2.15 &   16.53 &16.1 -- 26.5 &  2003Var \\
           &  23.64 &  7.25 &   39.12 &             & \\
$^{\rm nat}$Mg &  22.55 &  7.97 &   20.88 &15.1 -- 26.6 &  1965Wyc \\
$^{ 27}$Al &  20.73 &  7.45 &   38.99 &14.0 -- 24.1 &  1985Ahr \\
$^{ 28}$Si &  19.81 &  2.56 &   38.75 &16.7 -- 23.0 &  2003Var \\
           &  21.81 &  3.15 &   40.23 &             & \\
$^{ 29}$Si &  20.70 &  5.60 &   40.01 &14.2 -- 23.0 &  2002Ish \\
$^{ 30}$Si &  20.86 &  7.40 &   29.56 &14.2 -- 23.0 &  2002Ish \\
$^{\rm nat}$Si &  20.35 &  4.53 &   51.41 &16.4 -- 25.8 &  1975Ahr \\
           &  25.16 &  2.86 &   10.47 &             & \\
$^{ 32}$S  &  19.51 &  4.83 &   35.43 &14.7 -- 23.0 &  1968Dol \\
$^{ 34}$S  &  20.89 &  9.61 &   50.54 &12.0 -- 25.0 &  1986Ass \\
$^{\rm nat}$S  &  20.31 &  5.48 &   47.84 &17.2 -- 23.6 &  1965Wyc \\
$^{ 40}$Ar &  19.86 &  9.12 &   56.89 &10.5 -- 25.0 &  2002Ish \\
$^{\rm nat}$K  &  21.12 &  6.89 &   22.58 &16.0 -- 25.9 &  1974Ve1 \\
$^{ 40}$Ca &  20.58 &  6.23 &  104.78 &17.2 -- 23.7 &  1966Dol \\
$^{ 42}$Ca &  20.11 &  8.07 &   72.15 &15.2 -- 23.0 &  2003Ero \\
$^{ 44}$Ca &  19.60 & 11.33 &   63.70 &15.5 -- 26.0 &  2003Ero \\
$^{ 48}$Ca &  19.70 &  6.23 &  105.44 &17.9 -- 21.6 &  1987OKe \\
$^{\rm nat}$Ca &  20.06 &  4.89 &   94.13 &15.1 -- 24.0 &  1975Ahr \\
$^{ 46}$Ti &  19.96 &  6.92 &   78.94 &13.2 -- 25.0 &  2002Ish \\
\hline
\end{tabular}
\end{center}
\end{table}

\noindent TABLE~\ref{table:GDRAtlas1} continued.
\begin{center}
\begin{tabular}{p{1cm}|p{1cm}p{1cm}p{1cm}|c|c}
\hline
           & $E$ & $\Gamma$ & $\sigma$ & Range & Ref. \\
           & [MeV] & [MeV]      & [mb]      & [MeV] & \\
\hline
$^{ 48}$Ti &  19.78 &  8.42 &   63.74 &14.5 -- 23.0 &  2002Ish \\
$^{ 51}$V  &  17.90 &  4.55 &   60.32 &14.1 -- 22.9 &  1962Fu1 \\
           &  21.26 &  4.37 &   26.38 &             & \\
$^{ 52}$Cr &  19.16 &  6.19 &   81.34 &14.3 -- 23.0 &  2002Ish \\
$^{ 55}$Mn &  16.43 &  2.95 &   27.02 &14.0 -- 23.0 &  1979Al2 \\
           &  19.77 &  8.61 &   52.03 &             & \\
$^{ 54}$Fe &  19.35 &  5.50 &  147.00 &16.0 -- 23.0 &  1978Nor \\
$^{ 59}$Co &  16.43 &  2.73 &   28.28 &14.0 -- 20.9 &  1979Al2 \\
           &  18.64 &  7.31 &   57.16 &             & \\
$^{ 58}$Ni &  18.78 &  5.57 &   87.90 &14.1 -- 22.0 &  2003Var \\
$^{ 60}$Ni &  16.69 &  3.47 &   62.80 &12.1 -- 21.0 &  2003Var \\
           &  19.57 &  5.26 &   71.14 &             & \\
$^{ 63}$Cu &  16.43 &  4.84 &   79.79 &14.0 -- 21.0 &  2003Var \\
           &  20.15 &  5.52 &   49.39 &             & \\
$^{ 65}$Cu &  16.92 &  8.09 &   86.38 &14.2 -- 21.0 &  2003Var \\
$^{\rm nat}$Cu &  18.12 &  5.61 &   97.93 &14.4 -- 24.9 &  1965Wyc \\
$^{ 64}$Zn &  16.23 &  3.25 &   41.21 &14.0 -- 20.8 &  1976Ca1 \\
           &  19.16 &  5.91 &   54.90 &             & \\
$^{ 65}$Zn &  16.17 &  3.06 &   34.89 &12.0 -- 21.0 &  2003Rod \\
           &  19.04 &  6.50 &   55.53 &             & \\
$^{ 70}$Ge &  15.16 &  5.92 &  160.51 &10.0 -- 20.0 &  1975Mcc \\
$^{ 72}$Ge &  17.88 &  5.71 &  167.57 &10.0 -- 24.0 &  1975Mcc \\
$^{ 74}$Ge &  14.51 &  2.01 &   25.54 &13.1 -- 20.8 &  1976Ca1 \\
           &  17.03 &  7.97 &  100.80 &             & \\
$^{ 76}$Ge &  15.48 &  4.37 &   61.70 &13.1 -- 20.8 &  1976Ca1 \\
           &  18.87 & 10.99 &   71.09 &             & \\
$^{ 75}$As &  14.98 &  3.66 &   41.85 &13.1 -- 20.9 &  1969Be1 \\
           &  17.59 &  7.12 &   75.35 &             & \\
$^{ 76}$Se &  15.67 &  6.33 &  151.59 &13.1 -- 19.7 &  1978Gur \\
$^{ 78}$Se &  14.97 &  3.91 &   70.45 &13.1 -- 20.8 &  1976Ca1 \\
           &  18.42 &  6.19 &   79.42 &             & \\
$^{ 80}$Se &  16.60 &  6.80 &  137.82 &13.1 -- 20.0 &  2016Va1 \\
$^{ 82}$Se &  16.00 &  5.68 &  175.06 &13.1 -- 19.9 &  1978Gur \\
$^{\rm nat}$Rb &  16.73 &  4.25 &  190.04 &10.6 -- 17.9 &  1971Lep \\
$^{\rm nat}$Sr &  16.79 &  4.32 &  205.99 &10.9 -- 17.9 &  1971Lep \\
$^{ 89}$Y  &  16.74 &  4.23 &  224.56 &14.0 -- 19.0 &  1971Le1 \\
$^{ 90}$Zr &  16.82 &  3.99 &  253.58 &14.9 -- 18.5 &  2003Var \\
$^{ 91}$Zr &  16.58 &  4.17 &  183.17 &14.0 -- 18.9 &  1967Be2 \\
$^{ 92}$Zr &  16.26 &  4.64 &  164.72 &14.0 -- 18.9 &  1967Be2 \\
$^{ 94}$Zr &  16.21 &  5.25 &  159.83 &14.0 -- 18.9 &  1967Be2 \\
$^{ 93}$Nb &  16.58 &  4.95 &  200.25 &14.0 -- 19.0 &  1971Le1 \\
$^{ 92}$Mo &  17.16 &  4.68 &  239.77 &14.4 -- 19.0 &  2003Var \\
$^{ 94}$Mo &  16.53 &  5.12 &  192.92 & 9.6 -- 18.9 &  1974Be3 \\
$^{ 96}$Mo &  16.11 &  5.64 &  184.80 &13.2 -- 17.0 &  1974Be3 \\
$^{ 98}$Mo &  15.79 &  5.90 &  188.16 &13.2 -- 18.9 &  1974Be3 \\
$^{100}$Mo &  15.72 &  7.68 &  170.10 &12.1 -- 20.0 &  1974Be3 \\
$^{103}$Rh &  16.24 &  7.49 &  192.03 &13.1 -- 19.0 &  2003Var \\
$^{108}$Pd &  14.97 &  5.42 &  143.49 &10.0 -- 19.0 &  1969Dea \\
           &  18.13 &  4.03 &  103.37 &             & \\
$^{\rm nat}$Pd &  15.89 &  6.30 &  201.91 &10.2 -- 17.8 &  1971Lep \\
$^{107}$Ag &  15.83 &  6.49 &  185.05 & 9.5 -- 19.0 &  1969Ish \\
$^{109}$Ag &  13.54 &  3.49 &   80.46 &13.1 -- 19.0 &  1969Ish \\
           &  16.62 &  4.41 &  113.46 &             & \\
$^{\rm nat}$Ag &  16.06 &  7.33 &  197.91 &13.2 -- 18.9 &  1971Lep \\
$^{\rm nat}$Cd &  15.77 &  5.70 &  228.50 &10.2 -- 17.8 &  1971Lep \\
\hline
\end{tabular}
\end{center}

\noindent TABLE~\ref{table:GDRAtlas1} continued.
\begin{center}
\begin{tabular}{p{1cm}|p{1cm}p{1cm}p{1cm}|c|c}
\hline
           & $E$ & $\Gamma$ & $\sigma$ & Range & Ref. \\
           & [MeV] & [MeV]      & [mb]      & [MeV] & \\
\hline
$^{115}$In &  15.72 &  5.57 &  245.50 &13.2 -- 17.8 &  1974Le1 \\
$^{112}$Sn &  15.62 &  5.01 &  262.94 &10.9 -- 18.0 &  1974Sor \\
$^{114}$Sn &  15.82 &  6.08 &  243.01 &13.1 -- 18.0 &  1975Sor \\
$^{116}$Sn &  15.55 &  5.06 &  269.30 &13.1 -- 17.9 &  1974Le1 \\
$^{117}$Sn &  15.64 &  5.02 &  257.30 &13.2 -- 17.8 &  1974Le1 \\
$^{118}$Sn &  15.43 &  4.84 &  277.65 &13.1 -- 17.9 &  1974Le1 \\
$^{119}$Sn &  15.53 &  4.78 &  251.37 &13.0 -- 17.9 &  1969Fu1 \\
$^{120}$Sn &  15.37 &  5.08 &  284.00 &13.1 -- 17.9 &  1974Le1 \\
$^{122}$Sn &  15.34 &  4.73 &  266.89 &13.1 -- 18.0 &  1975Sor \\
$^{124}$Sn &  15.31 &  4.94 &  270.63 &13.1 -- 18.0 &  2003Var \\
$^{\rm nat}$Sb &  15.48 &  5.01 &  279.38 &10.2 -- 17.8 &  1971Lep \\
$^{124}$Te &  15.23 &  5.50 &  280.36 &12.0 -- 18.9 &  1976Le2 \\
$^{126}$Te &  15.15 &  5.36 &  295.55 &12.0 -- 18.9 &  1976Le2 \\
$^{128}$Te &  15.12 &  5.30 &  304.18 &12.0 -- 18.9 &  1976Le2 \\
$^{130}$Te &  15.11 &  4.98 &  319.24 &12.0 -- 18.9 &  1976Le2 \\
$^{127}$I  &  14.77 &  4.09 &  229.29 &12.0 -- 20.0 &  1999Bel \\
           &  17.30 &  3.69 &   55.30 &             & \\
$^{133}$Cs &  15.33 &  5.28 &  315.25 &12.0 -- 19.0 &  1974Le1 \\
$^{138}$Ba &  15.13 &  4.51 &  317.33 &12.1 -- 18.7 &  2016Va2 \\
$^{\rm nat}$Ba &  15.29 &  4.93 &  353.21 &10.1 -- 17.8 &  1971Be4 \\
$^{139}$La &  15.24 &  4.82 &  367.47 &12.0 -- 18.9 &  1972De1 \\
$^{140}$Ce &  15.03 &  4.39 &  381.89 &12.0 -- 18.9 &  1976Le2 \\
$^{142}$Ce &  14.85 &  5.08 &  331.25 &12.0 -- 18.9 &  1976Le2 \\
$^{141}$Pr &  15.19 &  4.23 &  342.68 &12.1 -- 16.9 &  1987Ber \\
$^{142}$Nd &  14.95 &  4.46 &  360.54 &13.1 -- 18.0 &  2003Var \\
$^{143}$Nd &  15.00 &  4.73 &  347.60 &12.0 -- 19.0 &  1971Ca1 \\
$^{144}$Nd &  15.04 &  5.25 &  315.51 &12.0 -- 18.9 &  1971Ca1 \\
$^{145}$Nd &  14.94 &  6.27 &  295.27 &12.0 -- 18.9 &  1971Ca1 \\
$^{146}$Nd &  14.73 &  5.74 &  309.04 &12.0 -- 18.9 &  1971Ca1 \\
$^{148}$Nd &  12.78 &  4.03 &  110.23 &10.8 -- 18.6 &  1971Ca1 \\
           &  15.49 &  5.22 &  215.89 &             & \\
$^{150}$Nd &  12.30 &  3.38 &  175.75 &10.8 -- 18.6 &  1971Ca1 \\
           &  16.03 &  5.12 &  220.50 &             & \\
$^{144}$Sm &  15.31 &  4.42 &  381.69 &12.1 -- 18.9 &  1974Ca5 \\
$^{148}$Sm &  14.82 &  5.06 &  337.78 &12.1 -- 18.9 &  1974Ca5 \\
$^{150}$Sm &  14.59 &  5.92 &  310.73 &12.1 -- 18.9 &  1974Ca5 \\
$^{152}$Sm &  12.39 &  2.99 &  176.80 &10.9 -- 18.8 &  1974Ca5 \\
           &  15.73 &  5.15 &  232.25 &             & \\
$^{154}$Sm &  12.17 &  2.80 &  181.90 &10.9 -- 18.6 &  1981Gur \\
           &  15.63 &  5.89 &  209.70 &             & \\
$^{151}$Eu &  13.88 &  4.69 &  254.17 &10.2 -- 18.0 &  1971Vas \\
           &  14.45 &  0.75 &   65.45 &             & \\
$^{153}$Eu &  12.33 &  2.77 &  155.86 &10.9 -- 18.7 &  1969Be8 \\
           &  15.78 &  5.76 &  219.21 &             & \\
$^{152}$Gd &  11.79 &  3.03 &  145.74 &10.2 -- 18.0 &  1971Vas \\
           &  14.72 &  3.16 &  253.50 &             & \\
$^{154}$Gd &  11.97 &  2.65 &  163.89 &10.2 -- 18.0 &  1971Vas \\
           &  15.05 &  3.35 &  245.22 &             & \\
$^{156}$Gd &  12.46 &  3.14 &  230.03 &10.9 -- 18.7 &  1981Gur \\
           &  15.79 &  4.56 &  215.94 &             & \\
$^{158}$Gd &  11.86 &  2.94 &  182.05 &10.2 -- 18.0 &  1971Vas \\
           &  15.16 &  3.31 &  245.62 &             & \\
$^{160}$Gd &  12.28 &  3.33 &  234.33 &10.9 -- 18.8 &  2003Var \\
           &  16.06 &  5.12 &  247.23 &             & \\
\hline
\end{tabular}
\end{center}

\noindent TABLE~\ref{table:GDRAtlas1} continued.
\begin{center}
\begin{tabular}{p{1cm}|p{1cm}p{1cm}p{1cm}|c|c}
\hline
           & $E$ & $\Gamma$ & $\sigma$ & Range & Ref. \\
           & [MeV] & [MeV]      & [mb]      & [MeV] & \\
\hline
$^{159}$Tb &  12.42 &  2.71 &  171.70 &11.1 -- 19.0 &  1976Gor \\
           &  15.86 &  5.98 &  295.78 &             & \\
$^{165}$Ho &  12.38 &  2.59 &  220.67 &11.1 -- 18.7 &  1981Gur \\
           &  15.48 &  4.05 &  226.69 &             & \\
$^{168}$Er &  12.09 &  3.66 &  237.40 &10.9 -- 18.8 &  1981Gur \\
           &  15.54 &  3.99 &  252.64 &             & \\
$^{174}$Yb &  12.50 &  3.41 &  339.31 &10.9 -- 18.7 &  1981Gur \\
           &  15.68 &  3.74 &  291.85 &             & \\
$^{175}$Lu &  12.32 &  2.59 &  218.42 &11.0 -- 18.7 &  1969Be6 \\
           &  15.47 &  4.64 &  284.83 &             & \\
$^{176}$Hf &  12.34 &  2.77 &  279.26 &10.9 -- 17.9 &  1977Gor \\
           &  15.67 &  4.72 &  275.10 &             & \\
$^{178}$Hf &  12.42 &  4.89 &  363.72 &10.8 -- 18.6 &  1981Gur \\
           &  15.70 &  3.13 &  234.94 &             & \\
$^{180}$Hf &  12.55 &  4.71 &  351.75 &10.8 -- 18.7 &  1981Gur \\
           &  15.61 &  3.27 &  243.23 &             & \\
$^{181}$Ta &  12.19 &  2.93 &  262.35 &10.8 -- 18.6 &  1981Gur \\
           &  14.99 &  5.13 &  317.52 &             & \\
$^{182}$W  &  11.98 &  3.91 &  283.99 &11.0 -- 18.8 &  1981Gur \\
           &  14.94 &  5.16 &  259.40 &             & \\
$^{184}$W  &  11.92 &  4.52 &  347.68 &11.0 -- 17.6 &  1981Gur \\
           &  15.05 &  3.87 &  233.17 &             & \\
$^{186}$W  &  13.04 &  6.60 &  410.29 &10.9 -- 18.7 &  1981Gur \\
           &  14.89 &  2.12 &   69.05 &             & \\
$^{185}$Re &  12.59 &  2.22 &  221.75 &10.2 -- 18.0 &  1973Gor \\
           &  15.17 &  6.02 &  328.79 &             & \\
$^{\rm nat}$Re &  14.11 &  6.52 &  470.27 &10.2 -- 18.0 &  1975Vey \\
$^{186}$Os &  12.73 &  2.34 &  214.24 &11.1 -- 18.9 &  2015Var \\
           &  14.73 &  4.08 &  355.57 &             & \\
$^{188}$Os &  12.81 &  2.83 &  281.11 &10.8 -- 18.9 &  2014Var \\
           &  14.93 &  3.83 &  388.79 &             & \\
$^{189}$Os &  12.92 &  3.07 &  329.04 &10.8 -- 18.9 &  2014Var \\
           &  15.02 &  3.94 &  299.55 &             & \\
$^{190}$Os &  13.10 &  3.34 &  372.87 &10.8 -- 18.9 &  2015Var \\
           &  15.12 &  3.80 &  247.98 &             & \\
$^{192}$Os &  12.59 &  2.13 &  173.74 &10.8 -- 18.9 &  2015Var \\
           &  14.32 &  4.60 &  426.27 &             & \\
$^{191}$Ir &  12.72 &  2.08 &  183.14 &11.0 -- 16.8 &  1978Go1 \\
           &  14.21 &  5.27 &  382.41 &             & \\
$^{193}$Ir &  12.86 &  1.90 &  229.81 &11.0 -- 16.8 &  1978Go1 \\
           &  14.30 &  5.62 &  356.07 &             & \\
$^{\rm nat}$Ir &  13.77 &  4.86 &  495.41 &10.2 -- 18.0 &  1975Vey \\
$^{194}$Pt &  13.42 &  3.61 &  453.02 &11.0 -- 17.8 &  1978Go1 \\
           &  15.97 &  6.16 &  111.63 &             & \\
$^{195}$Pt &  12.99 &  2.92 &  357.53 &11.0 -- 17.8 &  1978Go1 \\
           &  14.90 &  4.85 &  253.95 &             & \\
$^{196}$Pt &  13.28 &  3.10 &  345.43 &11.0 -- 17.8 &  1978Go1 \\
           &  14.81 &  7.51 &  192.99 &             & \\
$^{198}$Pt &  13.56 &  4.88 &  528.76 &11.0 -- 17.8 &  1978Go1 \\
$^{197}$Au &  13.58 &  5.32 &  522.88 &11.1 -- 17.0 &  1981Gur \\
$^{203}$Tl &  14.04 &  3.77 &  436.99 & 9.0 -- 17.9 &  1970Ant \\
$^{205}$Tl &  14.46 &  2.95 &  478.97 &10.5 -- 17.9 &  1970Ant \\
$^{206}$Pb &  13.58 &  3.83 &  512.45 &10.0 -- 17.0 &  1964Ha2 \\
$^{207}$Pb &  13.55 &  3.95 &  479.80 &10.0 -- 17.0 &  1964Ha2 \\
\hline
\end{tabular}
\end{center}

\noindent TABLE~\ref{table:GDRAtlas1} continued.
\begin{center}
\begin{tabular}{p{1cm}|p{1cm}p{1cm}p{1cm}|c|c}
\hline
           & $E$ & $\Gamma$ & $\sigma$ & Range & Ref. \\
           & [MeV] & [MeV]      & [mb]      & [MeV] & \\
\hline
$^{208}$Pb &  13.37 &  3.93 &  645.49 &10.9 -- 18.8 &  2003Var \\
$^{\rm nat}$Pb &  13.48 &  3.99 &  637.69 &12.1 -- 16.9 &  1985Ahr \\
$^{209}$Bi &  13.79 &  5.02 &  588.63 &10.9 -- 18.3 &  1976Gu2 \\
$^{232}$Th &  10.86 &  2.86 &  230.28 &10.2 -- 18.3 &  1976Gu1 \\
           &  13.74 &  4.74 &  382.21 &             & \\
$^{233}$U  &  10.98 &  1.53 &  140.38 & 9.4 -- 17.8 &  1986Be2 \\
           &  13.30 &  5.71 &  439.10 &             & \\
$^{234}$U  &  11.18 &  2.41 &  379.61 & 9.4 -- 17.8 &  1986Be2 \\
           &  14.03 &  4.46 &  398.77 &             & \\
$^{235}$U  &  10.82 &  3.88 &  302.02 & 9.5 -- 18.4 &  1976Gu1 \\
           &  13.80 &  4.54 &  319.35 &             & \\
$^{236}$U  &  11.04 &  2.65 &  279.96 & 9.5 -- 17.8 &  1980Ca1 \\
           &  13.92 &  4.77 &  407.27 &             & \\
$^{238}$U  &  11.06 &  2.95 &  283.53 & 9.2 -- 18.8 &  1976Gu1 \\
           &  14.26 &  4.80 &  347.16 &             & \\
$^{\rm nat}$U  &  10.73 &  2.47 &  301.11 &10.2 -- 17.9 &  1985Ahr \\
           &  13.72 &  5.04 &  405.91 &             & \\
$^{237}$Np &  10.99 &  2.20 &  310.02 & 9.4 -- 17.8 &  1986Be2 \\
           &  14.08 &  4.66 &  535.89 &             & \\
$^{239}$Pu &  11.07 &  3.29 &  227.77 & 9.3 -- 18.7 &  1976Gu1 \\
           &  14.00 &  5.51 &  358.18 &             & \\
\hline\hline
\end{tabular}
\end{center}

\begin{table}[!h]
\caption{Recommended experimental GDR parameters within the 
Simplified version of the modified Lorentzian (SMLO) approach.}
\label{table:GDRAtlas2}
\begin{center}
\begin{tabular}{p{1cm}|p{1cm}p{1cm}p{1cm}|c|c}
\hline\hline
           & $E$ & $\Gamma$ & $\sigma$ & Range & Ref. \\
           & [MeV] & [MeV]      & [mb]      & [MeV] & \\
\hline
$^{  6}$Li &  23.75 &  5.33 &    3.70 &21.5 -- 27.0 &  1986Var \\
$^{  7}$Li &  20.21 & 21.81 &    3.47 &13.2 -- 25.6 &  1985Ahr \\
$^{  9}$Be &  24.18 & 10.86 &    5.09 &17.5 -- 26.0 &  1975Ahr \\
$^{ 10}$B  &  23.31 & 17.51 &    4.12 & 8.5 -- 24.9 &  1987Ahs \\
$^{ 12}$C  &  22.90 &  3.69 &   21.21 &20.1 -- 25.0 &  1969Bez \\
$^{ 13}$C  &  25.02 & 10.72 &   11.65 &14.5 -- 29.0 &  2002Ish \\
$^{ 14}$C  &  15.69 &  6.51 &    7.67 &14.5 -- 30.0 &  2002Ish \\
           &  26.27 &  7.11 &    7.74 &             & \\
$^{\rm nat}$C  &  23.14 &  4.24 &   19.38 &19.5 -- 25.6 &  1985Ahr \\
$^{ 14}$N  &  23.19 &  7.09 &   22.95 &18.2 -- 28.0 &  1969Bez \\
$^{ 15}$N  &  26.29 & 19.45 &   12.85 &14.5 -- 28.0 &  2002Ish \\
$^{ 16}$O  &  23.78 &  5.67 &   27.70 &18.1 -- 26.0 &  1975Ahr \\
$^{ 17}$O  &  23.46 &  5.95 &   21.29 &18.5 -- 26.5 &  2002Ish \\
$^{ 18}$O  &  19.12 &  2.31 &    5.51 &18.5 -- 26.0 &  2002Ish \\
           &  24.19 &  5.24 &   13.38 &             & \\
$^{\rm nat}$O  &  23.64 &  5.88 &   27.13 &18.9 -- 27.9 &  1985Ahr \\
$^{ 19}$F  &  24.09 & 24.69 &   14.94 &10.0 -- 24.0 &  2002Ish \\
$^{ 23}$Na &  17.57 &  3.75 &   13.62 &14.2 -- 23.0 &  1981Ish \\
           &  21.26 &  4.35 &   25.45 &             & \\
$^{ 24}$Mg &  19.46 &  2.80 &   22.51 &16.5 -- 27.0 &  1966Dol \\
           &  24.30 &  9.80 &   24.18 &             & \\
$^{ 25}$Mg &  22.73 &  8.44 &   33.04 & 9.0 -- 24.2 &  2002Ish \\
$^{ 26}$Mg &  17.39 &  2.36 &   17.88 &16.1 -- 26.5 &  2003Var \\
           &  23.83 &  7.57 &   38.38 &             & \\
$^{\rm nat}$Mg &  22.32 &  8.72 &   21.08 &15.1 -- 26.6 &  1965Wyc \\
\hline
\end{tabular}
\end{center}
\end{table}

\pagebreak
\noindent TABLE~\ref{table:GDRAtlas2} continued.
\begin{center}
\begin{tabular}{p{1cm}|p{1cm}p{1cm}p{1cm}|c|c}
\hline
           & $E$ & $\Gamma$ & $\sigma$ & Range & Ref. \\
           & [MeV] & [MeV]      & [mb]      & [MeV] & \\
\hline
$^{ 27}$Al &  21.00 &  8.62 &   37.64 &14.0 -- 24.1 &  1985Ahr \\
$^{ 28}$Si &  19.88 &  2.91 &   38.41 &16.7 -- 23.0 &  2003Var \\
           &  21.84 &  3.39 &   36.60 &             & \\
$^{ 29}$Si &  20.97 &  6.88 &   37.83 &14.2 -- 23.0 &  2002Ish \\
$^{ 30}$Si &  21.32 &  9.16 &   28.52 &14.2 -- 23.0 &  2002Ish \\
$^{\rm nat}$Si &  20.45 &  4.85 &   51.22 &16.4 -- 25.8 &  1975Ahr \\
           &  25.24 &  2.10 &    8.02 &             & \\
$^{ 32}$S  &  19.57 &  5.10 &   34.93 &14.7 -- 23.0 &  1968Dol \\
$^{ 34}$S  &  21.66 & 13.43 &   47.29 &12.0 -- 25.0 &  1986Ass \\
$^{\rm nat}$S  &  20.42 &  5.74 &   47.22 &17.2 -- 23.6 &  1965Wyc \\
$^{ 40}$Ar &  19.91 & 11.50 &   54.19 &10.5 -- 25.0 &  2002Ish \\
$^{\rm nat}$K  &  21.28 &  7.40 &   22.28 &16.0 -- 25.9 &  1974Ve1 \\
$^{ 40}$Ca &  20.72 &  6.44 &  104.09 &17.2 -- 23.7 &  1966Dol \\
$^{ 42}$Ca &  20.53 &  9.75 &   68.87 &15.2 -- 23.0 &  2003Ero \\
$^{ 44}$Ca &  20.08 & 12.45 &   62.19 &15.5 -- 26.0 &  2003Ero \\
$^{ 48}$Ca &  19.90 &  6.42 &  104.95 &17.9 -- 21.6 &  1987OKe \\
$^{\rm nat}$Ca &  20.09 &  5.07 &   93.50 &15.1 -- 24.0 &  1975Ahr \\
$^{ 46}$Ti &  19.95 &  7.99 &   76.44 &13.2 -- 25.0 &  2002Ish \\
$^{ 48}$Ti &  20.13 &  9.98 &   61.66 &14.5 -- 23.0 &  2002Ish \\
$^{ 51}$V  &  18.18 &  5.29 &   63.92 &14.1 -- 22.9 &  1962Fu1 \\
           &  21.37 &  3.36 &   18.23 &             & \\
$^{ 52}$Cr &  19.16 &  6.70 &   79.20 &14.3 -- 23.0 &  2002Ish \\
$^{ 55}$Mn &  16.43 &  2.91 &   21.84 &14.0 -- 23.0 &  1979Al2 \\
           &  20.13 & 11.28 &   51.12 &             & \\
$^{ 54}$Fe &  19.39 &  5.69 &  145.35 &16.0 -- 23.0 &  1978Nor \\
$^{ 59}$Co &  16.44 &  2.42 &   20.11 &14.0 -- 20.9 &  1979Al2 \\
           &  18.68 &  8.63 &   59.76 &             & \\
$^{ 58}$Ni &  18.87 &  6.16 &   86.03 &14.1 -- 22.0 &  2003Var \\
$^{ 60}$Ni &  16.59 &  3.17 &   39.23 &12.1 -- 21.0 &  2003Var \\
           &  19.39 &  7.82 &   77.98 &             & \\
$^{ 63}$Cu &  16.79 &  5.76 &   87.60 &14.0 -- 21.0 &  2003Var \\
           &  20.37 &  4.56 &   35.92 &             & \\
$^{ 65}$Cu &  17.23 &  8.38 &   86.10 &14.2 -- 21.0 &  2003Var \\
$^{\rm nat}$Cu &  17.91 &  5.24 &   99.60 &14.4 -- 24.9 &  1965Wyc \\
$^{ 64}$Zn &  16.37 &  3.84 &   48.04 &14.0 -- 20.8 &  1976Ca1 \\
           &  19.49 &  5.82 &   47.80 &             & \\
$^{ 65}$Zn &  16.06 &  2.26 &   21.02 &12.0 -- 21.0 &  2003Rod \\
           &  18.87 &  8.62 &   60.27 &             & \\
$^{ 70}$Ge &  15.31 &  7.19 &  151.53 &10.0 -- 20.0 &  1975Mcc \\
$^{ 72}$Ge &  17.85 &  6.22 &  159.50 &10.0 -- 24.0 &  1975Mcc \\
$^{ 74}$Ge &  14.42 &  2.46 &   32.43 &13.1 -- 20.8 &  1976Ca1 \\
           &  17.47 &  8.37 &   96.06 &             & \\
$^{ 76}$Ge &  15.42 &  3.62 &   38.51 &13.1 -- 20.8 &  1976Ca1 \\
           &  18.69 & 13.81 &   82.19 &             & \\
$^{ 75}$As &  15.25 &  4.73 &   59.84 &13.1 -- 20.9 &  1969Be1 \\
           &  18.16 &  6.73 &   59.05 &             & \\
$^{ 76}$Se &  15.86 &  6.50 &  150.72 &13.1 -- 19.7 &  1978Gur \\
$^{ 78}$Se &  15.23 &  4.67 &   82.67 &13.1 -- 20.8 &  1976Ca1 \\
           &  18.76 &  5.68 &   66.04 &             & \\
$^{ 80}$Se &  16.84 &  7.45 &  135.22 &13.1 -- 20.0 &  2016Va1 \\
$^{ 82}$Se &  16.13 &  5.81 &  175.20 &13.1 -- 19.9 &  1978Gur \\
$^{\rm nat}$Rb &  16.92 &  4.93 &  185.68 &10.6 -- 17.9 &  1971Lep \\
$^{\rm nat}$Sr &  16.99 &  5.03 &  201.18 &10.9 -- 17.9 &  1971Lep \\
$^{ 89}$Y  &  16.82 &  4.42 &  222.48 &14.0 -- 19.0 &  1971Le1 \\
\hline
\end{tabular}
\end{center}

\noindent TABLE~\ref{table:GDRAtlas2} continued.
\begin{center}
\begin{tabular}{p{1cm}|p{1cm}p{1cm}p{1cm}|c|c}
\hline
           & $E$ & $\Gamma$ & $\sigma$ & Range & Ref. \\
           & [MeV] & [MeV]      & [mb]      & [MeV] & \\
\hline
$^{ 90}$Zr &  16.90 &  4.13 &  251.98 &14.9 -- 18.5 &  2003Var \\
$^{ 91}$Zr &  16.64 &  4.32 &  181.76 &14.0 -- 18.9 &  1967Be2 \\
$^{ 92}$Zr &  16.34 &  4.70 &  165.00 &14.0 -- 18.9 &  1967Be2 \\
$^{ 94}$Zr &  16.35 &  5.52 &  157.58 &14.0 -- 18.9 &  1967Be2 \\
$^{ 93}$Nb &  16.70 &  5.18 &  198.63 &14.0 -- 19.0 &  1971Le1 \\
$^{ 92}$Mo &  17.28 &  5.05 &  236.19 &14.4 -- 19.0 &  2003Var \\
$^{ 94}$Mo &  16.73 &  6.04 &  186.31 & 9.6 -- 18.9 &  1974Be3 \\
$^{ 96}$Mo &  16.42 &  6.52 &  182.00 &13.2 -- 17.0 &  1974Be3 \\
$^{ 98}$Mo &  15.96 &  6.17 &  186.76 &13.2 -- 18.9 &  1974Be3 \\
$^{100}$Mo &  16.02 &  8.44 &  167.02 &12.1 -- 20.0 &  1974Be3 \\
$^{103}$Rh &  16.59 &  8.44 &  187.85 &13.1 -- 19.0 &  2003Var \\
$^{108}$Pd &  16.15 &  8.26 &  169.74 &10.0 -- 19.0 &  1969Dea \\
           &  18.09 &  1.65 &   45.85 &             & \\
$^{\rm nat}$Pd &  16.24 &  7.63 &  195.79 &10.2 -- 17.8 &  1971Lep \\
$^{107}$Ag &  16.05 &  7.51 &  179.36 & 9.5 -- 19.0 &  1969Ish \\
$^{109}$Ag &  13.74 &  3.81 &   90.32 &13.1 -- 19.0 &  1969Ish \\
           &  16.76 &  4.17 &  103.59 &             & \\
$^{\rm nat}$Ag &  16.39 &  8.03 &  194.95 &13.2 -- 18.9 &  1971Lep \\
$^{\rm nat}$Cd &  16.03 &  6.69 &  222.13 &10.2 -- 17.8 &  1971Lep \\
$^{115}$In &  15.91 &  6.00 &  242.23 &13.2 -- 17.8 &  1974Le1 \\
$^{112}$Sn &  15.82 &  5.87 &  254.14 &10.9 -- 18.0 &  1974Sor \\
$^{114}$Sn &  16.08 &  6.74 &  238.62 &13.1 -- 18.0 &  1975Sor \\
$^{116}$Sn &  15.69 &  5.29 &  267.45 &13.1 -- 17.9 &  1974Le1 \\
$^{117}$Sn &  15.77 &  5.29 &  254.71 &13.2 -- 17.8 &  1974Le1 \\
$^{118}$Sn &  15.55 &  5.02 &  276.03 &13.1 -- 17.9 &  1974Le1 \\
$^{119}$Sn &  15.65 &  5.09 &  248.89 &13.0 -- 17.9 &  1969Fu1 \\
$^{120}$Sn &  15.50 &  5.26 &  282.55 &13.1 -- 17.9 &  1974Le1 \\
$^{122}$Sn &  15.45 &  4.98 &  262.09 &13.1 -- 18.0 &  1975Sor \\
$^{124}$Sn &  15.41 &  5.06 &  269.85 &13.1 -- 18.0 &  2003Var \\
$^{\rm nat}$Sb &  15.62 &  5.59 &  273.38 &10.2 -- 17.8 &  1971Lep \\
$^{124}$Te &  15.36 &  5.81 &  276.72 &12.0 -- 18.9 &  1976Le2 \\
$^{126}$Te &  15.27 &  5.62 &  292.05 &12.0 -- 18.9 &  1976Le2 \\
$^{128}$Te &  15.23 &  5.55 &  300.47 &12.0 -- 18.9 &  1976Le2 \\
$^{130}$Te &  15.21 &  5.19 &  316.19 &12.0 -- 18.9 &  1976Le2 \\
$^{127}$I  &  15.05 &  4.76 &  242.61 &12.0 -- 20.0 &  1999Bel \\
           &  17.33 &  1.83 &   22.56 &             & \\
$^{133}$Cs &  15.44 &  5.50 &  312.95 &12.0 -- 19.0 &  1974Le1 \\
$^{138}$Ba &  15.16 &  4.68 &  312.33 &12.1 -- 18.7 &  2016Va2 \\
$^{\rm nat}$Ba &  15.46 &  5.57 &  344.80 &10.1 -- 17.8 &  1971Be4 \\
$^{139}$La &  15.30 &  5.11 &  360.94 &12.0 -- 18.9 &  1972De1 \\
$^{140}$Ce &  15.09 &  4.51 &  378.93 &12.0 -- 18.9 &  1976Le2 \\
$^{142}$Ce &  14.95 &  5.24 &  328.39 &12.0 -- 18.9 &  1976Le2 \\
$^{141}$Pr &  15.33 &  4.49 &  340.06 &12.1 -- 16.9 &  1987Ber \\
$^{142}$Nd &  15.02 &  4.41 &  361.62 &13.1 -- 18.0 &  2003Var \\
$^{143}$Nd &  15.08 &  4.99 &  341.49 &12.0 -- 19.0 &  1971Ca1 \\
$^{144}$Nd &  15.17 &  5.56 &  311.38 &12.0 -- 18.9 &  1971Ca1 \\
$^{145}$Nd &  15.14 &  6.75 &  289.80 &12.0 -- 18.9 &  1971Ca1 \\
$^{146}$Nd &  14.88 &  6.04 &  304.19 &12.0 -- 18.9 &  1971Ca1 \\
$^{148}$Nd &  13.34 &  5.41 &  156.93 &10.8 -- 18.6 &  1971Ca1 \\
           &  15.79 &  4.57 &  164.60 &             & \\
$^{150}$Nd &  12.49 &  3.93 &  194.54 &10.8 -- 18.6 &  1971Ca1 \\
           &  16.23 &  4.85 &  197.33 &             & \\
$^{144}$Sm &  15.37 &  4.53 &  379.27 &12.1 -- 18.9 &  1974Ca5 \\
$^{148}$Sm &  14.91 &  5.15 &  336.69 &12.1 -- 18.9 &  1974Ca5 \\
\hline
\end{tabular}
\end{center}

\noindent TABLE~\ref{table:GDRAtlas2} continued.
\begin{center}
\begin{tabular}{p{1cm}|p{1cm}p{1cm}p{1cm}|c|c}
\hline
           & $E$ & $\Gamma$ & $\sigma$ & Range & Ref. \\
           & [MeV] & [MeV]      & [mb]      & [MeV] & \\
\hline
$^{150}$Sm &  14.76 &  6.01 &  309.77 &12.1 -- 18.9 &  1974Ca5 \\
$^{152}$Sm &  12.56 &  3.53 &  202.99 &10.9 -- 18.8 &  1974Ca5 \\
           &  15.97 &  4.77 &  206.96 &             & \\
$^{154}$Sm &  12.31 &  3.27 &  208.97 &10.9 -- 18.6 &  1981Gur \\
           &  15.95 &  5.53 &  186.51 &             & \\
$^{151}$Eu &  13.43 &  5.18 &  168.67 &10.2 -- 18.0 &  1971Vas \\
           &  14.52 &  2.28 &  158.70 &             & \\
$^{153}$Eu &  12.47 &  3.26 &  180.63 &10.9 -- 18.7 &  1969Be8 \\
           &  16.07 &  5.48 &  198.64 &             & \\
$^{152}$Gd &  11.96 &  3.49 &  161.00 &10.2 -- 18.0 &  1971Vas \\
           &  14.75 &  2.87 &  234.24 &             & \\
$^{154}$Gd &  12.09 &  3.08 &  176.27 &10.2 -- 18.0 &  1971Vas \\
           &  15.10 &  3.09 &  227.94 &             & \\
$^{156}$Gd &  12.63 &  3.61 &  252.40 &10.9 -- 18.7 &  1981Gur \\
           &  15.97 &  4.10 &  188.83 &             & \\
$^{158}$Gd &  12.00 &  3.37 &  193.78 &10.2 -- 18.0 &  1971Vas \\
           &  15.19 &  2.97 &  231.63 &             & \\
$^{160}$Gd &  12.47 &  3.85 &  258.30 &10.9 -- 18.8 &  2003Var \\
           &  16.26 &  4.69 &  218.92 &             & \\
$^{159}$Tb &  12.55 &  3.23 &  205.41 &11.1 -- 19.0 &  1976Gor \\
           &  16.16 &  5.74 &  270.81 &             & \\
$^{165}$Ho &  12.47 &  2.84 &  238.17 &11.1 -- 18.7 &  1981Gur \\
           &  15.59 &  3.78 &  209.10 &             & \\
$^{168}$Er &  12.33 &  4.18 &  258.17 &10.9 -- 18.8 &  1981Gur \\
           &  15.65 &  3.58 &  222.84 &             & \\
$^{174}$Yb &  12.73 &  3.95 &  367.77 &10.9 -- 18.7 &  1981Gur \\
           &  15.80 &  3.18 &  249.27 &             & \\
$^{175}$Lu &  12.44 &  2.99 &  245.80 &11.0 -- 18.7 &  1969Be6 \\
           &  15.65 &  4.32 &  260.78 &             & \\
$^{176}$Hf &  12.46 &  3.13 &  303.74 &10.9 -- 17.9 &  1977Gor \\
           &  15.86 &  4.44 &  247.79 &             & \\
$^{178}$Hf &  12.59 &  4.95 &  372.55 &10.8 -- 18.6 &  1981Gur \\
           &  15.65 &  3.00 &  218.91 &             & \\
$^{180}$Hf &  12.74 &  4.93 &  364.50 &10.8 -- 18.7 &  1981Gur \\
           &  15.59 &  3.04 &  220.92 &             & \\
$^{181}$Ta &  12.36 &  3.41 &  311.29 &10.8 -- 18.6 &  1981Gur \\
           &  15.26 &  4.71 &  273.41 &             & \\
$^{182}$W  &  13.08 &  7.29 &  380.79 &11.0 -- 18.8 &  1981Gur \\
           &  15.20 &  1.53 &   99.76 &             & \\
$^{184}$W  &  12.27 &  5.17 &  373.92 &11.0 -- 17.6 &  1981Gur \\
           &  15.10 &  3.15 &  186.15 &             & \\
$^{186}$W  &  13.01 &  6.25 &  381.26 &10.9 -- 18.7 &  1981Gur \\
           &  14.81 &  2.84 &  122.86 &             & \\
$^{185}$Re &  12.89 &  3.48 &  329.91 &10.2 -- 18.0 &  1973Gor \\
           &  15.84 &  4.37 &  252.98 &             & \\
$^{\rm nat}$Re &  14.30 &  7.05 &  462.90 &10.2 -- 18.0 &  1975Vey \\
$^{186}$Os &  13.00 &  3.09 &  309.47 &11.1 -- 18.9 &  2015Var \\
           &  15.09 &  3.39 &  274.49 &             & \\
$^{188}$Os &  13.14 &  3.59 &  382.50 &10.8 -- 18.9 &  2014Var \\
           &  15.22 &  2.98 &  294.24 &             & \\
$^{189}$Os &  13.18 &  3.64 &  416.27 &10.8 -- 18.9 &  2014Var \\
           &  15.32 &  2.95 &  214.75 &             & \\
$^{190}$Os &  13.21 &  3.59 &  427.44 &10.8 -- 18.9 &  2015Var \\
           &  15.28 &  2.93 &  206.88 &             & \\
\hline
\end{tabular}
\end{center}

\noindent TABLE~\ref{table:GDRAtlas2} continued.
\begin{center}
\begin{tabular}{p{1cm}|p{1cm}p{1cm}p{1cm}|c|c}
\hline
           & $E$ & $\Gamma$ & $\sigma$ & Range & Ref. \\
           & [MeV] & [MeV]      & [mb]      & [MeV] & \\
\hline
$^{192}$Os &  13.08 &  3.61 &  365.35 &10.8 -- 18.9 &  2015Var \\
           &  14.98 &  3.68 &  249.26 &             & \\
$^{191}$Ir &  13.16 &  3.75 &  418.50 &11.0 -- 16.8 &  1978Go1 \\
           &  15.24 &  3.69 &  170.79 &             & \\
$^{193}$Ir &  12.85 &  1.64 &  160.61 &11.0 -- 16.8 &  1978Go1 \\
           &  14.18 &  5.94 &  393.43 &             & \\
$^{\rm nat}$Ir &  13.81 &  4.97 &  492.98 &10.2 -- 18.0 &  1975Vey \\
$^{194}$Pt &  13.66 &  4.29 &  499.56 &11.0 -- 17.8 &  1978Go1 \\
           &  16.70 &  4.55 &   34.46 &             & \\
$^{195}$Pt &  13.28 &  3.71 &  464.49 &11.0 -- 17.8 &  1978Go1 \\
           &  15.44 &  3.56 &  137.09 &             & \\
$^{196}$Pt &  13.38 &  2.87 &  235.00 &11.0 -- 17.8 &  1978Go1 \\
           &  14.18 &  6.94 &  275.78 &             & \\
$^{198}$Pt &  13.62 &  4.83 &  533.11 &11.0 -- 17.8 &  1978Go1 \\
$^{197}$Au &  13.72 &  5.43 &  522.61 &11.1 -- 17.0 &  1981Gur \\
$^{203}$Tl &  14.06 &  3.95 &  435.91 & 9.0 -- 17.9 &  1970Ant \\
$^{205}$Tl &  14.47 &  2.93 &  482.88 &10.5 -- 17.9 &  1970Ant \\
$^{206}$Pb &  13.61 &  4.01 &  504.02 &10.0 -- 17.0 &  1964Ha2 \\
$^{207}$Pb &  13.57 &  4.22 &  467.03 &10.0 -- 17.0 &  1964Ha2 \\
$^{208}$Pb &  13.34 &  3.64 &  662.00 &10.9 -- 18.8 &  2003Var \\
$^{\rm nat}$Pb &  13.58 &  4.05 &  635.27 &12.1 -- 16.9 &  1985Ahr \\
$^{209}$Bi &  13.87 &  5.04 &  591.22 &10.9 -- 18.3 &  1976Gu2 \\
$^{232}$Th &  11.10 &  3.51 &  285.95 &10.2 -- 18.3 &  1976Gu1 \\
           &  13.97 &  4.22 &  334.57 &             & \\
$^{233}$U  &  11.01 &  2.23 &  209.80 & 9.4 -- 17.8 &  1986Be2 \\
           &  13.68 &  5.38 &  403.97 &             & \\
$^{234}$U  &  11.30 &  2.83 &  428.01 & 9.4 -- 17.8 &  1986Be2 \\
           &  14.23 &  4.04 &  353.66 &             & \\
$^{235}$U  &  11.11 &  4.52 &  345.36 & 9.5 -- 18.4 &  1976Gu1 \\
           &  13.92 &  3.88 &  269.50 &             & \\
$^{236}$U  &  11.25 &  3.34 &  340.25 & 9.5 -- 17.8 &  1980Ca1 \\
           &  14.18 &  4.25 &  351.15 &             & \\
$^{238}$U  &  11.24 &  3.54 &  324.58 & 9.2 -- 18.8 &  1976Gu1 \\
           &  14.45 &  4.26 &  305.65 &             & \\
$^{\rm nat}$U  &  10.86 &  2.83 &  353.45 &10.2 -- 17.9 &  1985Ahr \\
           &  13.96 &  4.69 &  368.18 &             & \\
$^{237}$Np &  11.10 &  2.72 &  366.61 & 9.4 -- 17.8 &  1986Be2 \\
           &  14.27 &  4.27 &  498.40 &             & \\
$^{239}$Pu &  11.47 &  4.50 &  303.03 & 9.3 -- 18.7 &  1976Gu1 \\
           &  14.37 &  4.76 &  278.70 &             & \\
\hline\hline
\end{tabular}
\end{center}

\vfill\null
\break

\end{document}